\documentclass[12pt]{article}
\pdfoutput=1
\usepackage{graphicx,psfrag,epsf,color}
\usepackage{amsmath,amssymb,amsfonts}
\usepackage{array}
\usepackage{cite}
\usepackage{rotating}
\usepackage{slashed, cancel}
\numberwithin{equation}{section}

\newcounter{MBQ}

\def\slash#1{#1 \hskip-0.45em /}
\def\Slash#1{#1 \hskip-0.59em /}

\newcommand{\be}{\begin{equation}}
\newcommand{\ee}{\end{equation}}
\newcommand{\bea}{\begin{eqnarray}}
\newcommand{\eea}{\end{eqnarray}}
\newcommand{\bi}{\begin{itemize}}
\newcommand{\ei}{\end{itemize}}
\newcommand{\ben}{\begin{enumerate}}
\newcommand{\een}{\end{enumerate}}
\newcommand{\bt}{\begin{tabular}}
\newcommand{\et}{\end{tabular}}

\newcommand{\nn}{\nonumber}

\newcommand{\nm}{n_-}
\newcommand{\np}{n_+}
\newcommand{\nms}{\slashed{n}_-}
\newcommand{\nps}{\slashed{n}_+}
\newcommand{\npshalf}{\frac{\slashed{n}_+}{2}}

\newcommand{\T}{{\bf T}}

\newcommand{\nnm}[1]{n_{#1-}}
\newcommand{\nnp}[1]{n_{#1+}}
\newcommand{\nnms}[1]{\slashed{n}_{#1-}}
\newcommand{\nnps}[1]{\slashed{n}_{#1+}}

\newcommand{\calA}{{\cal A}}
\newcommand{\als}{\alpha_s}

\def\Wc{W_{c}}

\global\long\def\order#1{\mathcal{O}\left(#1\right)}

\begin{document}
\allowdisplaybreaks

\begin{titlepage}

\begin{flushright}
{\small
TUM-HEP-1155/18\\
August 14, 2018
}
\end{flushright}

\vskip1cm
\begin{center}
{\Large \bf\boldmath Anomalous dimension of subleading-power\\[0.2cm] 
${N}$-jet operators II}
\end{center}

\vspace{0.5cm}
\begin{center}
{\sc Martin~Beneke, Mathias~Garny, Robert~Szafron, Jian~Wang} \\[6mm]
{\it Physik Department T31,\\
James-Franck-Stra\ss e~1, 
Technische Universit\"at M\"unchen,\\
D--85748 Garching, Germany\\
}
\end{center}

\vspace{0.6cm}
\begin{abstract}
\vskip0.2cm\noindent
We continue the investigation of the anomalous dimension of 
subleading-power $N$-jet operators. In this paper, we focus on the 
operators with fermion number one in each collinear direction, 
corresponding to quark (antiquark) initiated jets in QCD. We 
investigate the renormalization effects induced by the soft loop 
and compute the one-loop mixing of time-ordered products involving 
power-suppressed SCET Lagrangian insertions into $N$-jet currents 
through soft loops. We discuss fermion number conservation in 
collinear directions and provide explicit results for the collinear 
anomalous dimension matrix of the currents. The Feynman 
rules for the power-suppressed SCET interactions in the 
position-space formalism are collected in an appendix.
\end{abstract}
\end{titlepage}

\section{Introduction}
\label{sec:introduction}

The analysis of infrared (IR) divergences in QCD and gauge theories in 
general has always been a fertile field for exposing the universal 
structure of high-energy scattering amplitudes and performing all-order 
resummations of the perturbative expansion in the gauge coupling. What 
is commonly called {\em the} soft anomalous dimension of an amplitude of 
$N$ widely separated energetic particles refers in the framework of 
soft-collinear effective theory (SCET) to the simplest $N$-jet 
operator, where every jet is sourced by a single collinear gauge-invariant 
quark or gluon field~\cite{Becher:2009cu,Becher:2009qa}. The increasing 
sophistication of multi-loop calculations and the corresponding advance 
in precision has also triggered recent interest in subleading-power 
effects in the expansion in the scale $1/Q$ of the hard scattering~\cite{Bonocore:2014wua,Larkoski:2014bxa,Bonocore:2015esa,Boughezal:2015dva,Gaunt:2015pea,Bonocore:2016awd,Moult:2016fqy,Boughezal:2016zws,Feige:2017zci,DelDuca:2017twk,Ebert:2018lzn}. In a recent paper~\cite{Beneke:2017ztn,Beneke:2017mmf} we began the systematic 
investigation of the one-loop anomalous dimension matrix of these subleading 
power operators. Previous relevant work on anomalous dimensions of 
power-suppressed operators has been done in the context of heavy-quark 
decay \cite{Hill:2004if, Beneke:2005gs} and for thrust 
\cite{Freedman:2014uta,Goerke:2017lei}. All-order resummations 
of subleading-power logarithms can be found in  
Refs.~\cite{Hill:2004if, Beneke:2005gs,Penin:2014msa,Liu:2017vkm,Moult:2018jjd} covering cases with one or two collinear directions at the leading 
logarithmic order (next-to-leading for heavy quark decay to one jet). 
The purpose of our investigation is the complete analysis of 
one-loop infrared divergences of an arbitrary subleading-power 
$N$-jet operator. This provides one of the ingredients in resumming 
or generating at fixed order subleading-power next-to-leading order logarithms 
for amplitudes with any number of collinear directions or jets.

In the present paper, which follows upon Ref.~\cite{Beneke:2017ztn}, 
we extend the calculation of the one-loop anomalous dimension matrix 
from the case $|F|=2$ to those with odd $F$, where $F$ refers to the 
fermion number of the product of collinear fields in a given collinear 
direction. This contains as its simplest realizations the 
quark-antiquark initiated two-jet operators relevant to the 
subleading-power resummation of thrust and other event shape 
variables in $e^+ e^-$ annihilation, and the threshold resummation of 
Drell-Yan type processes in hadron-hadron collisions. In addition to 
the collinear renormalization kernels for the odd-fermion number 
operators in a collinear sector, we discuss and calculate for the first 
time the subleading-power soft contributions to the anomalous dimension 
matrix, which did not appear for the $|F|=2$ operators. This involves a 
new contribution, which is not of the eikonal type, and arises instead 
from the mixing of power-suppressed soft-collinear interactions in the 
SCET Lagrangian into power-suppressed $N$-jet operators with additional 
transverse derivatives or collinear fields. The anomalous dimensions 
discussed here can be used to sum subleading-power 
logarithms due to the evolution of 
the hard functions multiplying an $N$-jet operator. Physical observables 
in general contain further logarithms from the evolution of soft or 
jet functions already at the leading-logarithmic order (see the 
example of the thrust distribution~\cite{Moult:2018jjd}). The soft and 
collinear kernels in the present paper contribute to the next-to-leading 
logarithmic resummation, which yet has to be completed for an 
observable.

The outline of the paper is as follows. In Sec.~\ref{sec:setup} we set 
up notation and conventions, and discuss the form of the one-loop anomalous 
dimension matrix with respect to current and time-ordered product 
operators, and its collinear and soft one-loop 
contributions. The bulk of the paper is devoted to the calculation 
of the soft mixing contribution in Sec.~\ref{sec:soft}, and the 
collinear kernels in Sec.~\ref{sec:coll}. We summarize in 
Sec.~\ref{sec:result}. Translation rules from positive to negative $F$, 
master integrals, and some auxiliary  
expressions for collinear kernels are collected in Appendices. We 
particularly note App.~\ref{sec:feynman}, which gives a complete list 
of SCET Feynman rules in the position-space formalism 
\cite{Beneke:2002ph,Beneke:2002ni} up to the  second order in 
power-suppressed interactions and up to four-point vertices.

\section{Set-up of notation and conventions}
\label{sec:setup}

To make the paper self-contained we first review some notation 
from Ref.~\cite{Beneke:2017ztn} and then discuss the structure of the 
$N$-jet operator basis and the anomalous dimension matrix relevant 
to the present work.

\subsection{Operator basis}

We consider $N$ copies of the collinear SCET Lagrangian ${\cal L}_i$ 
\cite{Beneke:2002ni},
$i=1,\dots,N$, furnished with corresponding collinear fields $\psi_i$, 
as well as one set of soft fields $\psi_s$ that interact with all collinear 
fields and with themselves according to the soft Lagrangian ${\cal L}_s$, 
in total
\be
{\cal L}_{\rm SCET} = \sum_{i=1}^N {\cal L}_i(\psi_i,\psi_s)
+{\cal L}_s(\psi_s)\,.
\ee
The collinear fields are characterized by $N$ pairs of light-like reference 
vectors $n_{i\pm}$ with $\nnm{i}\cdot \nnp{i}=2$, 
$\nnm{i}\cdot\nnm{j}={\cal O}(1)$, defining $N$ widely separated directions. 
We are interested in current operators of the form
\be
 J = \int dt \, C(\{t_{i_k}\})\,J_s(0) 
\prod_{i=1}^N J_i(t_{i_1},t_{i_2},\dots)\,,
\ee
characterized by one soft and $N$ collinear contributions with certain 
transformation properties under soft- and collinear gauge 
transformations~\cite{Beneke:2017ztn}. The collinear contributions are 
composed of $n_i$ collinear \emph{building blocks} $\psi_{i_k}$,
\be
  J_i(t_{i_1},t_{i_2},\dots)=\prod_{k=1}^{n_i}\psi_{i_k}(t_{i_k}\nnp{i})\,,
\ee
that are offset along direction $\nnp{i}$ by an amount $t_{i_k}$ from the 
origin, which is chosen to be at the position of a hard interaction 
generating the $N$-jet current. Furthermore, $dt=\prod_{ik} dt_{i_k}$ and 
$C(\{t_{i_k}\})$ denotes a Wilson coefficient.

Up to ${\cal O}(\lambda^2)$, the soft building block $J_s(0)\equiv 1$ is 
trivial, and soft fields do not enter via $J_i$ as 
well~\cite{Beneke:2017ztn,Beneke:2017mmf} . Therefore, the most general 
current basis contains only collinear fields at this order. The complete 
operator basis can be constructed from the elementary collinear building 
blocks $\chi_i=W_i^\dag\xi_i$, its conjugate $\bar\chi_i$, and 
${\cal A}_{\perp i}^\mu=W_i^\dag [iD_{\perp i}^\mu W_i]$, as well as currents 
obtained by acting with one or several derivatives $i\partial_{\perp i}^\nu$
on the elementary building blocks~\cite{Beneke:2017ztn}. Here $\xi_i$ is the 
collinear quark field, and $W_i$ a collinear Wilson line. At the leading 
power only a single building block in each direction contributes (i.e. 
$n_i=1$ for $i=1,\dots,N$) and no extra derivatives appear. Each additional 
building block or  extra derivative supplies  a relative
power suppression of order $\lambda$.

Every collinear factor $J_i$ in the current operator $J$ can be characterized 
by its fermion number $F_i$, equal to the difference of the number of $\chi_i$
and $\bar\chi_i$ building blocks.\footnote{As we will see, $F_i$ is conserved 
under operator mixing up to ${\cal O}(\lambda^2)$ in the absence of a mass 
term for the quark field.} In the present paper we consider the case of 
collinear directions with odd fermion number, which implies $|F_i|=1,3$ at 
${\cal O}(\lambda^2)$. In the following, we write down explicitly the basis of 
operators for $F_i=1$. The leading power operator is 
$J_{\chi_\alpha}^{A0}(t_{i_1})=\chi_{i\alpha}(t_{i_1}\nnp{i})$, 
where we indicate the open Dirac index $\alpha$. 
At ${\cal O}(\lambda)$ there are two operators,
\bea
J_{\partial^\mu\chi_\alpha}^{A1}(t_{i_1}) &=& 
i\partial_{\perp i}^\mu \chi_{i\alpha}(t_{i_1}\nnp{i})\,, 
\nn\\
J_{{\cal A}^\mu\chi_\alpha}^{B1}(t_{i_1},t_{i_2}) &=& 
{\cal A}_{\perp i}^\mu(t_{i_1}\nnp{i}) \chi_{i\alpha}(t_{i_2}\nnp{i})\,.
\eea
At ${\cal O}(\lambda^2)$ there are further possibilities,
\bea
 J_{\partial^\mu\partial^\nu\chi_\alpha}^{A2}(t_{i_1}) &=& 
i\partial_{\perp i}^\mu i\partial_{\perp i}^\nu \chi_{i\alpha}(t_{i_1}\nnp{i})\,, 
\nn\\
  J_{{\cal A}^\mu\partial^\nu\chi_\alpha}^{B2}(t_{i_1},t_{i_2}) &=& 
{\cal A}_{\perp i}^\mu(t_{i_1}\nnp{i}) i\partial_{\perp i}^\nu \chi_{i\alpha}(t_{i_2}\nnp{i})\,,
\nn\\
  J_{\partial^\mu({\cal A}^\nu\chi_\alpha)}^{B2}(t_{i_1},t_{i_2}) &=& 
i\partial_{\perp i}^\mu \left({\cal A}_{\perp i}^\nu(t_{i_1}\nnp{i}) \chi_{i\alpha}(t_{i_2}\nnp{i})\right)\,,
\nn\\
  J_{{\cal A}^\mu{\cal A}^\nu\chi_\alpha}^{C2}(t_{i_1},t_{i_2},t_{i_3}) &=& 
{\cal A}_{\perp i}^\mu(t_{i_1}\nnp{i}) {\cal A}_{\perp i}^\nu(t_{i_2}\nnp{i}) \chi_{i\alpha}(t_{i_3}\nnp{i})\,,
\nn\\
  J_{\chi_\alpha\bar\chi_\beta\chi_\gamma}^{C2}(t_{i_1},t_{i_2},t_{i_3}) &=& 
\chi_{i\alpha}(t_{i_1}\nnp{i}) \bar\chi_{i\beta}(t_{i_2}\nnp{i}) \chi_{i\gamma}(t_{i_3}\nnp{i})\,.
\eea
The superscript denotes the number of building blocks ($A,B,C$ for one, two 
and three, respectively) and the power suppression relative to $A0$.

When expanding in powers of $\lambda$, time-ordered products of currents 
$J_i$ with subleading-power ${\cal O}(\lambda^n)$ Lagrangian insertions 
${\cal L}^{(n)}_{iV}$ need to be considered ($n>0$). In contrast to the 
current operators above, these insertions contain the soft field
explicitly. For each collinear sector we discriminate interactions involving 
only collinear quarks, both soft and collinear quarks, or no quarks, denoted 
by $V=\xi,\xi q,{\rm YM}$, respectively. Soft and collinear gluons 
may be contained in all three contributions.  The Lagrangians 
${\cal L}^{(n)}_{iV}$ are given in Ref.~\cite{Beneke:2002ni} for $n=1,2$, 
see also App.~\ref{sec:feynman}. Consequently, for $F_i=1$ the basis needs 
to be complemented by the time-ordered product operators 
\bea
J^{T1}_{\chi,V}(t_{i_1}) &=& 
i \int d^4 x \,T\left\{J^{A0}_{ \chi}(t_{i_1}), 
{\mathcal L}_{iV}^{(1)}(x) \right\},
\label{eq:Ti01}
\eea
at ${\cal O}(\lambda)$, and
\bea
J^{T2}_{\chi,V}(t_{i_1}) &=& 
i \int d^4 x \,T\left\{J^{A0}_{\chi}(t_{i_1}), 
{\mathcal L}_{iV}^{(2)}(x) \right\},
\nonumber \\
J^{T2}_{\partial\chi,V}(t_{i_1}) &=& 
i \int d^4 x \,T\left\{J^{A1}_{\partial \chi}(t_{i_1}), 
{\mathcal L}_{iV}^{(1)}(x) \right\},
\nonumber \\
J^{T2}_{\calA\chi,V}(t_{i_1},t_{i_2}) &=& 
i \int d^4 x \,T\left\{J^{B1}_{\calA \chi}(t_{i_1},t_{i_2}), 
{\mathcal L}_{iV}^{(1)}(x) \right\},
\nonumber \\
J^{T2}_{\chi,VW}(t_{i_1}) &=& 
\frac{i^2}{2} \int d^4 x \int d^4 y \,T\left\{J^{A0}_{\chi }(t_{i_1}), 
{\mathcal L}_{iV}^{(1)}(x),{\mathcal L}_{iW}^{(1)}(y) \right\},
\label{eq:Ti011}
\eea
at ${\cal O}(\lambda^2)$, where $V,W\in\{\xi,\xi q,{\rm YM}\}$.  
The open Dirac and Lorentz indices of the currents are left implicit here.

In momentum space associated with the collinear direction $\nnp{i}$, the 
basis operators $J_i$ depend on the total collinear momentum $P_i>0$ 
and $n_i$ momentum fractions $x_{i_k}$ that satisfy 
$\sum_{k=1}^{n_i} x_{i_k}=1$. Therefore, $J_i^{An}$, $J_i^{Bn}(x_{i_1})$ and 
$J_i^{Cn}(x_{i_1},x_{i_2})$ can be described by zero, one and two independent 
momentum fractions, respectively. Similarly, the time-ordered product 
$J^{T2}_{\calA\chi,V}(t_{i_1},t_{i_2})$ depends on one momentum fraction 
inherited from $J^{B1}_{\calA \chi}(t_{i_1},t_{i_2})$. 
In the following we write a generic $N$-jet operator in collinear momentum 
space as 
\be
  J_P(x)=\prod_{i=1}^N J_i(x_{i_1},\dots)
\ee
where $P$ collectively denotes the types of currents and/or time-ordered 
products in each of the directions labeled by $i$,\footnote{Explicitly, for 
the leading power $N$-jet operators, $P=(A0)_1 \ldots (A0)_N$ with the 
additional specification whether $(A0)_i$ refers to a quark, antiquark 
or gluon building block.} including Lorentz, Dirac 
and colour indices, and $x=\{x_{i_k}\}$ stands for the corresponding momentum 
fractions. At ${\cal O}(\lambda)$ one collinear direction, say $i$, may 
contain a current $J_i^{A1}$ or $J_i^{B1}$, or a time-ordered product 
$J_i^{T1}$, while all other directions $j\not=i$ are described by 
leading-power operators $J_j^{A0}$. At ${\cal O}(\lambda^2)$, either 
\emph{two} collinear directions, say $i$ and $j$, contain a single power 
suppression, of the form $J_i^{X1}J_j^{Y1}$ with $X,Y\in\{A,B,T\}$,
or a single direction contains a ${\cal O}(\lambda^2)$-suppressed operator 
$J_i^{X2}$ with $X\in\{A,B,C,T\}$.

\subsection{Anomalous dimension matrix}

The anomalous dimension matrix in the $\overline{\rm MS}$ scheme is 
defined by\footnote{We note a misprint in the corresponding Eq.~(32) 
in the published version of Ref.~\cite{Beneke:2017ztn}. The results for 
the $Z$-factors and and anomalous dimension are not affected by this 
misprint.} 
\be
{\bf \Gamma}=-\frac{d}{d\ln\mu}{\bf Z}\,{\bf Z}^{-1} = 
{\bf Z} \frac{d}{d\ln\mu}{\bf Z}^{-1},
\ee 
with $J_P = \sum_Q Z_{PQ} J_Q$, which implies
\be
\frac{d}{d\ln\mu} J_P =  - \sum_Q \Gamma_{PQ} J_Q
\ee
for the operators and 
\be
\frac{d}{d\ln\mu} C_P =   \sum_Q \Gamma_{QP} C_Q
\ee
for their coefficient functions.

The renormalization matrix ${\bf Z}={\bf 1}+{\bf \delta Z}$ is determined, 
at the one-loop order, by the condition
\bea
\label{eq:rencond}
{\rm finite} &=& \langle J_P(x)\rangle_{\rm 1-loop} \\
&& {} + \sum_Q \int dy\, \left[ \delta Z_{PQ}(x,y)+
\delta_{PQ}\delta(x-y)\left(\frac12 \sum_{\phi\in P}\delta Z_\phi+
\sum_{g\in P}\delta Z_g\right)\right]\langle J_Q(y)\rangle_{\rm tree}\,,
\nn
\eea
where $\delta(x-y)=\prod_i\prod_{k=2}^{n_i}\delta(x_{i_k}-y_{i_k})$ and 
$dy=\prod_i\prod_{k=2}^{n_i'}dy_{i_k}$ denotes integration over momentum 
fractions, with $n_i(n_i')$ being the number of collinear building blocks in 
direction $i$ contained in $J_P(J_Q)$.\footnote{We note that the definitions 
imply that $\int dy\,\delta(x-y)\not=1$, if $n_i'\not=n_i$. We shall 
come back to this subtlety below.} The sum over $Q$ includes the 
time-ordered product operators. We used the constraint 
$\sum_{i=1}^{n_i'}y_{i_k}=1$ to eliminate one of the momentum fractions 
(arbitrarily choosing the first one) and regard the appearing objects 
as functions of the remaining momentum fractions. We employ the 
convention that empty 
products are equal to unity to capture the case of $A$-type currents as well. 
Furthermore, $\delta Z_\phi$ and $\delta Z_g$ are the usual $\overline{\rm MS}$
renormalization factors for all fields and couplings contributing to $J_P$, 
given by 
\be
\frac12\delta Z_\chi=-\frac{\alpha_sC_F}{8\pi\epsilon}
\qquad \mbox{and} \qquad
\frac12\delta Z_A+\delta Z_{g_s}=-\frac{\alpha_sC_A}{4\pi\epsilon}
\ee 
for the collinear quark and gluon building blocks, respectively.
The one-loop renormalization matrix can be split according to
\be\label{eq:ZPQ}
\delta Z_{PQ}(x,y) = \sum_{i,j=1,\,i\not= j}^N 
\delta(x-y)\delta Z_{PQ}^{s,ij}(y) + 
\sum_{i=1}^N \delta^{[i]}(x-y)\delta Z_{PQ}^{c,i}(x,y)
\ee
where $\delta Z_{PQ}^{c,i}$ contains the divergent parts of collinear loops 
as well as field- and coupling renormalization in direction $i$, and
$\delta Z_{PQ}^{s,ij}(y)$ the divergent parts of soft loops connecting 
direction $i$ and $j$. The dependence on momentum fractions follows in each 
case from the structure of the one-loop amplitude for soft and collinear 
loops, and will be discussed in more detail below. We define the 
Dirac delta-function with respect to direction $i$ by
$\delta^{(i)}(x-y)=\prod_{k=2}^{n_i}\delta(x_{i_k}-y_{i_k})$, and with 
respect to all directions \emph{but} $i$
by $\delta^{[i]}(x-y)=\prod_{j\not=i}\delta^{(j)}(x-y)$, such that we can 
decompose $\delta(x-y)=\delta^{(i)}(x-y)\delta^{[i]}(x-y)$.

In order to extract the ultraviolet divergent part of loop amplitudes, 
we assume a small off-shellness $p_{i_k}^2$ for all external particles.
The soft and collinear contributions to the anomalous dimension are then of 
the form~\cite{Beneke:2017ztn}
\bea
&& \delta Z_{PQ}^{c,i}(x,y) =
-\delta_{PQ}\delta^{(i)}(x-y) \frac{\alpha_s}{4\pi}\sum_{k,l=1}^{n_i}
\T_{i_k}\cdot\T_{j_l}\left[\frac{2}{\epsilon^2}+
\frac{2}{\epsilon}\ln\left(\frac{\mu^2}{-p_{i_k}^2}\right)+
\delta_{lk}\frac{c_{i_k}}{\epsilon}\right]
\nn\\
&&\hspace*{2.5cm} +\,\frac{1}{\epsilon}\gamma_{PQ}^i(x,y)\,, 
\label{eq:Zc}
\\
&& \delta Z_{PQ}^{s,ij}(y) = 
-\delta_{PQ} \frac{\alpha_s}{4\pi}\sum_{k=1}^{n_i}\sum_{l=1}^{n_j}
\frac{\T_{i_k}\cdot\T_{j_l}}{2}\left[\frac{2}{\epsilon^2}+
\frac{2}{\epsilon}\ln\left(\frac{-\mu^2 y_{i_k}y_{j_l}s_{ij}}
{-p_{i_k}^2p_{j_l}^2}\right)\right] + 
\frac{1}{\epsilon}\frac{\gamma_{PQ}^{ij}(y)}{2}\,, 
\qquad\quad
\label{eq:Zs}
\eea
where $s_{ij}=\frac12(\nnm{i}\cdot\nnm{j})P_iP_j$, and $c_{i_k}=3/2$ for 
fermionic and $c_{i_k}=0$ for gluon building blocks. The last term in both 
expressions captures operator mixing, and will be discussed in 
detail in the following sections. The factors $1/2$ included in the soft 
contribution account for summation over both $i<j$ and $i>j$ in 
Eq.~\eqref{eq:ZPQ}.

When combining the soft and collinear $Z$-factor, the dependence on the 
off-shell regulators cancels, and the resulting anomalous dimension is given by
\bea\label{eq:GammaPQ}
\Gamma_{PQ}(x,y)&=&\delta_{PQ}\delta(x-y)\left[-\gamma_{\rm cusp}(\alpha_s)
\sum_{i<j}\sum_{k,l}\T_{i_k}\!\cdot\T_{j_l}
\ln\left(\frac{-s_{ij}x_{i_k}x_{j_l}}
{\mu^2}\right)\hspace*{-0.05cm}
+\hspace*{-0.05cm}\sum_i\sum_k\gamma_{i_k}(\alpha_s)\right] 
\nn\\
&& {} +2\sum_i\delta^{[i]}(x-y)\gamma_{PQ}^i(x,y) + 
2\sum_{i<j}\delta(x-y)\gamma_{PQ}^{ij}(y)\,,
\eea
where 
\be 
\gamma_{\rm cusp}(\alpha_s)=\frac{\alpha_s}{\pi}
\qquad \mbox{and} \qquad
\gamma_{i_k}(\alpha_s)=
\left\{\begin{array}{cl} 
-\displaystyle \frac{3\alpha_s C_F}{4\pi}\;\;\;& \mbox{(q)}\\[0.3cm]
\;\;0 &  \mbox{(g)}
\end{array}
\right.\
\ee
for the collinear quark (q) and gluon (g) building block. This general 
structure covers both currents and time-ordered products. 

For the further discussion of the structure of the matrix we momentarily 
restrict the indices $P,Q$ to the current operators, and label the 
time-ordered product operators that descend from current operators $P',Q'$ 
with indices  $T(P'), T(Q')$. The generic structure of the anomalous dimension 
can then be summarized as
\be
  {\bf \Gamma} = 
  \left( \begin{array}{cc}
  \Gamma_{PQ} & \Gamma_{PT(Q')}   \\ 
  \Gamma_{T(P')Q} & \Gamma_{T(P')T(Q')}
  \end{array}\right)
  =
  \left( \begin{array}{cc}
  \Gamma_{PQ} & 0   \\ 
  \Gamma_{T(P')Q} & \Gamma_{P'Q'}
  \end{array}\right)
\ee
Due to the non-renormalization property of the SCET 
Lagrangian~\cite{Beneke:2002ph}, the mixing of time-ordered products into 
themselves is directly inherited from the corresponding currents that appear 
inside of the time-ordered products, i.e. $\Gamma_{T(P')T(Q')}=\Gamma_{P'Q'}$. 
Mixing of currents into the time-ordered products is forbidden by locality, 
i.e. $\Gamma_{PT(Q')}=0$. However, time-ordered products can mix into 
currents, hence $\Gamma_{T(P')Q}\not=0$.

We now recall that the current operators do not contain soft fields, hence 
it is sufficient to consider matrix elements without external soft fields 
for the calculation of the anomalous dimension. Since the power-suppressed 
Lagrangian interactions ${\cal L}^{(n)}_{iV}$ always involve at least one 
soft field, soft fields must therefore be contracted to an 
\emph{internal} soft line, 
which exists only for soft loops. It follows that $\Gamma_{T(P')Q}$ arises 
entirely from soft loops, and is described by 
\be
  \Gamma_{T(P')Q} = 2\sum_{i<j}\delta(x-y)\gamma_{T(P')Q}^{ij}(y)\;.
\ee
Furthermore, the soft contribution $\gamma_{PQ}^{ij}(y)$ in 
Eq.~\eqref{eq:GammaPQ} vanishes for current-current 
mixing~\cite{Beneke:2017ztn}, 
so the above is the only soft mixing contribution. The collinear ($\gamma^i$)
and soft operator mixing ($\gamma^{ij}$) terms in the anomalous dimension 
matrix \eqref{eq:GammaPQ} therefore take the block forms
\be
  {\bf \gamma}^i = 
  \left( \begin{array}{cc}
  \gamma^i_{PQ} & 0   \\ 
  0 & \gamma^i_{P'Q'}
  \end{array}\right),
  \qquad
  {\bf \gamma}^{ij} =
  \left( \begin{array}{cc}
  0 & 0   \\ 
  \gamma^{ij}_{T(P')Q} & 0
  \end{array}\right)\,.
\label{eq:gammacs}
\ee
The soft, time-ordered product mixing matrix $\gamma^{ij}$ vanishes for 
operators with fermion number $F_i=\pm 2$ or 
$F_j=\pm 2$~\cite{Beneke:2017ztn}, but, as will be seen below, this is 
not the case for $F_i=\pm 1$ or $F_j=\pm 1$. This type of mixing therefore 
represents a qualitatively new feature compared to 
Ref.~\cite{Beneke:2017ztn}, discussed in detail in the following 
section.

Let us point out a subtlety concerning the dependence on collinear momentum 
fractions. The collinear one-loop renormalization matrix 
$\delta Z_{PQ}^{c,i}(x,y)$ is in general off-diagonal with respect to 
momentum fractions in direction $i$, and diagonal with respect to all other 
directions. Correspondingly, we extracted the factor $\delta^{[i]}(x-y)$ in 
Eq.~\eqref{eq:ZPQ}. The soft part $\delta Z_{PQ}^{s,ij}(y)$, on the other 
hand, is diagonal in momentum fractions for current-current mixing. For the 
mixing of time-ordered products into currents, this is strictly true only 
if the number of building blocks in $J_{T(P)}$ and $J_Q$ is equal, i.e. 
$n_i'=n_i$ and $n_j'=n_j$. As we will see below (see, e.g., 
Fig.~\ref{fig:soft_qqg_T1T1}) the cases $n_i'=n_i+1=2$ and/or $n_j'=n_j+1=2$ 
are also possible, in which case the number of collinear building 
blocks in $Q$ is larger than in $T(P')$. Then $\delta Z_{PQ}^{s,ij}(y)$ 
depends in addition on $y_{i_1}$ and/or $y_{j_1}$, which are not contained 
in $\delta(x-y)$ in this case, and are integrated over in \eqref{eq:rencond}. 
The first concrete example of this feature will appear in 
Eq.~\eqref{eq:T1xiT1xi_B1A1} below.

\section{Soft sector}
\label{sec:soft}

\begin{figure}
\begin{center}
  \includegraphics[width=0.8\textwidth]{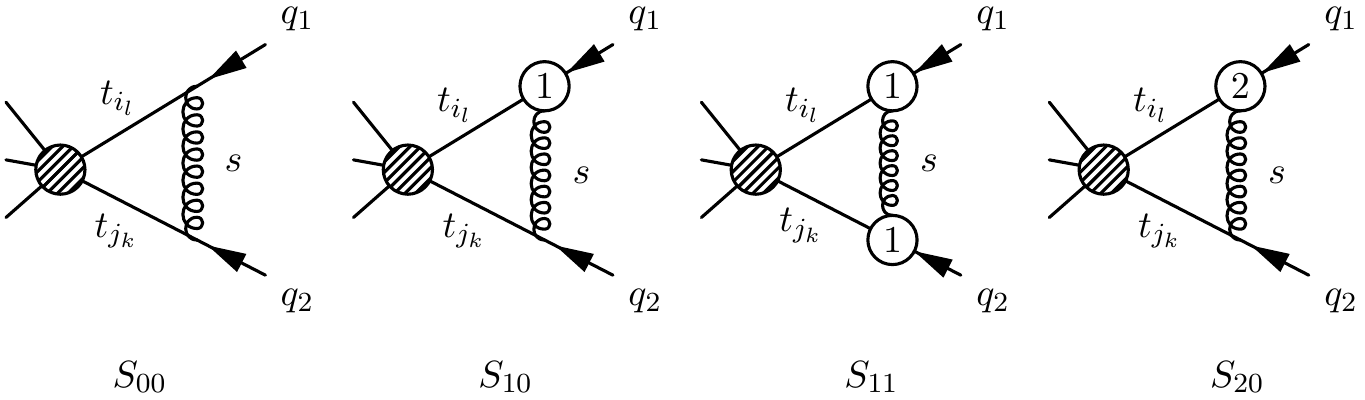}
\end{center}
\caption{\label{fig:soft_qq}
Sample diagrams for a soft gluon connecting two fermionic building blocks in 
collinear direction $i$ and $j$, respectively. The numbered vertices 
indicate insertions of vertices derived from the subleading power 
SCET Lagrangian ${\cal L}^{(n)}$ with $n=1,2$.
}
\end{figure}

In this section we discuss the contribution to the anomalous dimension from
soft loops. Its general structure at the one-loop order is described by the 
renormalization matrix $\delta Z_{PQ}^{s,ij}(y)$ defined in 
Eqs.~\eqref{eq:ZPQ} and \eqref{eq:Zs}. The first term on the right-hand side 
of Eq.~\eqref{eq:Zs} captures the contribution from soft loops for which all 
interaction vertices are derived from the leading-power SCET Lagrangian,
and was discussed in Ref.~\cite{Beneke:2017ztn}. An example is diagram 
$S_{00}$ shown in Fig.~\ref{fig:soft_qq}. In this case the power suppression 
arises from the currents themselves.
Employing colour space operator notation allows us to write the 
anomalous dimension for arbitrary combinations of
quark and gluon building blocks in a unified way, see 
Eq.~\eqref{eq:GammaPQ}. 

We will therefore focus on soft one-loop diagrams that contain at least one 
power-suppressed interaction vertex derived from ${\cal L}^{(n)}$ with 
$n=1,2$. Some examples are $S_{10}$, $S_{11}$ and $S_{20}$ shown in 
Fig.~\ref{fig:soft_qq}. Such loops may feature divergences proportional to 
current operators, which describe the mixing of time-ordered products into 
currents, captured by the second term on the right-hand side of 
Eq.~\eqref{eq:Zs}, i.e. by $\gamma^{ij}$ of Eq.~\eqref{eq:gammacs}.

Before we systematically investigate the anomalous dimension in the soft 
sector, we present an explicit example with a non-zero result. We do so to 
illustrate an unfamiliar feature of the power-suppressed SCET interactions in 
the position space formalism -- the momentum-space Feynman rules contain 
derivatives of  momentum conserving Dirac delta-functions. In our example, 
diagram $S_{11}$ shown in Fig.\,\ref{fig:soft_qq}, we demonstrate how to 
treat this objects in practical computations. The two insertions
of ${\cal L}_\xi^{(1)}$ lead to ${\cal O}(\lambda^2)$ power suppression. 
Therefore, we can take the collinear building blocks to be leading-power 
fermionic operators $\propto \chi_i\chi_j$. Diagram $S_{11}$ describes the 
possible mixing of $\left(J^{T1}_{\chi,\xi}\right)_i\left(J^{T1}_{\chi,\xi}
\right)_j$ into $A$-type currents. We consider a matrix element with two 
outgoing antiquarks,
\bea
\lefteqn{ \langle \bar q_j(q_2)\bar q_i(q_1)|\chi_{i\alpha}\chi_{j\beta}|0
\rangle_{{\rm 1-loop},\ S_{11}} }
\nn\\
&=&
\tilde\mu^{2\epsilon}\int_{l,\tilde q_1,\tilde q_2}\, 
\left(\frac{-i\nnp{i}\tilde q_1\frac{\nnms{i}}{2}}{\tilde q_1^2+i\varepsilon} 
ig_st^a X_{\perp i}^\rho \nnm{i}^\nu (-l_\rho g_{\nu\mu}+l_\nu g_{\rho\mu})
\frac{\nnps{i}}{2}v_i(q_1)\right)_{\!\alpha} 
\nn\\
 && {}\times \frac{-ig^{\mu\mu'}}{l^2+i\varepsilon}\, \times\,
\left(\frac{-i\nnp{j}\tilde q_2\frac{\nnms{j}}{2}}{\tilde q_2^2+i\varepsilon} 
ig_st^a X_{\perp j}^{\rho'} \nnm{j}^{\nu'} (l_{\rho'} g_{\nu'\mu'}-
l_{\nu'} g_{\rho'\mu'})\frac{\nnps{j}}{2}v_j(q_2)\right)_{\!\beta} \,,
\label{eq:softexample}
\eea
where $v_i(q)$ denotes the $i$-collinear antiquark spinor satisfying 
$\nnms{i} v_i(q)=0$, and we use the shorthand notation 
\be 
\int_{l_1,l_2,\ldots} \equiv\int\frac{d^dl_1}{(2\pi)^d}
\frac{d^dl_1}{(2\pi)^d}\ldots
\ee
for the loop integration measure in dimensional regularization with 
$d=4-2\epsilon$. The power-suppressed vertices arise from the 
interaction 
\be
S_{i\xi}^{(1)} = \int d^dx\, 
\bar{\chi}_i(x) \left( x_{\perp i}^\mu \nnm{i}^\nu 
\,g_s F_{\mu\nu}^{\rm s}(x_{i-}) 
\right) \frac{\slashed{n}_{i+}}{2} \chi_i(x),
\label{eq:Sixi}
\ee
with $x_{i-} \equiv (\nnp{i}x)\frac{\nnm{i}}{2}$, and the corresponding term 
for the direction $j$. Due to the explicit appearance 
of the space-time point $x^\mu$ in the interaction, the momentum-space 
vertex contains the derivative of the momentum-conserving delta functions,
\bea
X_{\perp i}&=&\frac{\partial}{\partial \tilde q_{1\perp i}}
(2\pi)^d\delta^{(d)}(\tilde q_1-q_1-(\nnm{i}l)\nnp{i}/2) \,,
\nn\\
X_{\perp j}&=&\frac{\partial}{\partial \tilde q_{2\perp j}}
(2\pi)^d\delta^{(d)}(\tilde q_2-q_2+(\nnm{j}l)\nnp{j}/2)\,.
\eea 
Momentum conservation at the vertex can be imposed only after evaluation the 
derivative with respect to the momenta $\tilde q_1$ and $\tilde q_2$ of the
fermion lines attached to the $i$- and $j$-collinear building blocks of the 
current, respectively. Note that, according to the SCET Feynman rules, only 
the $\nnm{i} l$ projection of the loop momentum $l$ enters in the delta 
function derived from ${\cal L}_{i\xi}^{(1)}$, and analogously for direction 
$j$. The reason is that the soft field is multipole expanded around $x_{i-}$
in ${\cal L}_i$ as can be seen in Eq.~\eqref{eq:Sixi}. After partial 
integration the derivatives can be eliminated. Using the collinear projection 
property of the external spinors we obtain
\bea\label{eq:ME_T1T1}
\lefteqn{ \langle \bar q_j(q_2)\bar q_i(q_1)|\chi_{i\alpha}\chi_{j\beta}|0
\rangle_{{\rm 1-loop},\ S_{11}} }
\nn\\
&=&
\tilde\mu^{2\epsilon}\int_l \Big( {\cal A}_{i\rho\alpha}^{a} \, 
(-\nnm{i}^\mu g_{\perp i}^{\rho\nu}+\nnm{i}^\nu g_{\perp i}^{\rho\mu})l_\nu 
\Big) \, \frac{-ig_{\mu\mu'}}{l^2+i\varepsilon}\, 
\Big( (\nnm{j}^{\mu'} g_{\perp j}^{\rho'\nu'}-\nnm{j}^{\nu'} 
g_{\perp j}^{\rho'\mu'})l_{\nu'} \, {\cal A}_{j\rho'\beta}^{a}\Big),
\eea
where
\bea\label{eq:Ai_noem}
{\cal A}_{i\rho\alpha}^{a} &=& -\frac{\partial}
{\partial \tilde q_{1\perp i}^\rho}\frac{\nnp{i}\tilde q_1}
{\tilde q_1^2+i\varepsilon}\Bigg|_{\tilde q_1=q_1+\frac12(\nnm{i}l)
\nnp{i}} g_s t^a v_{i\alpha}(q_1) \,,
\nn\\
{\cal A}_{j\rho'\beta}^{a} &=& -\frac{\partial}
{\partial \tilde q_{2\perp j}^{\rho'}}\frac{\nnp{j}\tilde q_2}
{\tilde q_2^2+i\varepsilon}\Bigg|_{\tilde q_2=q_2-\frac12(\nnm{j}l)
\nnp{j}} g_s t^a v_{j\beta}(q_2) \,.
\eea
The integrand has the structure $l^\nu l^{\nu'} F(\nnm{i}l,\nnm{j}l,l^2)$ 
with some scalar function $F$. There is no explicit dependence on 
$l_\perp$, since after the derivative with respect to 
$\tilde q_{1\perp i}^\rho$ ($\tilde q_{2\perp j}^{\rho'}$) is carried 
out $\tilde q_1^2$ ($\tilde q_2^2$) is set to $ q_1^2 + \nnm{i}l 
\nnp{i} q_1$ ($ q_2^2 -\nnm{j}l \nnp{j} q_2$). Therefore the integral can be 
decomposed, after integration, into terms proportional to the tensors 
$\nnm{i}^\nu\nnm{i}^{\nu'}$, $\nnm{j}^\nu\nnm{j}^{\nu'}$, 
$\nnm{i}^\nu\nnm{j}^{\nu'}$, $\nnm{j}^\nu\nnm{i}^{\nu'}$, and $g^{\nu\nu'}$. 
By making an ansatz of a linear combination of these tensors, and contracting 
with them, one finds that one can replace inside of the integrand
\bea
\label{eq:tensor_decomp}
l^\nu l^{\nu'} &\mapsto& \frac{1}{d-2}\left(l^2-2\frac{\nnm{i}l\,\nnm{j}l}
{\nnm{i}\nnm{j}}\right)g^{\nu\nu'} 
- \frac{1}{d-2}\left(l^2-d\frac{\nnm{i}l\,\nnm{j}l}{\nnm{i}\nnm{j}}\right)
\frac{\nnm{i}^\nu\nnm{j}^{\nu'}+\nnm{i}^{\nu'}\nnm{j}^\nu}{\nnm{i}\nnm{j}} 
\nn\\
&& {} + \left(\frac{\nnm{j}l}{\nnm{i}\nnm{j}}\right)^2
\nnm{i}^\nu\nnm{i}^{\nu'} + \left(\frac{\nnm{i}l}{\nnm{i}\nnm{j}}\right)^2
\nnm{j}^\nu\nnm{j}^{\nu'}\,.
\eea
The terms $\propto l^2$ vanish, since the gluon propagator $1/(l^2+i\varepsilon)$ 
is cancelled, resulting in an infrared finite and quadratically ultraviolet 
divergent scaleless transverse momentum integral, which does not contribute 
to the anomalous dimension. Furthermore, since $\nu$ is contracted with 
either a $\perp$ vector in the $i$ direction or with $\nnm{i}$, terms 
$\propto \nnm{i}^\nu$ vanish. Similarly, terms $\propto \nnm{j}^{\nu'}$ 
vanish. Inserting this decomposition in Eq.~\eqref{eq:ME_T1T1} yields
\bea
\lefteqn{ \langle \bar q_j(q_2)\bar q_i(q_1)|\chi_{i\alpha}\chi_{j\beta}|0
\rangle_{{\rm 1-loop},\ S_{11}} }
\nn\\
&=&
- 4ig_s^2 \tilde\mu^{2\epsilon}\,\frac{4-d}{d-2}\,
\frac{(q_{1\perp i}q_{2\perp j})(\nnm{i}\nnm{j})
-(\nnm{i}q_{2\perp j})(\nnm{j}q_{1\perp i})}{\nnm{i}\nnm{j}} 
\Big(t^av_{i\alpha}(q_1)\Big) \Big(t^av_{j\beta}(q_2)\Big)
\nn\\
&& \times \int_{l}\, \frac{\nnp{i}q_1\nnm{i}l}{(q_1^2-\nnp{i}q_1\nnm{i}l
+i\varepsilon)^2} \, \frac{1}{l^2+i\varepsilon}\, \frac{\nnp{j}q_2\nnm{j}l}
{(q_2^2+\nnp{j}q_2\nnm{j}l+i\varepsilon)^2} 
\nn\\
&= & -\frac{1}{\epsilon}\,\frac{2\alpha_s}{\pi}\,
\frac{(q_{1\perp i}q_{2\perp j})(\nnm{i}\nnm{j})-(\nnm{i}q_{2\perp j})
(\nnm{j}q_{1\perp i})}{\nnp{i}q_1\nnp{j}q_2(\nnm{i}\nnm{j})^2} 
\Big(t^av_{i\alpha}(q_1)\Big) \Big(t^av_{j\beta}(q_2)\Big)
+{\cal O}(\epsilon^0)\,,
\nn\\[-0.2cm]
\eea
where we used the master integral from App.~\ref{sec:integrals} in the last 
step. Note the explicit factor $4-d=2\epsilon$ in the first line, which 
cancels the double pole from the integral. The remaining  $1/\epsilon$ 
divergence can be absorbed by a counterterm proportional to the current 
$\left(J^{A1}_{\partial^\mu\chi}\right)_i\left(J^{A1}_{\partial^\nu\chi}
\right)_j=[i\partial_{\perp i}^\mu\chi_i][i\partial_{\perp j}^\nu\chi_j]$, 
resulting in the non-vanishing entry
\be\label{eq:T1xiT1xi_A1A1}
  \gamma^{ij}_{\left(J^{T1}_{\chi,\xi}\right)_i
\left(J^{T1}_{\chi,\xi}\right)_j,\left(J^{A1}_{\partial^\mu\chi}\right)_i
\left(J^{A1}_{\partial^\nu\chi}\right)_j} = 
{} \frac{2\alpha_s}{\pi}\,\T_i\cdot\T_j \,G^{\mu\nu}_{ij}\,,
\ee
of the anomalous dimension matrix, where
\be
G^{\mu\nu}_{ij} \equiv \left(g^{\mu\nu}
-\frac{\nnm{i}^\nu\nnm{j}^\mu}{\nnm{i}\nnm{j}}\right)
\frac{1}{(\nnm{i}\nnm{j})P_iP_j}\,.
\label{eq:gtensor}
\ee
Here we omitted the building blocks belonging to the $N-2$ collinear 
directions different from $i$ and $j$, which remain unchanged, as well as 
Dirac indices, since the above anomalous dimension is diagonal in them in each 
collinear direction. This computation also implies that
\be
\label{eq:T1T1_A2A0}
\gamma^{ij}_{\left(J^{T1}_{\chi,\xi}\right)_i\left(J^{T1}_{\chi,\xi}\right)_j,
\left(J^{A2}_{\partial^\mu\partial^\nu\chi}\right)_i
\left(J^{A0}_{\chi}\right)_j} = {} 0\,,
\ee
and analogously for $J^{A2}\leftrightarrow J^{A0}$. The computation above 
was done in Feynman gauge. Independence on the gauge-fixing parameter 
in general covariant gauge is easily seen from Eq.~(\ref{eq:softexample}), 
since replacing $g^{\mu\mu'}\to l^\mu l^{\mu'}/l^2$ produces zero upon 
contracting $l^\mu$ with the vertex factor. 
The reason for this is that the  soft-gluon vertex in 
${\cal L}^{(1)}_\xi$ comes from the field strength tensor. We also 
note that with our conventions off-diagonal elements of the anomalous 
dimension matrix need not be dimensionless as is apparent from 
the result (\ref{eq:gtensor}). In this way we avoid putting 
explicit factors of the hard scale into the generic operator basis. 
For the special case of back-to-back directions, $\nnm{j}=\nnp{i}$, 
and, since $\mu$, $\nu$ are contracted with transverse vectors, 
$G_{ij}^{\mu\nu}$ simplifies to $g_\perp^{\mu\nu}/(2 Q^2)$, where 
$Q^2=P_i P_j$ is the invariant mass of the particles in the 
back-to-back directions, resulting in a simple expression for the 
soft mixing anomalous dimension (\ref{eq:T1xiT1xi_A1A1}).

In order to determine the mixing into $B$- and $C$-type currents, as well as 
of different types of time-ordered products, we need to consider matrix 
elements with one or two additional collinear emissions in direction $i$ or 
$j$. To simplify this computation, we consider the general structure of soft 
loops, depending on the number and type of insertions of power-suppressed 
interactions. We first consider the case of a single ${\cal L}^{(1)}$ 
insertion, then a double ${\cal L}^{(1)}$ insertion, and finally a single 
${\cal L}^{(2)}$ insertion, for the case $F_i=F_j=+1$.
The results for fermion number $(F_i,F_j)=(+1,+1)$ can be 
generalized straightforwardly to operators with $\bar\chi$ instead of 
$\chi$ building blocks, i.e. $(F_i,F_j)=(-1,-1), (1,-1), (-1,1)$,  
see App.~\ref{app:cc}. 

\subsection{Single insertion of ${\cal L}^{(1)}$}

In~Ref.~\cite{Beneke:2017ztn} it has been shown that soft loops with a 
\emph{single} insertion of ${\cal L}^{(1)}$ vanish for the case 
$F_i=\pm 2$. We now generalize this result to currents with arbitrary 
fermion number (such as in diagram $S_{10}$ in Fig.~\ref{fig:soft_qq}). We 
consider an operator containing a single time-ordered product involving an 
insertion of ${\cal L}^{(1)}$ along direction $i$. This operator can 
potentially mix into current operators containing, instead of the 
time-ordered product, either an extra transverse derivative or an extra 
collinear building block along one of the $N$ collinear directions. To 
determine this mixing it is therefore sufficient to consider a soft loop 
diagram with a soft line connecting direction $i$ with any other direction 
$j$, and with up to one extra collinear emission. The vertex to which the 
soft line is attached along direction $j$ is a leading-power interaction. 
Since soft quarks do not interact with collinear particles at leading power, 
only the case of a soft gluon line needs to be considered. This implies in 
turn that the power-suppressed interaction along direction $i$ has to contain 
a soft gluon as well, i.e. we need to consider only ${\cal L}_\xi^{(1)}$ or  
${\cal L}_{\rm YM}^{(1)}$. Operators containing a single time-ordered product 
involving ${\cal L}_{\xi q}^{(1)}$ cannot mix into currents.

\begin{figure}
\begin{center}
  \includegraphics[width=0.2\textwidth]{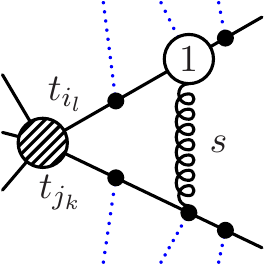}
\end{center}
\caption{\label{fig:soft_gen} 
Generic soft one-loop diagram featuring a single insertion of  
${\cal L}_\xi^{(1)}$ or  ${\cal L}_{\rm YM}^{(1)}$ 
along collinear direction $i$, a leading-power interaction along direction 
$j$, and a soft gluon exchanged between them. 
Solid lines without arrow can be either collinear quarks or gluons. Dotted 
lines illustrate possible additional collinear emissions (additional 
emissions directly off the current are not shown). The relevant diagrams are 
obtained by choosing \emph{one} emission out of the dashed lines, or none 
for the case without extra emission.}
\end{figure}

A generic soft loop diagram of the type described above is shown in 
Fig.~\ref{fig:soft_gen}. Here dotted lines illustrate possible attachments of 
an extra collinear emission. As stated in the previous paragraph, only 
\emph{one} extra emission needs to be considered. This emission can be 
attached to either direction $i$ or $j$. In addition, it can be either off 
an internal or external propagator, or off one of the vertices involving the 
soft gluon. Extra emission directly off the current is also possible, but
 not shown for brevity. 

It turns out that the loop amplitudes for all relevant diagrams can be 
written in a generic form. Let us assume that the soft gluon line carries 
momentum $l$. The leading-power interaction is proportional to 
$\nnm{j}^{\mu'}$ for all vertices involving a soft gluon, since
soft gluons enter only via the projection $\nm D$ of the covariant derivative 
into the leading-power collinear Lagrangian. The power suppressed interaction 
vertices derived from both ${\cal L}_\xi^{(1)}$ or  ${\cal L}_{\rm YM}^{(1)}$ 
contribute the factor 
\be
\label{eq:L1factor}
\nnm{i} l\, g_{\perp i}^{\rho\mu} - \nnm{i}^\mu l_{\perp i}^{\rho} = 
(\nnm{i}^\nu g_{\perp i}^{\rho\mu}-\nnm{i}^\mu g_{\perp i}^{\rho\nu})l_\nu\,,
\ee 
where the index $\rho$ is contracted either with a derivative with respect 
to a $\perp$ momentum or, for the 4-gluon vertex, with  the $\perp$ component 
of some collinear momentum. The reason is that at ${\cal O}(\lambda)$ the 
soft field enters the collinear SCET Lagrangian only via the soft field 
strength tensor projected in the $-$ and $\perp$ directions,
 $\nm^\nu F^{s}_{\nu\kappa_\perp}$. Restricting for the moment to the 
time-ordered product with a leading-power current, the loop amplitude in 
Feynman gauge can be written in the form
\be
\langle f|\chi_{i\alpha}\chi_{j\beta}|0\rangle_{\,{\rm single}\ 
{\cal L}^{(1)}\ {\rm insertion}} = \tilde\mu^{2\epsilon}\int_l 
\Big( {\cal A}_{i\rho\alpha}^{a} \, 
(\nnm{i}^\nu g_{\perp i}^{\rho\mu}-\nnm{i}^\mu g_{\perp i}^{\rho\nu})l_\nu 
\Big) \, \frac{-ig_{\mu\mu'}}{l^2+i\varepsilon}\, 
\Big( \nnm{j}^{\mu'}{\cal A}_{j\beta}^a\Big)\,,
\ee
where ${\cal A}_{i\rho\alpha}^{a}$ and ${\cal A}_{j\beta}^a$ denote the 
pieces of the amplitude involving propagators and vertices along direction 
$i$ and $j$, respectively, and $a$ refers to the colour index of the soft 
gluon. Furthermore, $\langle f|\in\{\langle \bar q_j\bar q_i|, 
\langle \bar q_j g_j\bar q_i|,  \langle \bar q_j \bar q_i g_i|\}$ stands for 
states with either two outgoing antiquarks, or with an extra collinear gluon 
in direction $i$ or $j$. For example, for the particular case without 
additional collinear emission ${\cal A}_{i\rho\alpha}^{a}$ is given by 
Eq.~\eqref{eq:Ai_noem} and 
\be
\label{eq:Aj_noem}
{\cal A}_{j\beta}^a\Big|_{\rm no\ extra\ emission} \ = \ 
\frac{\nnp{j}q_2}{q_2^2-\nnp{j}q_2\nnm{j}l+i\varepsilon}
g_st^av_{j\beta}(q_2)\,.
\ee
Note that, also in the general case, 
${\cal A}_{i\rho\alpha}^{a}$ depends on the loop momentum only via 
$\nnm{i}l$, and ${\cal A}_{j\beta}^a$ only via $\nnm{j}l$, due to the 
multipole expansion.

The loop integrand has the structure $l^\nu F(\nnm{i}l,\nnm{j}l,l^2)$, such 
that by analogous reasoning as for the example above, the integral can 
be decomposed in contributions proportional to $\nnm{i}^\nu$ and $\nnm{j}^\nu$.
Inside the loop integrand one may replace
\be
\label{eq:vector_decomp}
l^\nu \mapsto \frac{\nnm{j}l}{\nnm{j}\nnm{i}}\nnm{i}^\nu 
+ \frac{\nnm{i}l}{\nnm{j}\nnm{i}}\nnm{j}^\nu\;.
\ee
The former term on the right-hand side vanishes since $\nnm{i}^2=0$ and 
$g_{\perp i}^{\rho\nu}{\nnm{i}}_\nu=0$. The latter vanishes
since $(\nnm{i}^\nu g_{\perp i}^{\rho\mu}-\nnm{i}^\mu g_{\perp i}^{\rho\nu})
{\nnm{j}}_\nu{\nnm{j}}_\mu=0$. 
Therefore, operators containing a single time-ordered product involving an 
${\cal L}^{(1)}$ insertion do not mix into local currents.
This generalizes the argument of Ref.~\cite{Beneke:2017ztn} to operators 
with $F_i=F_j=+1$, and implies at ${\cal O}(\lambda)$
\be
\gamma^{ij}_{\left(J^{T1}_{\chi,V}\right)_i\left(J^{A0}_\chi\right)_j,
\left(J^{A1}_{\partial\chi}\right)_i\left(J^{A0}_\chi\right)_j}
= \gamma^{ij}_{\left(J^{T1}_{\chi,V}\right)_i\left(J^{A0}_\chi\right)_j,
\left(J^{B1}_{{\cal A}\chi}\right)_i\left(J^{A0}_\chi\right)_j} = 0 \,,
\ee
for $V\in\{\xi,{\rm YM},\xi q\}$.
Adding transverse derivatives or collinear building blocks to 
the operator does not affect the general form of the soft loop integral,
since after the transverse momentum derivatives from the vertices 
have been done, the denominator of the integrand depends only on 
the $\nnm{i} l$ and $\nnm{j}l$ components of the loop momentum, 
and since at most one factor of $l_\perp$ can appear in the 
numerator. Therefore, we conclude that at ${\cal O}(\lambda^2)$
\be
\gamma^{ij}_{\left(J^{T2}_{\partial\chi,V}\right)_i\left(J^{A0}_\chi\right)_j,
J'_iJ'_j}
= \gamma^{ij}_{\left(J^{T2}_{{\cal A}\chi,V}\right)_i
\left(J^{A0}_\chi\right)_j,J'_iJ'_j}
= \gamma^{ij}_{\left(J^{T1}_{\chi,V}\right)_i
\left(J^{A1}_{\partial\chi}\right)_j,J'_iJ'_j}
= \gamma^{ij}_{\left(J^{T1}_{\chi,V}\right)_i
\left(J^{B1}_{{\cal A}\chi}\right)_j,J'_iJ'_j} = 0\,,
\label{eq:lam2mixing1}
\ee
where $J'_iJ'_j$ is a product of arbitrary local currents in directions $i$ 
and $j$. Once again, gauge invariance is a trivial consequence of the 
structure of the SCET power-suppressed soft-gluon vertices.

We also note that the loop amplitude has the same structure
for operators containing gluon building blocks. 
Thus Eq.~(\ref{eq:lam2mixing1}) remains true, when $\chi$ in 
$J^{T1}$, $J^{T2}$ and/or $J^{A0}_\chi$ is replaced by ${\cal A}$  
-- the single insertions with ${\cal L}^{(1)}$ never contribute to 
the one-loop anomalous dimension matrix to $\mathcal{O}(\lambda^2)$.

\subsection{Double insertion of ${\cal L}^{(1)}$}

There are two types of operators containing a double insertion of 
${\cal L}^{(1)}$. Either both insertions belong to the same collinear 
direction (involving $J^{T2}_{\chi,VW}$), or the two insertions 
belong to different directions (involving $J^{T1}_{\chi,V}J^{T1}_{\chi,W}$).
We consider the two cases in turn.

\subsubsection{Double insertion in a single collinear direction}

Diagrams containing a double time-ordered product along a single collinear 
direction, say $i$, can potentially mix into local currents at the one-loop 
order by connecting the two insertions with a soft line. This leads to 
loop integrals of the form
\be
\int_l\, \frac{l^\mu l^\nu \cdots}{l^2+i\varepsilon}\prod_a 
\frac{1}{p_a^2-\nnp{i} p_a \nnm{i} l+i\varepsilon}\,,
\label{eq:generaldoubleins}
\ee
where $p_a$ are linear combinations of collinear momenta in direction $i$, 
and $p_a^2\not =0$ due to the off-shell regularization. The momentum 
derivatives contained in the SCET Feynman rules do not change this general 
structure. If some of the $p_a$ are identical, higher powers of the 
propagator occur, which can be related to an integral with a single power, 
differentiated with respect to $p_a^2$.

We now show that the above integrals always vanish. Note that the 
integrand has only a single pole at 
$\nnp{i} l = -(l_{\perp i}^2+i\varepsilon)/\nnm{i}l$, but closing the contour 
in the half plane that does not contain this pole, does not 
allow us to conclude that the integral is zero, since the integral 
over the half-circle at infinity is not convergent. On the other hand, 
inspection of the diagrams shows that an additional factor of $l^\mu$ is 
accompanied by an additional denominator containing $\nnm{i} l$ 
such that the $\nnm{i} l$ integral over the infinite circle is always 
zero. Hence, 
when $\nnp{i} l>0$  we pick up the residues in the upper $\nnm{i} l$ half 
plane at $\nnm{i} l =p_a^2/\nnp{i}p_a+i\varepsilon$ ($\nnp{i}p$ is positive), 
while the integral vanishes for $\nnp{i} l<0$, when all poles lie in 
the positive half plane. This converts the loop integral into a sum of 
terms of the form
\be
\int_0^\infty d\nnp{i}l 
\int d^{d-2}l_{\perp i} \,\frac{(\nnp{i}l)^b \,l_{\perp i}^\mu 
l_{\perp i}^\nu\ldots}
{\frac{\nnp{i}l}{\nnp{i}p_a}\,p_a^2 +l_{\perp i}^2
+i\varepsilon}
\label{eq:softononedirection}
\ee
with non-negative $b$. Performing the dimensionally regulated transverse 
momentum integral results in $\nnp{i}l$ integrals of the form
\be
 \int_0^\infty  d\nnp{i}l \left(\nnp{i} l\right)^{b-\epsilon},\;\; b\geq 0  
\ee
with some other non-negative $b$ and neglecting any $l$ independent 
prefactors. These scaleless integrals are IR finite 
and develop power-like divergences in the UV region. Even though the 
integral over $l_\perp$ may generate a $\frac{1}{\epsilon}$ pole, the 
result is zero due to the vanishing scaleless $\nnp{i}l$ integral as 
was to be shown.\footnote{An alternative derivation proceeds by 
performing the $l_\perp$ integral first. Since in 
Eq.~\eqref{eq:generaldoubleins} only the soft gluon 
propagator $1/(l^2+i \varepsilon)$ carries a dependence on $l_\perp$ 
and $\nnp{i}l$ in the denominator, this results in a $\nnp{i}l$ 
integral of the form $\int d\nnp{i}l\,(-\nnm{i} l\nnp{i}l-i\varepsilon)^{
b-\epsilon}$.
The cut can be avoided by closing the contour in the hemisphere 
opposite to the cut. The integral over the circle is now regulated 
dimensionally and therefore can be set to zero.} 
Hence, soft one-loop diagrams within a single collinear direction do not 
contribute to the anomalous dimension at any power of $\lambda$. 
In particular,
\be
\gamma^{ij}_{\left(J^{T2}_{\chi,VW}\right)_i
\left(J^{A0}_\chi\right)_j,J'_iJ'_j} = 0
\ee
for $V,W\in\{\xi,{\rm YM},\xi q\}$, and by the same argument 
\be
\gamma^{ij}_{T(P,{\cal L}^{(1)}_k,{\cal L}^{(1)}_k),Q} = 0
\ee
for $i,j,k=1,\dots,N$ and arbitrary local $N$-jet operators $P$, $Q$.

\subsubsection{Double insertion in different collinear directions}

\begin{figure}
\begin{center}
  \includegraphics[width=0.2\textwidth]{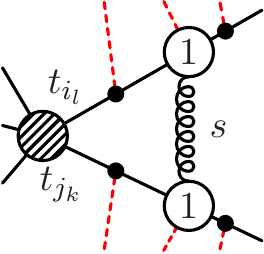}\\[0.5cm]
  \includegraphics[width=0.95\textwidth]{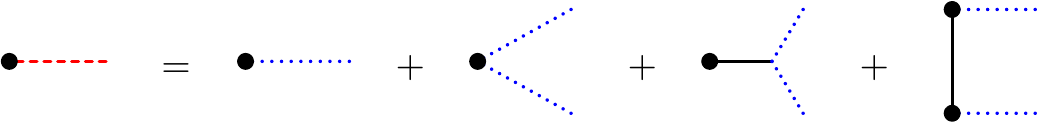}
\end{center}
\caption{\label{fig:soft_gen2} 
Generic soft one-loop diagram featuring a double insertion of  
${\cal L}_\xi^{(1)}$ or  ${\cal L}_{\rm YM}^{(1)}$ 
along collinear directions $i$ and $j$, respectively. 
Solid lines without arrow can be either collinear quarks or gluons. 
Dashed lines illustrate possible sub-diagrams with additional collinear 
emissions (additional emissions directly off the current are not shown). 
In the second line, we show possible sub-diagrams. Only diagrams that 
contain at most two dotted lines, i.e. two collinear emissions, need to 
be considered.}
\end{figure}

We next consider a double time-ordered product operator with one 
insertion of ${\cal L}^{(1)}_i$ in direction $i$ and one
of ${\cal L}^{(1)}_j$ in another direction $j$. One-loop mixing with a current 
operator can occur when both insertions are connected by
a soft line. For the moment we focus on the case of a soft gluon line 
and refer to Sec.\,\ref{sec:xiq} for the case of soft quark mixing. 
Therefore we consider insertions of ${\cal L}_\xi^{(1)}$ or 
${\cal L}_{\rm YM}^{(1)}$. The corresponding operators contain two 
time-ordered products 
$\left(J^{T1}_{\chi,V}\right)_i\left(J^{T1}_{\chi,W}\right)_j$ with 
$V,W\in\{\xi,{\rm YM}\}$, which can mix into currents with two additional 
$\perp$ derivatives, or two additional building blocks, or one $\perp$ 
derivative together with one extra building block. Accordingly, we 
analyze diagrams with up to \emph{two} additional collinear emissions. 
Such diagrams are summarized in Fig.\,\ref{fig:soft_gen2}. 
No more than two of the dashed lines should be replaced by one of the dotted 
subdiagrams shown in the second line, such that the resulting diagram 
contains at most two dotted lines. Emissions directly off the current are 
not shown for simplicity. 
The ${\cal L}^{(1)}$ insertion in direction $i$ 
contributes a factor of the form \eqref{eq:L1factor}, and a corresponding
factor arises for direction $j$ with opposite sign since the direction of 
the soft momentum is reversed. The amplitude for all relevant diagrams
can therefore be written in the form
\bea
\label{eq:ME_T1T1_general}
\lefteqn{ \langle f|\chi_{i\alpha}\chi_{j\beta}|0\rangle_{\,{\rm double}\ 
{\cal L}^{(1)}\ {\rm insertion}} }\nn\\
&=&
\tilde\mu^{2\epsilon}\int_l \Big( {\cal A}_{i\rho\alpha}^{a} \, 
(-\nnm{i}^\mu g_{\perp i}^{\rho\nu}+\nnm{i}^\nu g_{\perp i}^{\rho\mu})l_\nu 
\Big) \, \frac{-ig_{\mu\mu'}}{l^2+i\varepsilon}\, 
\Big( (\nnm{j}^{\mu'} g_{\perp j}^{\rho'\nu'}-\nnm{j}^{\nu'} 
g_{\perp j}^{\rho'\mu'})l_{\nu'} \, {\cal A}_{j\rho'\beta}^{a}\Big)\,,
\qquad
\eea
where ${\cal A}_{i\rho\alpha}^{a}$ ( ${\cal A}_{j\rho'\beta}^{a}$) contains 
collinear propagators and vertices in direction $i$ ($j$). For the case 
without extra collinear emission they are given by Eq.~\eqref{eq:Ai_noem}, 
and we recover Eq.~\eqref{eq:ME_T1T1}. The external states for up to two 
additional collinear emissions are
\be
\label{eq:flist}
\langle f|\in\{\langle \bar q_j\bar q_i|, \langle \bar q_jg_j\bar q_i|, 
\langle \bar q_j\bar q_ig_i|, 
\langle \bar q_jg_j\bar q_ig_i|, \langle \bar q_jg_jg'_j\bar q_i|,  
\langle \bar q_j\bar q_ig_ig_i'|,  \langle \bar q_j q_j'\bar q_j''\bar q_i|,
\langle \bar q_j \bar q_i q_i'\bar q_i''|\,\}\,.
\ee
The four-fermion states $\langle \bar q_j q_j'\bar q_i \bar q_i'|$ and 
$\langle \bar q_j \bar q_j'\bar q_i q_i'|$ could contribute only in 
conjunction with soft fermion exchange diagrams and ${\cal L}^{(1)}_{\xi q}$ 
insertions, which will be discussed separately in Sec.\,\ref{sec:xiq}. 
Since ${\cal A}_{i\rho\alpha}^{a}$ and ${\cal A}_{j\rho'\beta}^{a}$ depend on the 
loop momentum only via $\nnm{i}l$ and $\nnm{j}l$, respectively, the loop 
integral can be decomposed using Eq.~\eqref{eq:tensor_decomp} and we obtain
\bea
\label{eq:T1T1_gen}
\lefteqn{ \langle f|\chi_{i\alpha}\chi_{j\beta}|0\rangle_{\,{\rm double}\ 
{\cal L}^{(1)}\ {\rm insertion}} }\nn\\
&=&
-ig_s^2 \tilde\mu^{2\epsilon}\,\frac{4-d}{d-2}\,\left(g_{\lambda\lambda'}
-\frac{{\nnm{i}}_{\lambda'}{\nnm{j}}_\lambda}{\nnm{i}\nnm{j}}\right)
g_{\perp i}^{\lambda\rho}g_{\perp j}^{\lambda'\rho'} \, 
\int_{l}\, {\cal A}_{i\rho\alpha}^{a} \, 
\frac{\nnm{i}l\nnm{j}l}{l^2+i\varepsilon}\, {\cal A}_{j\rho'\beta}^{a} \,.
\eea
For $d\to 4$ the loop integral generically has a double $1/\epsilon^2$ pole, 
but due to the prefactor the complete amplitude has only a single pole. 
A large number of a priori possible mixings can now be 
eliminated by two general considerations:
\begin{itemize}
\item The loop integral depends on two additional Lorentz indices 
$\rho_{\perp i}$ and $\rho'_{\perp j}$, projected along the $\perp$ directions
with respect to $i$ and $j$. The possible $\perp$ vectors entering 
${\cal A}_{i\rho\alpha}^{a}$ are a) the external momenta of $i$-collinear 
particles, b) the polarization vectors of $i$-collinear gluons, or 
c) $\gamma_{\perp i}^\rho$. Case a) implies mixing into an $A1$ operator 
$J_i^{A1}$, b) a $B1$ operator $J_i^{B1}$ and the same is true 
for case c) for the following reason: Inspecting the Feynman rules 
Eqs.~\eqref{eq:Vxixic}, 
\eqref{eq:Vxixicc} and \eqref{eq:Vccs}, one finds that 
$\gamma_{\perp i}^\rho$ may enter in ${\cal A}_{i\rho\alpha}^{a}$ only in 
connection to vertices involving extra collinear emissions in direction~$i$. 
Analogous properties hold for direction $j$. Together with power counting 
in $\lambda$, this implies that the divergent part can be absorbed by a 
counterterm containing either an extra derivative or an extra building 
block in both the $i$ as well as the $j$ direction.
This means that only mixings of the form 
\be
\left(J^{T1}_{\chi,V}\right)_i\left(J^{T1}_{\chi,W}\right)_j 
\to J^{X1}_i J^{Y1}_j
\ee
are possible, i.e. into a product of ${\cal O}(\lambda)$ currents in both
directions (with $X,Y=A,B$), but not mixing into e.g. $J^{X2}_i J^{A0}_j$ 
($X=A,B,C$). This generalizes the result obtained earlier in 
Eq.~\eqref{eq:T1T1_A2A0}.

\item The  insertion ${\cal L}_{\rm YM}^{(1)}$ can give non-zero matrix 
elements only if an additional collinear gluon appears in the final state 
(see diagrams in Fig.~\ref{fig:soft_YM} below). 
To renormalize such a diagram by a current without gluon building block, 
such as $[\partial_{\perp i}\chi_i][\partial_{\perp j}\chi_j]$, would
require a collinear emission from a Wilson line. There are two arguments 
why this cannot happen: (i) in the light-cone gauge such diagrams do not exist. 
(ii) suppose one would have to introduce a counterterm proportional to 
$[\partial_{\perp i}\chi_i][\partial_{\perp j}\chi_j]$
to renormalize a one-loop diagram containing 
$(J^{T1}_{\chi,{\rm YM}})_i(J^{T1}_{\chi,{\rm YM}})_j$. Then one could 
compute the corresponding diagram without extra emission. In this case the 
tree-level diagram with the counterterm is non-zero, while the one-loop
diagram with $(J^{T1}_{\chi,{\rm YM}})_i(J^{T1}_{\chi,{\rm YM}})_j$ vanishes. 
This contradicts the property that, once the counterterm is fixed,
\emph{all} possible matrix elements have to be finite. 
This argument implies also that 
\be\label{eq:YMtochi}
 \left( J^{Tn}_{\chi,{\rm YM}}\right)_i J_j \to 
\left[\partial_{\perp i}^{\mu_1}\ldots \partial_{\perp i}^{\mu_n}
\chi_i\right]J_j'
\ee
vanishes, where $J_j,J_j'$ can be arbitrary currents or time-ordered products.
For example, $\left(J^{T1}_{\chi,{\rm YM}}\right)_i \left(J^{T1}_{\chi,{\rm YM}}
\right)_j \to [\partial_{\perp i} \chi_i] ({\cal A}\chi)_j$ mixing does not 
occur.
\end{itemize}

\begin{figure}
\begin{center}
  \includegraphics[width=0.25\textwidth]{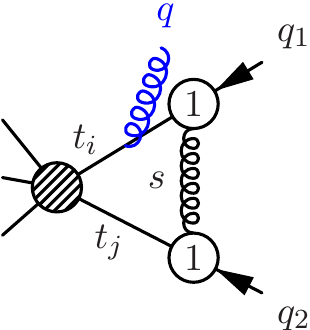}
  \raisebox{-5mm}{\includegraphics[width=0.25\textwidth]{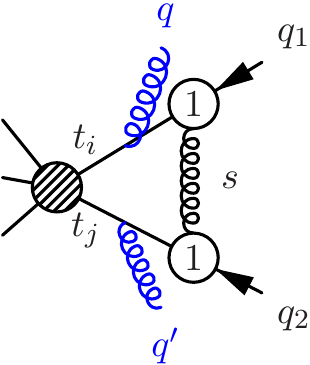}}
\end{center}
\caption{\label{fig:soft_qqg_T1T1}
Relevant diagrams for the mixing $\left(J_{\chi,\xi}^{T1}\right)_i
\left(J_{\chi,\xi}^{T1}\right)_j\to J^{B1}_i J^{A1}_j$ (left)
and $\left(J_{\chi,\xi}^{T1}\right)_i\left(J_{\chi,\xi}^{T1}\right)_j\to 
J^{B1}_i J^{B1}_j$ (right).}
\end{figure}

The only possible mixings of two time-ordered products into
currents with $F_i=F_j=1$ are therefore
\begin{itemize}
\item $\left(J_{\chi,\xi}^{T1}\right)_i\left(J_{\chi,\xi}^{T1}\right)_j
\to J^{X1}_i J^{Y1}_j$\,,
\item $\left(J_{\chi,{\rm YM}}^{T1}\right)_i\left(J_{\chi,\xi}^{T1}\right)_j
\to J^{B1}_i J^{Y1}_j$\,,
\item $\left(J_{\chi,{\rm YM}}^{T1}\right)_i\left(J_{\chi,{\rm YM}}^{T1}\right)_j
\to J^{B1}_i J^{B1}_j$\,,
\end{itemize}
with $X,Y=A,B$ (in the middle line the case with $i\leftrightarrow j$ is 
analogous). This reduces the number of states $\langle f|$ that need to be 
considered to compute the anomalous dimension to the first four 
in Eq.~\eqref{eq:flist}. The mixing $\left(J_{\chi,\xi}^{T1}\right)_i
\left(J_{\chi,\xi}^{T1}\right)_j\to J^{A1}_i J^{A1}_j$
has been treated already, and the result is given in 
Eq.~\eqref{eq:T1xiT1xi_A1A1}. For $\left(J_{\chi,\xi}^{T1}\right)_i
\left(J_{\chi,\xi}^{T1}\right)_j\to J^{B1}_i J^{A1}_j$
we consider the  matrix element with $\langle f|=\langle\bar q_j\bar q_i g_i|$.
To extract the operator mixing it is sufficient to let the gluon have  
$\perp$ polarization, and assume $(q_1)_{\perp i}=q_{\perp i}=0$. Then the 
only non-zero diagram is shown in Fig.\,\ref{fig:soft_qqg_T1T1} (left). 
The diagram with gluon emission off the power-suppressed vertex vanishes 
for $\perp$ polarization, and with emission off the external quark line 
because $(q_1)_{\perp i}=q_{\perp i}=0$. In this case 
${\cal A}_{j\rho'\beta}^{a}$ 
is given by Eq.~\eqref{eq:Ai_noem} and
\bea
{\cal A}_{i\rho\alpha}^{a} 
 &=& -\frac{\partial}{\partial \tilde q_{1\perp i}^\rho}
\Bigg[\frac{\nnp{i}(\tilde q_1+q)}{(\tilde q_1+q)^2+i\varepsilon} 
\nn\\
&& {} \times \left(\frac{\slashed{\tilde q}_{1\perp i}
\slashed{\epsilon}^*_{\perp i}}{\nnp{i}(\tilde q_1+q)}
+\frac{\slashed{\epsilon}^*_{\perp i}\slashed{\tilde q}_{1\perp i}}
{\nnp{i}\tilde q_1}\right)
\frac{\nnp{i}\tilde q_1}{\tilde q_1^2+i\varepsilon} 
g_s^2 t^b t^a v_{i}(q_1)\Bigg]_{\alpha}
\Bigg|_{\tilde q_1=q_1+\frac12(\nnm{i}l)\nnp{i}} \,.
\eea
For $q_{1\perp i}=0$ a non-zero contribution arises when the derivative acts 
on $\slashed{\tilde q}_{1\perp i}$. In collinear momentum space, we obtain 
using Eq.~\eqref{eq:T1T1_gen} and App.~\ref{sec:integrals}
\bea
\langle\bar q_j\bar q_i g_i^b|  \left(J_{\chi_{\alpha},\xi}^{T1}\right)_i 
\left(J_{\chi_{\beta},\xi}^{T1}\right)_j | 0 \rangle
 &=& P_i\delta(P_i-\nnp{i}(q_1+q))P_j\delta(P_j-\nnp{j} q_2) 
\nn\\
 && {} \times {\cal M}_{\alpha\beta,\gamma\delta}^{\mu\nu}
 \epsilon_{\mu\perp i}^*(q)  \left(t^bt^av_{i\gamma}(q_1)\right) 
(q_2)_{\nu\perp j} (t^av_{j\delta}(q_2))\,,\qquad 
\eea
with
\be
{\cal M}_{\alpha\beta,\gamma\delta}^{\mu\nu} =
\frac{g_s^3}{4\pi^2\epsilon} \frac{(\nnm{i}\nnm{j})g^{\lambda\nu}
-\nnm{i}^\nu\nnm{j}^\lambda}{\nnp{j} q_2 (\nnm{i}\nnm{j})^2}  
\left(\frac{\gamma_{\lambda\perp i}\gamma_{\perp i}^\mu}{\nnp{i}(q_1+q)}+
\frac{\gamma_{\perp i}^\mu\gamma_{\lambda\perp i}}
{\nnp{i}q_1}\right)_{\alpha\gamma} \delta_{\beta\delta}
+{\cal O}(\epsilon^0)\,. 
\ee
The tree-level matrix element of the 
$({\cal A}\chi)_i i\partial_{\perp j} \chi_j$ operator is
\bea
\langle\bar q_j\bar q_i g_i^a | \left(J^{B1}_{{\cal A}^{\mu }_b
\chi_{\gamma}}(y)\right)_i\, \left(J^{A1}_{\partial^\nu \chi_\delta}
\right)_j|0 \rangle_{\rm tree} 
&=& P_i^2\delta(yP_i-\nnp{i} q)\delta(\bar yP_i-\nnp{i} q_1) \, 
P_j\delta(P_j-\nnp{j} q_2) 
\nn\\
&& {} \times g_s\delta_{ab} \epsilon_{\perp i}^{*\mu}(q) v_{i\gamma}(q_1)\,  
(-q_2^\nu)_{\perp j} v_{j\delta}(q_2)\,,
\eea
where $y$ is the collinear momentum fraction carried by the first building 
block in direction $i$ (i.e. the gluon) and $\bar y=1-y$. Comparing the 
two expressions, we find the operator mixing 
\bea
\left(J_{\chi_{\alpha},\xi}^{T1}\right)_i \left(J_{\chi_{\beta},\xi}^{T1}
\right)_j 
&\to&  \frac{\alpha_s}{\pi\epsilon}\int_0^1 dy \, \T^b_{i} \, 
(\T_{i}\cdot\T_j) G_{\lambda\nu}^{ij}  
\nn\\
&& {}\times \left(\gamma_{\perp i}^\lambda\gamma_{\perp i}^\mu
+ \frac{\gamma_{\perp i}^\mu\gamma_{\perp i}^\lambda}{\bar y}
\right)_{\alpha\gamma} \delta_{\beta\delta} 
\left(J^{B1}_{{\cal A}_{\mu }^b\chi_{\gamma}}(y)\right)_i\, 
\left(J^{A1}_{\partial^\nu \chi_\delta}\right)_j\,.
\eea
Here $ G_{\lambda\nu}^{ij}$ is defined in Eq.~\eqref{eq:gtensor}. It is 
understood that we take the 
$\langle\bar q_ig_i\bar q_j|(\cdots)|0\rangle$ matrix element on both sides, 
and keep only the divergent part for $d\to 4$. Furthermore, 
we converted to colour operator notation which gives a minus sign. This 
yields the following result for the anomalous dimension,
\bea
\label{eq:T1xiT1xi_B1A1}
\gamma^{ij}_{\left(J_{\chi_{\alpha},\xi}^{T1}\right)_i
\left(J_{\chi_{\beta},\xi}^{T1}\right)_j,
\left(J^{B1}_{{\cal A}_b^\mu\chi_\gamma}\right)_i
\left(J^{A1}_{\partial^\nu\chi_\delta}\right)_j}(y_{i_1})
&=&  - \frac{\alpha_s}{\pi} \,\T^b_{i} \, (\T_{i}\cdot\T_j) G^{\lambda\nu}_{ij}  
\nn\\
&& {} \times \left(\gamma_{\lambda\perp i}\gamma_{\perp i}^\mu+
\frac{\gamma_{\perp i}^\mu\gamma_{\lambda\perp i}}{\bar y}
\right)_{\alpha\gamma} \delta_{\beta\delta}\,,
\eea
where $y$ corresponds to $y_{i_1}$ in the general notation. 

We note that although soft mixing does not transfer momentum between 
the two collinear directions $i$, $j$, the anomalous dimension above  
acquires a dependence on the momentum fractions of the collinear 
building blocks in the B1 current. This happens because in the left 
Fig.~\ref{fig:soft_qqg_T1T1} the divergent part of the diagram 
depends on how the gluon and quark with momentum $\nnp{i} q=y P_i$ 
and $\nnp{i} q_1 = \bar{y} P_i$, respectively, share the total momentum 
$P_i$. For the case at hand, Eq.~\eqref{eq:rencond} takes the 
form 
\be
\label{eq:rencondexample}
{\rm finite} = \langle J_P(x)\rangle_{\rm 1-loop} 
+ \sum_Q \int dy\, \delta(x-y)\,\delta Z^s_{PQ}(y)
\,\langle J_Q(y)\rangle_{\rm tree}\,.
\ee
Since in direction $i$, $n_i=1$ and $n_i'=2$, the delta function is 
empty, and $dy = dy_{i_1}$. This is consistent with the fact that 
after applying the constraint that momentum fractions in a given 
collinear direction must sum to 1, there is no dependence on 
momentum fraction $x$ for $n_i=1$, while for the B1 operator 
contained in $Q$, the single momentum fraction $y=y_{i_1}$ is 
integrated in the above equation. 

\begin{figure}
\begin{center}
 \includegraphics[width=0.65\textwidth]{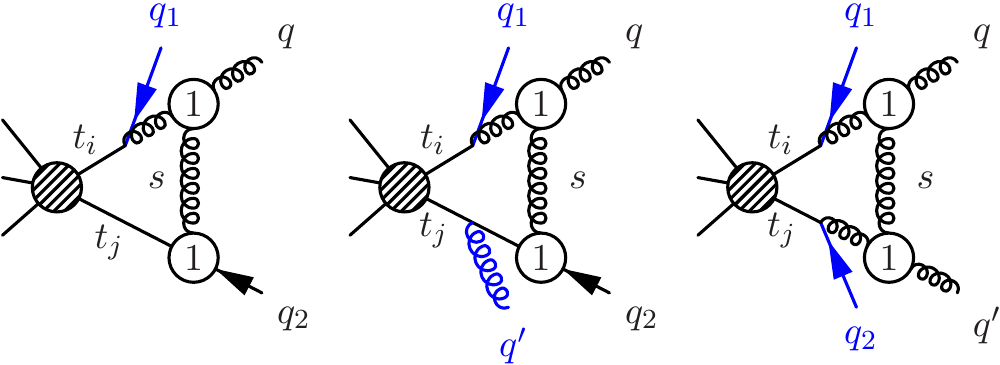}
\end{center}
\caption{\label{fig:soft_YM}
Relevant diagrams for the mixing $\left(J_{\chi,{\rm YM}}^{T1}\right)_i
\left(J_{\chi,\xi}^{T1}\right)_j\to J^{B1}_i J^{A1}_j$ (left),
$\left(J_{\chi,{\rm YM}}^{T1}\right)_i\left(J_{\chi,\xi}^{T1}\right)_j
\to J^{B1}_i J^{B1}_j$ (middle)
and $\left(J_{\chi,{\rm YM}}^{T1}\right)_i\left(J_{\chi,{\rm YM}}^{T1}\right)_j
\to J^{B1}_i J^{B1}_j$ (right).}
\end{figure}

In order to determine the mixing into two $B$-type currents, we consider 
the external state $\langle f|=\langle \bar q_j g_j \bar q_i g_i|$. Restricting 
to gluons with $\perp$ polarization and momenta with vanishing 
$\perp$ component, the only possible diagram is shown in 
Fig.~\ref{fig:soft_qqg_T1T1} (right). We find that all mixings of the 
type $\left(J_{\chi,\xi}^{T1}\right)_i\left(J_{\chi,\xi}^{T1}\right)_j
\to J^{X1}_i J^{Y1}_j$ can be summarized as
\bea
\label{eq:L1L1toA1A1}
\left(J_{\chi_{\alpha},\xi}^{T1}\right)_i 
\left(J_{\chi_{\beta},\xi}^{T1}\right)_j 
&\to& - \frac{2\alpha_s}{\pi\epsilon} \,G_{\lambda\kappa}^{ij}  
\nn\\
&& \times \left[ \T_i^a J^{A1}_{\partial^\lambda\chi_\alpha} 
-\frac12 \int dy  \left(\gamma_{\perp i}^\lambda\gamma_{\perp i}^\mu
+\frac{\gamma_{\perp i}^\mu\gamma_{\perp i}^\lambda}{\bar y}\right)_{\alpha
\gamma}\T_i^b\T_i^a J^{B1}_{{\cal A}_{\mu }^b\chi_{\gamma}}(y) \right]_i 
\nn\\
&& \times \left[\T_j^a J^{A1}_{\partial^\kappa\chi_\beta} 
-\frac12 \int dy'  \left(\gamma_{\perp j}^\kappa\gamma_{\perp j}^\nu
+\frac{\gamma_{\perp j}^\nu\gamma_{\perp j}^\kappa}{\bar{y}'}
\right)_{\beta
\delta}\T_j^c\T_j^a J^{B1}_{{\cal A}_{\nu }^c\chi_{\delta}}(y') \right]_j \,.
\nn\\
\eea
Similarly, for time-ordered products involving ${\cal L}^{(1)}_{\rm YM}$ we 
find (see Fig.~\ref{fig:soft_YM})
\bea\label{eq:L1YML1}
\left(J_{\chi_{\alpha},{\rm YM}}^{T1}\right)_i 
\left(J_{\chi_{\beta},\xi}^{T1}\right)_j 
&\to& 
-\frac{2\alpha_s}{\pi\epsilon}  G_{\lambda\kappa}^{ij}  
\nn\\
&& \times \,\frac{if^{bda}}{2} \int dy  \, 
\left( \frac{2g_{\perp i}^{\mu\lambda}}{y} 
- \gamma_{\perp i}^\lambda\gamma_{\perp i}^\mu \right)_{\alpha\gamma} 
 \T_i^d \left(J^{B1}_{{\cal A}_{\mu }^b\chi_{\gamma}}(y)\right)_i 
\nn\\
 && \times \left[\T_j^a J^{A1}_{\partial^\kappa\chi_\beta} 
-\frac12 \int dy'  \left(\gamma_{\perp j}^\kappa\gamma_{\perp j}^\nu
+ \frac{\gamma_{\perp j}^\nu\gamma_{\perp j}^\kappa}{\bar y'}
\right)_{\beta\delta}\T_j^c\T_j^a J^{B1}_{{\cal A}_{\nu }^c
\chi_{\delta}}(y') \right]_j\,, 
\nn\\
\\[-0.2cm]
\left(J_{\chi_{\alpha},{\rm YM}}^{T1}\right)_i 
\left(J_{\chi_{\beta},{\rm YM}}^{T1}\right)_j &\to& 
- \frac{2\alpha_s}{\pi\epsilon} G_{\lambda\kappa}^{ij}  
\nn\\
&& \times \,\frac{if^{bda}}{2} \int dy  \, \left( \frac{2g_{\perp i}^{\mu
\lambda}}{y} - \gamma_{\perp i}^\lambda\gamma_{\perp i}^\mu \right)_{\alpha
\gamma} \T_i^d \left(J^{B1}_{{\cal A}_{\mu }^b\chi_{\gamma}}(y)\right)_i 
\nn\\
 && \times \,\frac{if^{cea}}{2} \int dy'  \, \left( \frac{2g_{\perp j}^{\nu
\kappa}}{y'} - \gamma_{\perp j}^\kappa\gamma_{\perp j}^\nu \right)_{\beta
\delta}  \T_j^e \left(J^{B1}_{{\cal A}_{\nu }^c\chi_{\delta}}(y')\right)_j 
\,. 
\label{eq:L1YML1YM}
\eea
The corresponding anomalous dimension matrix entries read
\bea\label{eq:T1xiT1xi_B1B1}
\lefteqn{ \gamma^{ij}_{\left(J_{\chi_{\alpha},\xi}^{T1}\right)_i
\left(J_{\chi_{\beta},\xi}^{T1}\right)_j,
\left(J^{B1}_{{\cal A}_b^\mu\chi_\gamma}\right)_i
\left(J^{B1}_{{\cal A}^\nu_c\chi_\delta}\right)_j}(y_{i_1},y_{j_1}) }
\nn\\
&=&   \frac{\alpha_s}{2\pi} \,\T_i^b \T_j^c (\T_i\cdot\T_j)\,
G_{\lambda\kappa}^{ij}  
\left(\gamma_{\perp i}^\lambda\gamma_{\perp i}^\mu
+\frac{\gamma_{\perp i}^\mu\gamma_{\perp i}^\lambda}{\bar y_{i_1}}
\right)_{\alpha\gamma}
\left(\gamma_{\perp j}^\kappa\gamma_{\perp j}^\nu
+\frac{\gamma_{\perp j}^\nu\gamma_{\perp j}^\kappa}{\bar y_{j_1}}
\right)_{\beta\delta}\,, 
\\[0.2cm]
\label{eq:T1YMT1xi_B1A1}
\lefteqn{ \gamma^{ij}_{\left(J_{\chi_{\alpha},{\rm YM}}^{T1}\right)_i
\left(J_{\chi_{\beta},\xi}^{T1}\right)_j,\left(J^{B1}_{{\cal A}_b^\mu
\chi_\gamma}\right)_i\left(J^{A1}_{\partial^\nu\chi_\delta}\right)_j}
(y_{i_1}) }
\nn\\
&=&\frac{\alpha_s}{\pi} \,(\T_i\times\T_j)^b\,G_{\lambda\kappa}^{ij}  
\left( \frac{2g_{\perp i}^{\mu\lambda}}{y_{i_1}} - 
\gamma_{\perp i}^\lambda\gamma_{\perp i}^\mu \right)_{\alpha\gamma}
g^{\kappa\nu}\delta_{\beta\delta}\,, 
\\[0.2cm]
\label{eq:T1YMT1xi_B1B1}
\lefteqn{ \gamma^{ij}_{\left(J_{\chi_{\alpha},{\rm YM}}^{T1}\right)_i
\left(J_{\chi_{\beta},\xi}^{T1}\right)_j,\left(J^{B1}_{{\cal A}_b^\mu
\chi_\gamma}\right)_i\left(J^{B1}_{{\cal A}^\nu_c\chi_\delta}\right)_j}
(y_{i_1},y_{j_1}) }
\nn\\
&=& -\frac{\alpha_s}{2\pi} \,\T_j^c (\T_i\times\T_j)^b\,G_{\lambda\kappa}^{ij}  
\left( \frac{2g_{\perp i}^{\mu\lambda}}{y_{i_1}} 
- \gamma_{\perp i}^\lambda\gamma_{\perp i}^\mu \right)_{\alpha\gamma}
\left(\gamma_{\perp j}^\kappa\gamma_{\perp j}^\nu
+\frac{\gamma_{\perp j}^\nu\gamma_{\perp j}^\kappa}{\bar y_{j_1}}
\right)_{\beta\delta}\,, 
\\[0.2cm]
\label{eq:T1YMT1YM_B1B1}
\lefteqn{ \gamma^{ij}_{\left(J_{\chi_{\alpha},{\rm YM}}^{T1}\right)_i
\left(J_{\chi_{\beta},{\rm YM}}^{T1}\right)_j,
\left(J^{B1}_{{\cal A}_b^\mu\chi_\gamma}\right)_i
\left(J^{B1}_{{\cal A}^\nu_c\chi_\delta}\right)_j}(y_{i_1},y_{j_1}) }
\nn\\
&=&   -\frac{\alpha_s}{2\pi} \,f^{bda}f^{cea}\,
\T_i^d\T_j^e \,G_{\lambda\kappa}^{ij}  
\left( \frac{2g_{\perp i}^{\mu\lambda}}{y_{i_1}} 
- \gamma_{\perp i}^\lambda\gamma_{\perp i}^\mu \right)_{\alpha\gamma}
\!\left( \frac{2g_{\perp j}^{\nu\kappa}}{y_{j_1}} 
- \gamma_{\perp j}^\kappa\gamma_{\perp j}^\nu \right)_{\beta\delta} \,.
\eea
Here we defined the colour operator  cross product via 
$({\bf T}_{i_1} \times {\bf T}_{i_2})^a\equiv i f^{abc}{\bf T}_{i_1}^b{\bf T}_{i_2}^c$.

\subsection{Single insertion of ${\cal L}^{(2)}$}

\begin{figure}
\begin{center}
  \includegraphics[width=0.2\textwidth]{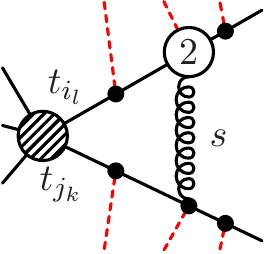}
\end{center}
\caption{\label{fig:soft_gen_T2} 
As Fig.\,\ref{fig:soft_gen}, but for an insertion of ${\cal L}^{(2)}_{\xi}$ 
or ${\cal L}^{(2)}_{\rm YM}$. The relevant diagrams are obtained by choosing 
up to \emph{two} emissions. Dashed lines should be replaced with subdiagrams 
according to the lower panel of Fig.\,\ref{fig:soft_gen}. In addition, 
also emission directly off the current is possible.}
\end{figure}

In this section we consider the possible mixing of single insertions 
of ${\cal O}(\lambda^2)$ SCET interactions, that is, $J^{T2}_{\chi,V}$, into 
current operators. It is sufficient to take $V=\xi,{\rm YM}$, because
at ${\cal O}(\lambda^2)$, the time-ordered product $J^{T2}_{\chi,\xi q}$ 
cannot mix into currents. The reason is that for diagrams with only 
collinear external lines, the soft quark field from ${\cal L}^{(2)}_{\xi q}$ 
would have to be contracted with another subleading-power Lagrangian.
 
The class of one-loop diagrams to be considered is illustrated exemplarily 
in Fig.\,\ref{fig:soft_gen_T2}, where the dashed lines represent again 
possible additional collinear emissions. To capture all possible mixings 
into currents at ${\cal O}(\lambda^2)$ we need to consider up to \emph{two} 
additional collinear emissions. 

We begin with the case of no extra collinear emission off the 
power-suppressed vertex from the ${\cal L}_\xi^{(2)}$ insertion (in direction 
$i$). The soft loop momentum $l$ carried by the soft gluon propagator 
is assumed to flow outwards of the ${\cal L}_\xi^{(2)}$-vertex, and 
the single external collinear momentum in the $i$ direction is denoted 
by $p$. Furthermore, we assume that the internal collinear line attached 
to the ${\cal L}_\xi^{(2)}$-vertex has momentum $\tilde p-l$, and we 
keep $\tilde p\not= p$ until derivatives are taken. The loop amplitude 
can then be written as
\bea
\langle f|\chi_{i}\chi_{j}|0\rangle_{{\rm single}\ {\cal L}^{(2)}_{\xi}\ 
{\rm insertion}} &=& \tilde\mu^{2\epsilon} \int_l  
\bigg( S^{\rho\nu}(-l,-p,\tilde p+l) \, {\cal A}_{i}^a \, 
(-l_\rho g_{\nu\mu}+l_\nu g_{\rho\mu} ) \bigg)_i \bigg|_{\tilde p=p} 
\nn\\
&& \times \frac{-ig_{\mu\mu'}}{l^2+i\varepsilon} \times 
\Big( \nnm{j}^{\mu'} {\cal A}_{j}^a\Big)_j
\label{eq:l2single}
\eea
where $S^{\rho\nu}$ is given by Eq.~\eqref{eq:Srnonu} and arises from the 
${\cal L}_\xi^{(2)}$ insertion. As before, ${\cal A}_{i}^a$ (${\cal A}_{j}^a$) 
contains the part of the amplitude involving $i$-collinear ($j$-collinear) 
propagators, vertices, polarization vectors and external spinors, and 
$a$ denotes the colour index of the soft gluon. We suppress Dirac indices 
for brevity. For the diagram without any extra emissions, 
$q_1\equiv p$, $\tilde q_1=\tilde p-l$,
\bea
\label{eq:ampnoem}
{\cal A}_{i}^a\Big|_{\rm no\ extra\ emission} &=& 
\frac{-i\nnp{i}\tilde q_1}{\tilde q_1^2+\nnp{i}\tilde q_1\nnm{i} l} 
ig_st^a v_{i}(q_1) \,,
\eea
and ${\cal A}_{j}^a$ is given by Eq.~\eqref{eq:Aj_noem}.

In general ${\cal A}_i^a$ and ${\cal A}_j^a$ contain several propagators 
involving various combinations of external momenta, and vertex factors 
that may depend polynomially on $\nnm{i} l$ and $\nnm{j}l$, respectively.
By partial fractioning the integrand ${\cal A}_i$ can be brought into the 
generic form
\be
\label{eq:partialfrac}
{\cal A}_i^a = \sum_{p_i,a_i,b_i} C^a(p_{i};a_i,b_i)\, (\nnm{i}l)^{b_i} 
\left(\frac{\nnp{i}p_{i}}{p_{i}^2+\nnp{i}p_{i}\nnm{i}l}\right)^{a_i}
\ee
where  $p_{i}$ are (linear combinations of) collinear momenta in direction 
$i$, including $p$ and $\tilde p$. We also use $p_i$ in the above 
equation to label the sum of terms that arises from the partial 
fractioning. The coefficients $C^a(p_{i};a_i,b_i)$ may 
depend on the collinear momenta $p_i$, but not on $l$.
${\cal A}_j^a$ can be decomposed analogously. From the explicit form 
of $S^{\rho\nu}$ together with Eqs.~\eqref{eq:tensor_decomp} and 
\eqref{eq:vector_decomp} we obtain
\bea
\label{eq:Dsoft}
\lefteqn{ \langle f|\chi_{i\alpha}\chi_{j\beta}|0\rangle_{{\rm single}\ 
{\cal L}^{(2)}_{\xi}\ {\rm insertion}} }
\nn\\
 &=& -i\tilde\mu^{2\epsilon} \int_l \,\left[\left(  \frac{\partial}
{\partial\nnp{i}\tilde p} -\frac{1}{d-2} \frac{\partial}
{\partial\tilde p_{\perp i}}\cdot\frac{\partial}{\partial\tilde p_{\perp i}} 
\nnm{i}l\right){\cal A}_i^a \right]_{\tilde p=p}\times 
\frac{\nnm{j} l}{l^2}{\cal A}_j^a\,.
\eea
After inserting Eq.~\eqref{eq:partialfrac}, the loop integral takes the form 
of the master integral~\eqref{eq:master_scalar}. A peculiar property of this 
integral is that it factors into two terms, each of which depends only on 
quantities related to a single collinear direction, here $i$ and $j$. 
This property is manifest in a frame where directions $i$ and $j$ are 
back-to-back. In the back-to-back frame, the $l_\perp$ integral can be 
performed first and the resulting expression is a product of two integrals 
that depend only on $\nm l$ or on $\np l$. Such a boost to the 
back-to-back frame can always be performed. Hence, the soft loop 
diagram factorizes into 
\be
\langle f|\chi_{i}\chi_{j}|0\rangle_{{\rm single}\ 
{\cal L}^{(2)}_{\xi}\ {\rm insertion}} = 
-F_\epsilon \times {\cal D}_i^a \times {\cal D}_j^a 
\label{eq:factorsoft}
\ee
with $F_\epsilon$ given by Eq.~\eqref{eq:Fepsilon}, and
\bea
{\cal D}_i^a &=& 
\sum_{p_i a_i b_i} \bigg(D(a_i,b_i,\epsilon)\frac{\partial}
{\partial\nnp{i}\tilde p} 
\nn\\
&& {} +\frac{D(a_i,b_i+1,\epsilon)}{2-2\epsilon} 
\frac{\partial}{\partial\tilde p_{\perp i}}\cdot\frac{\partial}
{\partial\tilde p_{\perp i}} \frac{p_{i}^2}{\nnp{i} p_{i}}\bigg) 
\frac{C^a(p_i;a_i,b_i)}
{(p_{i}^2/\nnp{i}p_{i})^{a_i-b_i-1+\epsilon}}\bigg|_{\tilde p=p},
\eea
where the numerical coefficients $D(a,b,\epsilon)$ are defined 
in Eq.~\eqref{eq:softcoeff}, and
\bea
\label{eq:Dj}
 {\cal D}_j^a &=&  
- \sum_{p_j, a_j, b_j}  \frac{C^a(p_j;a_j,b_j) D(a_j,b_j+1,\epsilon)}
{(p_{j}^2/\nnp{j}p_{j})^{a_j-b_j-2+\epsilon}}\,.
\eea

The diagrams with extra emission off the subleading-power vertex can be 
treated analogously, and lead to an integral of similar form, proportional
to the same factor ${\cal D}_j^a$. Similarly, diagrams involving an insertion 
of ${\cal L}_{\rm YM}^{(2)}$ in the $i$ direction can be shown to factorize 
into a product involving the same ${\cal D}_j^a$. This is a consequence 
of the fact that the $j$-direction involves only leading-power soft 
interactions, which are of eikonal type, and hence identical for 
quarks and gluons except for the colour factor. From the following 
discussion it will become clear that this is the relevant property.

\paragraph{No extra emissions in direction $j$:}
In this case the amplitude $\mathcal{A}^a_j$ is given by 
Eq.~\eqref{eq:Aj_noem}. There is only a single collinear propagator with 
$p_j=q_2$, $a_j=1$, $b_j=0$, $C^a(p_j;a_j,b_j)=g_st^a$ (where $a$ is the 
colour index of the soft gluon), $D(1,1,\epsilon)=-1$, which gives
\be
({\cal D}_j^{a})_{{\rm no\,}j-{\rm coll.\,em.}} = 
g_st^a\mu^\epsilon\left(\frac{q_2^2}{\nnp{j}q_2}\right)^{1-\epsilon}\,.
\ee
Hence, the factor ${\cal D}_j^{a}$ vanishes when the off-shell infrared 
regulator is removed, $q_2^2\to 0$. 
Due to integral factorization ${\cal D}_i$ cannot depend on 
the $j$-collinear momentum $q_2$, and therefore the complete diagram 
vanishes in this limit. Together with the factorization property, this 
finding also proves that all diagrams with extra collinear emissions 
in direction $i$, but no emissions in direction $j$, vanish,
because they are all proportional to 
$({\cal D}_j^{a})_{{\rm no\,}j-{\rm coll.\,em.}}$.

\paragraph{Single extra emission in direction $j$:}

\begin{figure}
\begin{center}
  \includegraphics[width=0.25\textwidth]{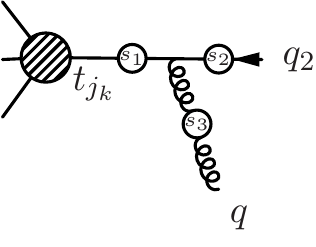}
\end{center}
\caption{\label{fig:soft_T2_single}
Diagram with a single collinear emission in direction $j$. The 
insertions $s_k$ show possible attachments of the soft line.
The part of the diagram in direction $i$ is not shown.
}
\end{figure}

The factorization property \eqref{eq:factorsoft} extends to 
the sum of all diagrams with 
soft attachments to a given collinear splitting pattern in
direction $j$. This can be captured by the expression ${\cal D}_j^a$ 
from Eq.~\eqref{eq:Dj} by including in the sum on the right-hand side 
the sum over all soft attachments. 
In the following, it turns out to be sufficient to consider only 
the $j$-collinear direction, which contains the leading-power soft 
interactions.

To be specific, 
consider a collinear quark in direction $j$ with external momentum $q_2$, 
and an emission of a collinear gluon off the quark line (momentum $q$, 
colour $b$, polarization tensor $\epsilon$), as shown in 
Fig.~\ref{fig:soft_T2_single}. The part of the diagram along direction $i$, 
which involves the ${\cal L}_\xi^{(2)}$ insertion, is not shown, because it 
is irrelevant for the computation of ${\cal D}_j^{a}$ 
due to the factorization property. The insertions $s_k$ mark 
attachments of the soft gluon (momentum $l$, colour $a$) to either the 
internal quark propagator ($s_1$), the external quark ($s_2$) or
the gluon ($s_3$). The corresponding amplitudes are given by
\bea
{\cal A}_j^{a(s_1)} &=& 
\frac{-i\nnp{j}\hat q}{\hat q^2-\nnp{j}\hat q\nnm{j} l}
\,ig_st^a\frac{-i\nnp{j}\hat q}{\hat q^2}
\,ig_st^bC_j v_{j}(q_2) \,,
\nn\\
{\cal A}_j^{a(s_2)} &=& 
\frac{-i\nnp{j}\hat q}{\hat q^2-\nnp{j}\hat q\nnm{j} l}
\,ig_st^bC_j \frac{-i\nnp{j}q_2}{q_2^2-\nnp{j}q_2\nnm{j}l} 
\,ig_st^a v_{j}(q_2) \,,
\nn\\
{\cal A}_j^{a(s_3)} &=& 
\frac{-i\nnp{j}\hat q}{\hat q^2-\nnp{j}\hat q\nnm{j} l}
\,ig_st^cC_j v_{j}(q_2)\,(-g_sf^{abc}\nnp{j} q)
\frac{-i}{q^2-\nnp{j} q\nnm{j} l} \,,
\eea
where $\hat q\equiv q_2+q$, and
\be
C_j \equiv \frac{\slashed{\hat q}_{\perp j}\slashed{\epsilon}_{\perp j}}
{\nnp{j}\hat q}+\frac{\slashed{\epsilon}_{\perp j}\slashed{q}_{2\perp j}}
{\nnp{j} q_2}\,.
\ee 
We have used that the three-gluon vertex involving two collinear  
and one soft gluon is diagonal in the Lorentz indices of the collinear fields. 
The factor $\nnm{j}^\mu$ that is contained in the soft leading-power vertex 
is not part of the above amplitudes, since it was taken out in the 
defining Eq.~\eqref{eq:l2single}. The sum of the three amplitudes can be 
expanded using partial fractioning as (we omit the label for 
$a_j=1, b_j=0$  on the $C^a$ coefficients common to all terms for brevity 
in the following equation) 
\be
\sum_{k=1}^3 {\cal A}_j^{a(s_k)} = C^a(\hat q)\frac{\nnp{j}\hat q}
{\hat q^2-\nnp{j}\hat q\nnm{j} l}
+ C^a(q_2)\frac{\nnp{j} q_2}{ q_2^2-\nnp{j} q_2\nnm{j} l}
+ C^a(q)\frac{\nnp{j} q}{ q^2-\nnp{j} q\nnm{j} l} \,,
\ee
with coefficients
\bea
C^a(\hat q) &=& 
g_s^2 \bigg(t^at^b\frac{\nnp{j}\hat q}{\hat q^2} 
+t^bt^a\Delta_{q_2\hat q}+if^{abc}t^c\Delta_{q\hat q}\bigg) C_j v_{j}(q_2) \,,
\nn\\
C^a(q_2) &=& 
g_s^2 t^bt^a\Delta_{\hat q q_2} C_j v_{j}(q_2) \,,
\nn\\[0.15cm]
C^a(q) &=& 
g_s^2 if^{abc}t^c\Delta_{\hat qq} C_j v_{j}(q_2) \,,
\eea
and
\be
\Delta_{qp} \equiv \left(\frac{q^2}{\nnp{j} q}-\frac{p^2}{\nnp{j}p}
\right)^{\!-1} = -\Delta_{pq} \,.
\ee
Using Eq.~\eqref{eq:Dj} with $a_j=1$, $b_j=0$, this gives
\bea
({\cal D}_j^a)_{{\rm single\, }j-{\rm coll.\, em.}} &=& 
\Bigg\{C^a(\hat q)\left(\frac{\hat q^2}{\nnp{j} \hat q}
\right)^{1-\epsilon} +
C^a(q_2)\left(\frac{q_2^2}{\nnp{j} q_2}\right)^{1-\epsilon} +
C^a(q)\left(\frac{q^2}{\nnp{j} q}\right)^{1-\epsilon} \Bigg\}
\nn\\
&=& g_s^2\mu^\epsilon\Bigg\{ t^at^b\left(\frac{\hat q^2}{\nnp{j} \hat q}
\right)^{-\epsilon}
+ t^bt^a \Delta_{q_2\hat q}\left(\left(\frac{\hat q^2}{\nnp{j} \hat q}
\right)^{1-\epsilon} -\left(\frac{q_2^2}{\nnp{j} q_2}\right)^{1-\epsilon}
\right)
\nn\\
&& {} +if^{abc}t^c\Delta_{q\hat q}\left(\left(\frac{\hat q^2}{\nnp{j} 
\hat q}\right)^{1-\epsilon} -\left(\frac{q^2}{\nnp{j} q}\right)^{1-\epsilon}
\right)\Bigg\}\,C_j u_{j}(q_2) 
\nn\\
&=& g_s^2\left\{ t^at^b-t^bt^a-if^{abc}t^c\right\}\left(\frac{\hat q^2}
{\nnp{j} \hat q}\right)^{-\epsilon}C_j u_{j}(q_2)  + {\cal O}(q^2,q_2^2) 
\nn\\
&=& {\cal O}(q^2,q_2^2) \,,
\eea
where in the last two lines we have expanded in the small off-shell 
regulators $q_2^2$ and $q^2$ of the external quark and gluon, respectively.
Therefore, in the on-shell limit $q^2,q_2^2\to 0$ when the regulators 
are removed, also all contributions with a single emission in direction $j$ 
vanish.

This result is not unexpected: 
It is well known that in the eikonal limit the coupling of a soft gluon 
to a pair of partons from collinear splitting is equal to the coupling 
to the parent parton. The above considerations proves that this holds true 
when the amplitude is first regulated by a small off-shellness, which is 
then removed. In the SCET framework the standard eikonal cancellation 
in the absence of the off-shell regulator is reflected in the 
decoupling transformation \cite{Bauer:2001yt}, which removes soft-gluon 
interactions from the leading-power Lagrangian. In the on-shell limit, 
the soft interaction is then described by a soft Wilson line evaluated at 
the position of the current (which we choose to be $x=0$). This gives 
\[
{\cal A}^a_j = \sum_{k=1}^3 {\cal A}^{a(s_k)}_j =
- \frac{g_st^a}{\nnm{j} l} \times\frac{-i\nnp{j}\hat{q}}{\hat{q}^2}\,
i g_s C_j v_j(q_2)\,,
\] 
which is the product of the Wilson line (eikonal) factor and the 
collinear splitting amplitude. Together with the explicit factor 
$\nnm{j} l$ in the numerator in Eq.~(\ref{eq:Dsoft}) this implies 
that the loop integral does not depend on the 
$\nnm{j}$ direction, and therefore vanishes. The above shows that 
in the present case the naive argument based on unregulated on-shell 
amplitudes remains valid as the limiting case of an off-shell regulated 
amplitude.

\paragraph{Double extra emission in direction $j$:}

\begin{figure}
\begin{center}
  \includegraphics[width=0.95\textwidth]{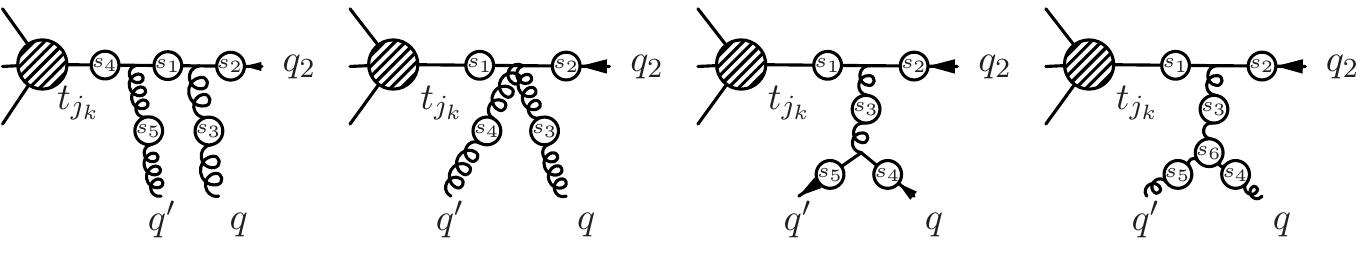}
\end{center}
\caption{\label{fig:soft_T2_double}
Diagrams with double collinear emission in direction $j$. The insertions 
$s_k$ show possible attachments of the soft line.
The part of the diagram in direction $i$ is not shown. 
Also the crossed diagram corresponding to the first one with 
momenta $q$ and $q'$ interchanged in not shown.
}
\end{figure}

The diagrams in Fig.~\ref{fig:soft_T2_double} show collinear splittings in 
the $j$ direction involving two extra emissions, and possible positions 
$s_k$ for attachment of the soft line in each case. One needs to sum up 
the amplitudes ${\cal A}_j^{a(s_k)}$ for all $s_k$. For each of the four 
classes of diagrams that are indicated in the figure, we find that, 
following the same steps as above, the 
${\cal D}_j^a$ part of the soft loop amplitude vanishes when the 
off-shell regulators are removed. Once again, this is a consequence 
of the SCET version of the leading-power eikonal-type couplings  
of soft gluons to collinear lines. Thus,
\bea
({\cal D}_j^a)_{{\rm double\, }j-{\rm coll.\, em.}} &=& 
{\cal O}(q_2^2,q^2,{q'}^2)
\eea
Altogether, this implies that the time-ordered products $J^{T2}_{\chi,V}$ do 
\emph{not} mix into current operators, i.e.
\be
\gamma^{ij}_{T(P,{\cal L}_k^{(2)}),Q} = 0\,,
\ee
where $i,j,k=1,\dots,N$ and $P, Q$ are $N$-jet current operators with fermion 
number one. 

\subsection{Soft-quark exchange}
\label{sec:xiq}

\begin{figure}[t]
\begin{center}
  \includegraphics[width=0.2\textwidth]{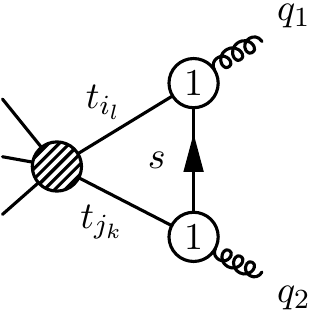}
\end{center}
\caption{\label{fig:soft_qqbar}
Example for a soft fermion exchange one-loop diagram.
}
\end{figure}

For the mixed cases $(F_i,F_j)=(+1,-1)$ and 
$(F_i,F_j)=(-1,+1)$, an additional class of diagrams with soft-quark exchange 
from  two ${\cal L}_{\xi q}^{(1)}$ 
insertions exists. An example is shown in Fig.~\ref{fig:soft_qqbar}.
We show that these diagrams vanish, if the mass of the soft quark 
can be neglected.

Even for an arbitrary number of collinear emissions attached to any 
internal or external propagator, or vertex, the soft loop momentum $l$ 
enters the integral only via $\nnm{i} l$ and $\nnm{j} l$, except for the 
soft (massless) quark propagator $\slashed{l}/l^2$.
This gives a loop integral of the form
\be
 {\cal I}_{\xi q} = -i\tilde\mu^{2\epsilon}\int \frac{d^dl}{(2\pi)^d}\, 
\slashed{l} F(\nnm{i} l,\nnm{j} l,l^2) = C_1\nnms{i}+C_2\nnms{j}\,,
\label{eq:Ixiq}
\ee
with some function $F$ and coefficients $C_k$. The vertex from 
${\cal L}_{\xi q}^{(1)}$ contracts the soft quark propagator indices 
with collinear quarks. Therefore we may insert a collinear
projector with respect to $i$ and $j$ to the left and to the right, 
respectively. This results in 
\be
\frac{\nnps{i}\nnms{i}}{4}{\cal I}_{\xi q}\frac{\nnms{j}\nnps{j}}{4}
=\frac{\nnps{i}\nnms{i}}{4}(C_1\nnms{i}+C_2\nnms{j})
\frac{\nnms{j}\nnps{j}}{4}=0\,.
\label{eq:fermionprojection}
\ee
Therefore, generally, there is no one-loop soft-quark exchange mixing
\be
  J^{T1}_{\psi,\xi q}J^{T1}_{\psi',\xi q} \to 0\,,
\ee
for $\psi,\psi'=\chi,\bar\chi,{\cal A}$.
Since operators involving only a single ${\cal L}_{\xi q}^{(1)}$ insertion 
vanish trivially (the soft quark cannot be connected to different collinear 
directions at leading power), this generalizes to
\be\label{eq:softquark}
  J^{T1}_{\psi,\xi q}J^{X1} \to 0\,,
\ee
where $J^{X1}$ can be an arbitrary  ${\cal O}(\lambda)$ current or time-ordered product.
This can be summarized by
\be
\gamma^{ij}_{T(P,{\cal L}^{(1)}_{k,\xi q}),Q} =  
\gamma^{ij}_{T(P,{\cal L}^{(1)}_{k,\xi q},{\cal L}^{(1)}_{l,\xi q}),Q} 
=0\,,
\ee
for $i,j,k,l=1,\dots,N$ and arbitrary current operators $P, Q$. The absence 
of diagrams with a soft quark line implies that fermion number is conserved 
in each collinear sector separately up to one-loop and $\order{\lambda^2}$, 
which allows us to classify the next-to-leading power anomalous dimension 
according to collinear sectors with definite fermion number.

The vanishing of mixing from soft-quark exchange holds only 
for massless fermions as assumed throughout this paper. 
As an aside, we note that when the fermion mass $m$ is parametrically of order 
of the soft scale, $\slashed{l}\to\slashed{l}+m$ in Eq.~\eqref{eq:Ixiq}, 
which adds a term $C_3 m$ to the right-hand side of this equation. This 
term is not projected to zero in Eq.~\eqref{eq:fermionprojection}. 
An explicit example of the relevance of soft-fermion exchange can be 
found in Ref.~\cite{Beneke:2017vpq}, where it contributes to the 
leading logarithm of a power-enhanced electromagnetic effect in the 
rare $B$-meson decay $B_s\to\ell^+\ell^-$. Technically, the basis of 
${\cal O}(\lambda^2)$ suppressed operators must be extended by mass-suppressed 
operators $J_i^{A2} = m J^{A0}_i$, and the non-zero mixing 
is of the form 
\be
  J^{T1}_{\psi_i,\xi q}J^{T1}_{\psi_j^\prime,\xi q} \to m J^{A0}_iJ^{A0}_j\,.
\ee

\section{Collinear sector}
\label{sec:coll}

In the collinear sector it is sufficient to consider a single collinear 
direction, say $i$, since collinear fields corresponding to different 
directions do not interact with each other. We categorize different cases 
by their fermion number $F_i$ and power suppression $\lambda^n$. Results for 
$F_i=-1$ can be obtained from $F_i=+1$ by hermitian conjugation 
(see App.~\ref{app:cc} for details). The case $F_i=2$ was treated in 
Ref.~\cite{Beneke:2017ztn}. Note that $|F_i| \leq n+1$ since each additional 
fermionic building block costs a power of $\lambda$ relative to the leading 
power. In the following we consider the cases $F_i=1$ and $F_i=3$. 
Since the time-ordered product operators inherit their collinear anomalous 
dimension from the current operators, see Eq.~\eqref{eq:gammacs}, 
we give only the current-current part of the anomalous 
dimension matrix $\gamma^i_{PQ}$. 

\subsection{\boldmath${\mathcal{O}(\lambda)}$}
\label{sec:colllambda1}

At ${\cal O}(\lambda)$ $F_i=3$ is not possible. For $F_i=1$ we find 
for the collinear anomalous dimension $\gamma^i_{PQ}$ in 
Eq.~\eqref{eq:GammaPQ}, 
\be
\label{eq:gammacoll_lambda}
  \gamma^i_{PQ} \quad = \quad \begin{array}{c||c|c}
                 & J^{A1}_{\partial\chi} & J^{B1}_{{\cal A}\chi}  \\ \hline\hline
  J^{A1}_{\partial\chi} & 0 & 0 \\ \hline
  J^{B1}_{{\cal A}\chi} & 0 & \gamma^i_{{\cal A}\chi,{\cal A}\chi} \\ 
  \end{array}
\ee
where the non-zero entry is given in App.~C of Ref.~\cite{Beneke:2017ztn}.
The $A$-type operator $J^{A1}_{\partial\chi}$ has matrix 
elements identical to the leading power operator $J^{A0}_{\chi}$, up to 
overall factors of external momenta due to the total derivative.
This implies $\gamma^i_{J^{A1}J^{B1}}=0$. The diagonal anomalous dimension
of $J^{A1}_{\partial\chi}$ is already accounted for by the first line of 
Eq.~\eqref{eq:GammaPQ}, and therefore also $\gamma^i_{J^{A1}J^{A1}}=0$.
To show $\gamma^i_{J^{B1}J^{A1}}=0$ we compute the matrix element of 
$J^{B1}_{{\cal A}\chi}$ for an external state with a single fermion, 
and find that it vanishes.

\subsection{\boldmath${\mathcal{O}(\lambda^2)}$, overview}
\label{sec:colllambda2}

At ${\cal O}(\lambda^2)$, we find for $F_i=1$
\bea\label{eq:gammacoll}
 \gamma^i_{PQ} \quad = \quad \begin{array}{c||c|cc|ccc} 
  & J^{A2}_{\partial\partial\chi} & J^{B2}_{{\cal A}\partial\chi} & 
J^{B2}_{\partial({\cal A}\chi)} & J^{C2}_{{\cal AA}\chi} & J^{C2}_{\chi\bar\chi\chi} 
\\ \hline\hline
  J^{A2}_{\partial\partial\chi} & 0 & 0  &0 &0 &0 \\ \hline
  J^{B2}_{{\cal A}\partial\chi} & 0  & \eqref{eq:gamma_Adelchi_Adelchi} & \eqref{eq:gamma_Adelchi_delAchi} & \eqref{eq:gamma_Adelchi_AAchi} & \eqref{eq:gamma_Adelchi_chichibarchi} \\ 
  J^{B2}_{\partial({\cal A}\chi)} & 0  & 0 & \eqref{eq:gamma_delAchi_delAchi} &0 &0  \\ \hline
  J^{C2}_{{\cal AA}\chi} & 0  & 0 & 0 & \eqref{eq:gamma_AAchi_AAchi} & \eqref{eq:gamma_AAchi_chichibarchi}   \\ 
  J^{C2}_{\chi\bar\chi\chi} & 0  & 0 & 0 & \eqref{eq:gamma_chichibarchi_AAchi} & \eqref{eq:gamma_chichibarchi_chichibarchi}   \\ 
  \end{array}
\eea
The first row vanishes, which follows from an argument analogous to the 
${\cal O}(\lambda)$ case. The non-zero entries of $\gamma^i_{PQ}$ 
point to the equation numbers of the 
corresponding results given below. They can be divided into three cases: 
first, mixing of $B$-type currents into $B$-type currents (middle block);
second, mixing of $B$-type currents into $C$-type currents (last two columns of second row); and 
third, mixing of $C$-type currents into $C$-type currents (lower right block). 
In the following we discuss these three cases in turn, see 
Secs.~\ref{sec:BB} to \ref{sec:CC}.

Let us briefly comment on the remaining zero entries. For the first column, 
second row, we compute a matrix element of $J^{B2}_{{\cal A}\partial\chi}$ 
with a single fermion of momentum $p$ and find that the result is proportional 
to the off-shell regulator $p^2$. This implies that the corresponding 
anomalous dimension vanishes in the on-shell limit.\footnote{However, a 
related one-particle reducible diagram with a collinear emission off the 
external fermion contributes to the $B$-to-$B$ mixing, see Sec.~\ref{sec:BB} 
and diagram $(d)$ in Fig.~\ref{fig:qg}.}
The renormalization of $J^{B2}_{\partial({\cal A}\chi)}$ is identical
to $J^{B1}_{{\cal A}\chi}$ at ${\cal O}(\lambda)$ due to the total derivative, 
which implies the  zero entries in the third row. Finally, as will be 
discussed in Sec.~\ref{sec:CC}, at the one-loop order considered here, the 
renormalization of $C$-type currents can be related to the one of $B$-type 
currents at ${\cal O}(\lambda)$. From Eq.~\eqref{eq:gammacoll_lambda} 
together with corresponding results found in Ref.~\cite{Beneke:2017ztn} this 
implies the zero entries in the last two rows.

For $F_i=3$ only the single operator $J^{C2}_{\chi\chi\chi}$ exists at 
${\cal O}(\lambda^2)$. The corresponding anomalous dimension is given
in Sec.~\ref{sec:CC}, see Eq.~\eqref{eq:gamma_chichichi_chichichi}.

Before turning to the explicit computation, we comment on the mixing 
into operators with gluon building blocks, which requires the calculation 
of matrix element with external gluons. We implement the transversality 
condition $\epsilon\cdot q=0$ of the polarization vector of a gluon with 
momentum $q$ by eliminating $\nnm{i}\epsilon$ through the 
identity
\begin{equation}
\nnm{i}\epsilon = -\frac{\nnp{i}\epsilon \nnm{i}q + 
2 \epsilon_{\perp i}\cdot q_{\perp i}}{\nnp{i} q}\,.
\end{equation}
This is consistent with the fact that we do not consider operators 
containing the building block $\nnm{i}{\cal A}$, which can be eliminated 
by an equation-of-motion identity~\cite{Beneke:2017ztn}.\footnote{Note that 
if we first included $\nm{\cal A}$ explicitly and then eliminated it
using the equation of motion at the operator level, this would also give a 
contribution to the mixing into $C$-type operators $J^{C2}_{{\cal AA}\chi}$ 
and $J^{C2}_{\chi\bar\chi\chi}$. 
Here we prefer \emph{not} to use the building block $\nm{\cal A}$ and its 
equation of motion explicitly. Instead, the contribution to mixing into
$C$-type operators that would arise from first introducing and then 
eliminating $\nm{\cal A}$ is,  in our computation, included in the 1PR 
diagrams contributing to $B$-to-$C$ mixing.}
In practice, the above equation can be simplified. Knowing that 
$\nnp{i} A$ appears only within collinear Wilson lines, it is never 
necessary to consider diagrams with external $\nnp{i} A$ gluons, hence 
we can replace 
\begin{equation}\label{eq:nmepsilon}
\nnm{i}\cdot\epsilon \to 
-\frac{2 \epsilon_{\perp i}\cdot q_{\perp i}}{\nnp{i} q}\,.
\end{equation}
Whenever the operator under consideration does not contain transverse 
derivatives, the external transverse momentum $q_{\perp i}$ may be 
set to zero, in which case the calculation can be performed from the 
beginning assuming $n_{i \pm}\cdot \epsilon =0$.

\subsection{Mixing of $B$-type currents into $B$-type currents}\label{sec:BB}

\begin{figure}
\begin{center}
  \includegraphics[width=0.75\textwidth]{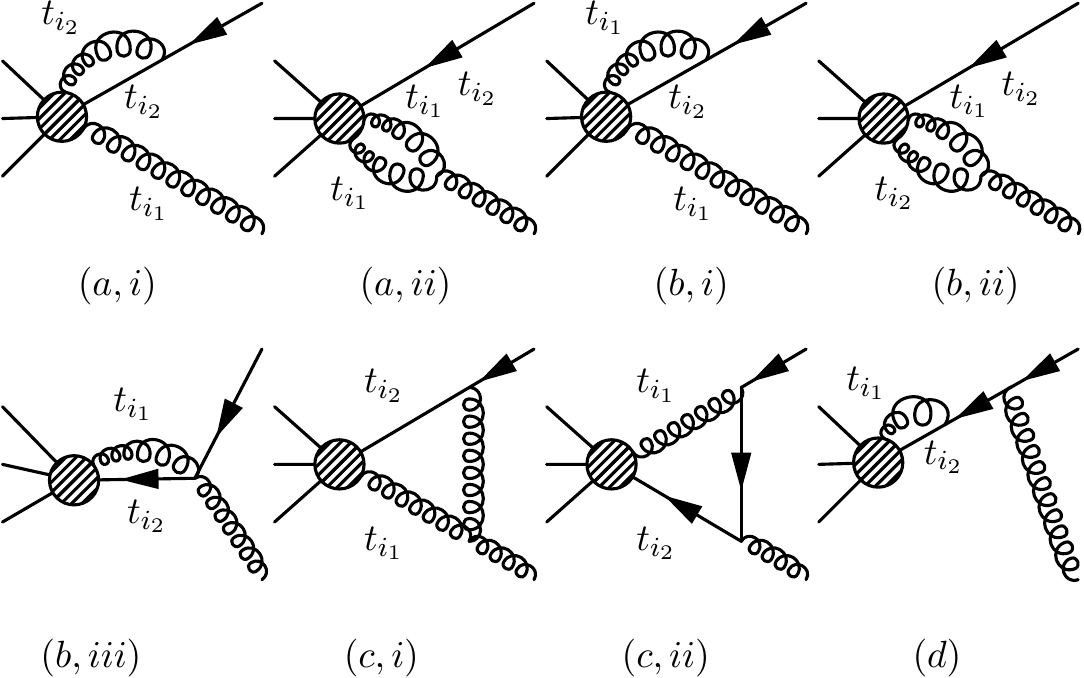}
\end{center}
\caption{\label{fig:qg}
Diagrams contributing to the mixing of $B$-type currents into $B$-type 
currents in the collinear sector with fermion number $F_i=1$.
}
\end{figure}

We start with the second row of Eq.~\eqref{eq:gammacoll}, i.e. the entry for 
$J^{B2}_{{\cal A}\partial\chi}(x)\to J^{B2}_{{\cal A}\partial\chi}(y)$
and the mixing $J^{B2}_{{\cal A}\partial\chi}(x)\to J^{B2}_{\partial({\cal A}
\chi)}(y)$. $B$-type currents depend only on a single independent collinear 
momentum fraction, which we denote by $x\equiv x_{i_1}$ ($y\equiv y_{i_1}$ 
for the operator mixed into.) The momentum fraction carried by the second 
(in the present case, fermion) building block is $\bar x=1-x$ and 
$\bar y=1-y$, respectively. In order to extract the anomalous dimension, we 
consider a matrix element of $J^{B2}_{{\cal A}\partial\chi}$ with an outgoing 
antiquark and a gluon. The corresponding collinear one-loop diagrams are 
shown in Fig.~\ref{fig:qg}. In our convention, $F_i=+1$ corresponds to fermion 
flow directed towards the current, as indicated by the arrows. 
As discussed above, we assume that the polarization vector for the external gluon
satisfies $\nnp{i}\epsilon=0$, and replace $\nnm{i}\epsilon$ according to 
Eq.\,\eqref{eq:nmepsilon}.\footnote{One may wonder what is the simplest possible choice for the polarization vector that
allows one to uniquely extract the anomalous dimension. In many cases a polarization vector for which $n_{i \pm}\epsilon=0$ is
sufficient. However, employing a polarization vector with non-zero $\nnm{i}\epsilon$ projection turns out to be necessary
for the present calculation. In particular, in order to be able to extract the
anomalous dimension, it is necessary to choose a matrix element such that the tree-level matrix elements
of the operators $J^{B2}_{{\cal A}^\rho\partial^\sigma\xi}$ and
$J^{B2}_{\partial^\sigma({\cal A}^\rho\xi)}$ are non-zero and linearly independent
for all possible values of $\rho$ and $\sigma$. This is only the case if we allow for
a non-zero value of $\nnm{i}\epsilon$.}

The diagrams can be classified as in Ref.~\cite{Beneke:2017ztn}. Loops 
involving only internal lines attached to a \emph{single} collinear building 
block ($(a, i)$ and $(a, ii)$ in Fig.~\ref{fig:qg}) are responsible for the 
contributions to $\delta Z_{PQ}^{c,i}(x,y)$ with $k=l$ in Eq.~\eqref{eq:Zc}, 
that encompass a double pole  $1/\epsilon^2$ and are diagonal with respect to 
collinear momentum fractions $x$ and $y$. They do not contribute to the part 
$\gamma^i_{PQ}$ that is off-diagonal with respect to the momentum fractions, 
and proportional to a single power of $1/\epsilon$. For completeness we report 
the result for the sum of diagrams $(a,i)$ and $(a,ii)$, added to the 
tree-level result, and adding also the contributions from the right-hand side 
of the renormalization condition \eqref{eq:rencond} that involves field 
renormalization factors. We denote this particular sum of terms by the 
subscript $(a)$,
\be
\langle \bar q(p) g(q)|J^{B2}_{{\cal A}\partial\chi}(x)|0
\rangle_{(a)} = J_q(p^2)J_g(q^2)\int dy \delta(x-y)\langle \bar q(p)\bar g(q)|
J^{B2}_{{\cal A}\partial\chi}(y)|0\rangle_{\rm tree}\,,
\ee
where
\bea\label{eq:Jqg}
J_q(p^2) &=& 1 + \frac{\als C_F}{4\pi}
\left[ \frac{2}{\epsilon^2}+\frac{2}{\epsilon}
\ln\left(\frac{\mu^2}{-p^2}\right)+\frac{3}{2\epsilon}\right]
+{\cal O}(\epsilon^0)\,,\nn\\
J_g(q^2) &=& 1 + \frac{\als C_A}{4\pi}
\left[ \frac{2}{\epsilon^2}+\frac{2}{\epsilon}
\ln\left(\frac{\mu^2}{-q^2}\right)\right]
+{\cal O}(\epsilon^0)
\eea
coincide with the leading-power collinear contributions 
from a single fermionic or gluonic building block\footnote{Note the different 
normalization of the gluon building block ${\cal A}_{\perp i}^\mu$ compared
to Ref.~\cite{Becher:2014oda} which explains the different coefficient of the 
$1/\epsilon$ term in $J_g$.} \cite{Becher:2009qa, Becher:2014oda}.
We also introduced an integration over $y$ in order to stress that 
contributions from $(a, i)$, $(a,ii)$  and field renormalization are diagonal 
with respect to the momentum fractions.

Loops involving internal lines that are attached to the two different building 
blocks may change momentum fractions and therefore contribute to 
$\gamma^i_{PQ}$. In addition, $(b, i)$ and $(b, ii)$ in Fig.~\ref{fig:qg} also yield diagonal contributions proportional to $\delta(x-y)$
that provide the terms with $k\not=l$ in \eqref{eq:Zc}. At 
${\cal O}(\lambda^2)$, as considered here, the one-particle reducible (1PR) 
diagram $(d)$  needs to be taken into account. The loop itself is proportional to the sum of all external momenta squared, 
which cancels the 1PR propagator, and yields a non-zero contribution.

Adding all contributions, we find for $J^{B2}_{{\cal A}\partial\chi}\to 
J^{B2}_{{\cal A}\partial\chi}$ (all results for collinear contributions $\gamma^i_{PQ}$ refer to collinear direction $i$; we omit
the label $i$ of the corresponding light-cone basis vectors and $\perp$ projections for brevity here and below)
\bea\label{eq:gamma_Adelchi_Adelchi}
  \gamma^{i}_{{\cal A}^\mu\partial^\nu\chi, {\cal A}^\rho\partial^\sigma\chi}(x,y) &=& 
   {}   g_{\perp }^{\mu\rho}g_{\perp }^{\nu\sigma}\frac{\alpha_s {\bf T}_{i_1}\cdot{\bf T}_{i_2}}{2\pi} \Bigg\{ \theta(x-y)\left[\frac{1}{x-y}\right]_+ + \theta(y-x)\left[\frac{1}{y-x}\right]_+ \nn\\
  && {} -\theta(x-y)\frac{\bar x+\bar y}{\bar y^2} -\theta(y-x)\frac{x+2y}{2y^2} \Bigg\}\nn\\
  && {} + \frac{\alpha_s {\bf T}_{i_1}\cdot{\bf T}_{i_2}}{8\pi} M^{\mu\nu,\rho\sigma}(x,y)  - \frac{\alpha_s( {\bf C_F}+{\bf T}_{i_1}\cdot{\bf T}_{i_2})}{8\pi} N^{\mu\nu,\rho\sigma}(x,y) \nn\\
  && {} + \frac{\alpha_s {\bf C_F} }{8\pi}\frac{\bar x}{\bar y} (2g_\perp^{\mu\nu}-x\gamma_\perp^\mu\gamma_\perp^\nu)
  \left(\gamma_\perp^\rho\gamma_\perp^\sigma +\frac{2\bar y}{y}g_\perp^{\rho\sigma}\right) \,,
\eea
where, in colour operator notation, ${\bf C_F} \equiv \frac16(1-3(\T_{i_1}+
{\bf D}_{i_1})\cdot \T_{i_2})$, see Ref.~\cite{Beneke:2017ztn}.\footnote{The 
colour operator ${\bf D}^b|a\rangle = d^{abc}|c\rangle$ involves the symmetric 
$d^{abc}$ symbol related to the anticommutator of $SU(3)$ Gell-Mann matrices 
$\{t^a,t^b\} = \frac{1}{3}\delta^{ab}+d^{abc}t^c$.}. The terms in curly 
brackets arise from diagrams $(b, i)$ and $(b, ii)$, while $(b, iii)$ gives 
no contribution. The parts obtained from diagrams $(c, i)$ and $(c, ii)$ are 
lengthy expressions encapsulated in the coefficients 
$M^{\mu\nu,\rho\sigma}(x,y)$ and $N^{\mu\nu,\rho\sigma}(x,y)$, respectively, 
see App.~\ref{app:coll}. The last line arises from the 1PR diagram $(d)$.
The anomalous dimension also features a non-trivial Dirac structure, with 
spinor indices $(\dots)_{\alpha\beta}$ corresponding to 
$\gamma^{i}_{{\cal A}^\mu\partial^\nu\chi_\alpha, {\cal A}^\rho\partial^\sigma\chi_\beta}$ left implicit. Products of four transverse Dirac matrices could 
be reduced to expressions with at most two Dirac matrices up to 
$\mathcal{O}(\epsilon)$ terms that correspond to a finite mixing into 
evanescent operators. However, we will not perform such simplifications 
of the anomalous dimension matrix and do not make use of identities 
valid only in four dimensions here and below. 

For the operator mixing $J^{B2}_{{\cal A}\partial\chi}\to J^{B2}_{\partial({\cal A}\chi)}$ we find
\bea\label{eq:gamma_Adelchi_delAchi}
  \gamma^{i}_{{\cal A}^\mu\partial^\nu\chi, \partial^\sigma({\cal A}^\rho\chi)}(x,y) &=& 
  g_{\perp }^{\mu\rho}g_{\perp }^{\nu\sigma}\frac{\alpha_s  {\bf T}_{i_1}\cdot{\bf T}_{i_2}}{4\pi}  \frac{\theta(y-x)}{y^2}(y+x) \nn\\
  && {} + \frac{\alpha_s  {\bf T}_{i_1}\cdot{\bf T}_{i_2}}{8\pi} \hat M^{\mu\nu,\rho\sigma}(x,y)  - \frac{\alpha_s ( {\bf C_F}+{\bf T}_{i_1}\cdot{\bf T}_{i_2})}{8\pi} \hat N^{\mu\nu,\rho\sigma}(x,y) \nn\\
  && {} +\frac{\alpha_s {\bf C_F} }{8\pi} \bar x \left[ 2\bar x
        g_\perp^{\nu\sigma}\gamma_\perp^\mu\gamma_\perp^\rho
  +  (2g_\perp^{\mu\nu}-x\gamma_\perp^\mu\gamma_\perp^\nu)\left(\gamma_\perp^\sigma\gamma_\perp^\rho -\frac{2}{y}g_\perp^{\rho\sigma}\right) \right]\,.\nn\\
\eea
Here diagram $(c, ii)$ yields a contribution that has a pole $\propto 1/(\bar x
-y)$, which cancels when combining with the part of diagram $(b, iii)$ that is 
proportional to ${\bf C_F}+{\bf T}_{i_1}\cdot{\bf T}_{i_2}$. The result is 
collected in the coefficient $\hat N^{\mu\nu,\rho\sigma}(x,y)$ given in 
App.~\ref{app:coll}, together with $\hat M^{\mu\nu,\rho\sigma}(x,y)$
obtained from diagram $(c, i)$. The last line contains the remaining contribution 
from diagram $(b, iii)$, as well as the contribution from diagram $(d)$.

Let us now turn to the third row of Eq.~\eqref{eq:gammacoll}, related to the 
renormalization of $J^{B2}_{\partial({\cal A}\chi)}$. Due to the total 
derivative, all matrix elements of this operator are identical to those 
containing $J^{B1}_{{\cal A}\chi}$ up to an overall factor containing the sum 
of external momenta. This property holds both at tree and loop 
level. Therefore, as mentioned above, the corresponding anomalous dimensions 
are related. From Eq.~\eqref{eq:gammacoll_lambda}
we find that the only non-zero contribution is given by
\be\label{eq:gamma_delAchi_delAchi}
  \gamma^{i}_{\partial^\nu({\cal A}^\mu\chi), \partial^\sigma({\cal A}^\rho\chi)}(x,y)
  = g_\perp^{\nu\sigma}\gamma^{i}_{{\cal A}^\mu\chi, {\cal A}^\rho\chi}(x,y)\,.
\ee

\subsection{Mixing of $B$-type currents into $C$-type 
currents}\label{sec:BC}

As discussed before, only the $B$-type current $J^{B2}_{{\cal A}\partial\chi}$ 
can mix into $C$-type currents with three collinear building blocks. We first 
discuss mixing into $J^{C2}_{{\cal AA}\chi}$, then into $J^{C2}_{\chi\bar\chi
\chi}$. The $B$-type current $J^{B2}_{{\cal A}\partial\chi}$ can be described 
by the single collinear momentum fraction $x\equiv x_{i_1}$ of the gluon 
building block, with $\bar x=1-x=x_{i_2}$ for the fermion then being fixed.
The $C$-type currents are parameterized by two independent momentum 
fractions, denoted by $y_1\equiv y_{i_1}$ and $y_2\equiv y_{i_2}$. The 
momentum fraction of the last building block is $y_3=1-y_1-y_2$.
According to the renormalization condition Eq.~\eqref{eq:rencond}, the 
anomalous dimension is then a function of $x, y_1$ and $y_2$.

\subsubsection{Mixing $J^{B2}_{{\cal A}\partial\chi}(x)\to 
J^{C2}_{{\cal AA}\chi}(y_1,y_2)$ }

We consider the matrix element of $J^{B2}_{{\cal A}\partial\chi}$
with one outgoing antiquark and two gluons. It is sufficient to consider 
external momenta with vanishing $\perp$ component (up to a subtlety for 1PR 
diagrams, that we will discuss below). As mentioned above, and in contrast 
to the mixing into $B$-type operators, the anomalous dimension can be extracted uniquely
when using gluon polarization vectors with $n_{i\pm}\epsilon=0$. This is the 
simplest choice that leads to a non-zero overlap with $J^{C2}_{{\cal AA}\chi}$.
Then the tree-level matrix element of $J^{B2}_{{\cal A}\partial\chi}$
vanishes, because for each diagram the $\perp$ derivative contained in the 
current leads to terms involving some linear combination of external 
transverse momenta, which are set to zero here.
A similar argument implies that we do not have to consider diagrams 
containing counterterms other than the one we are interested in.
In addition, all loops attached to a \emph{single} collinear building block (called type-$(a)$ in our notation) vanish,
\be\label{eq:BtoCatype}
  \langle g(q_1)g(q_2)\bar q(p) | J^{B2}_{{\cal A}\partial\chi} |  0 \rangle_{(a)} =  0\,,
\ee
because all propagators that belong to the loop are attached to a single 
building block. The derivative contained in the current is then again turned  
into a linear combination of \emph{external} momenta, and therefore 
$i\partial_\perp \to 0$.

\begin{figure}
\begin{center}
  \includegraphics[width=0.75\textwidth]{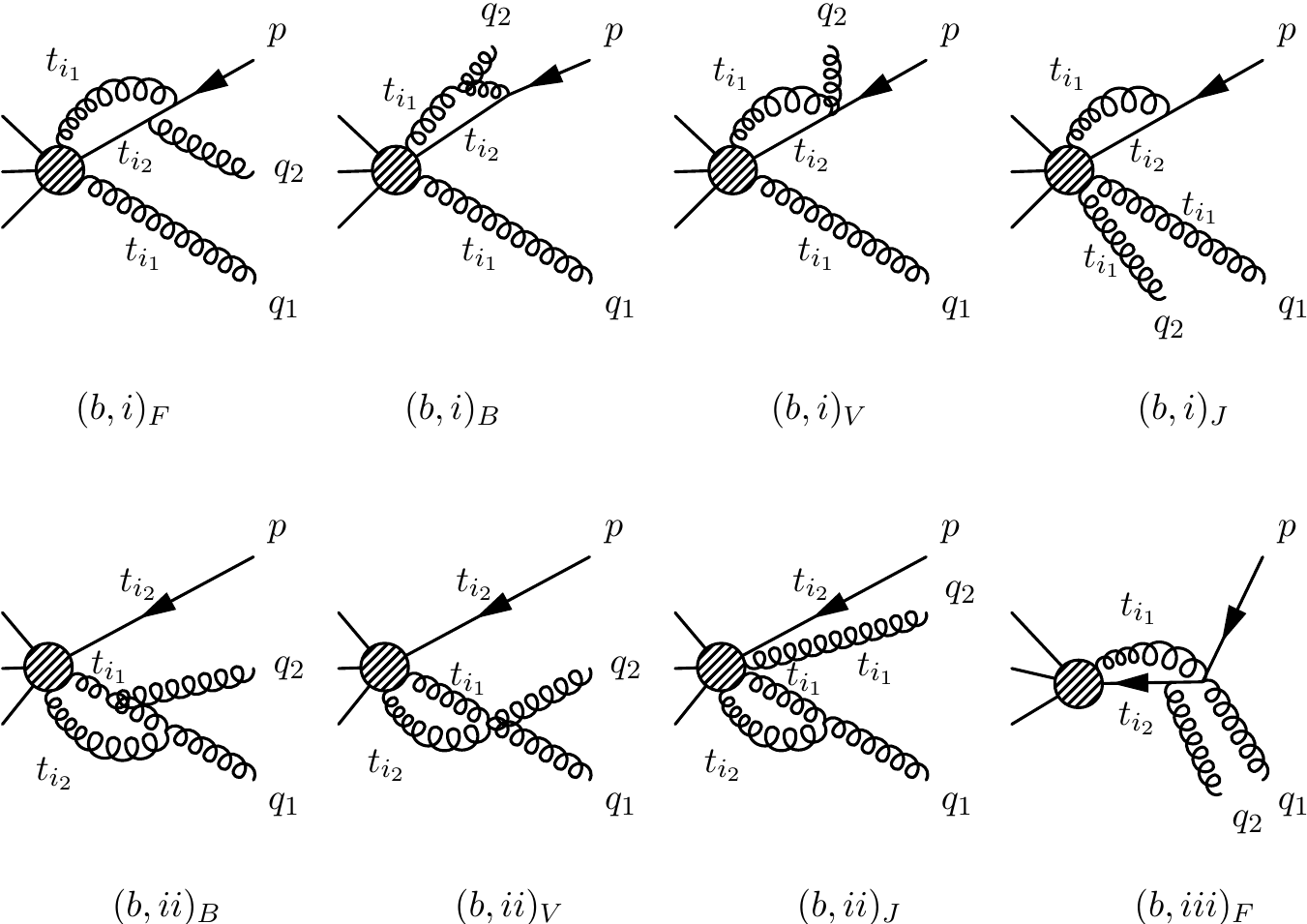}\\[3ex]
  \includegraphics[width=0.75\textwidth]{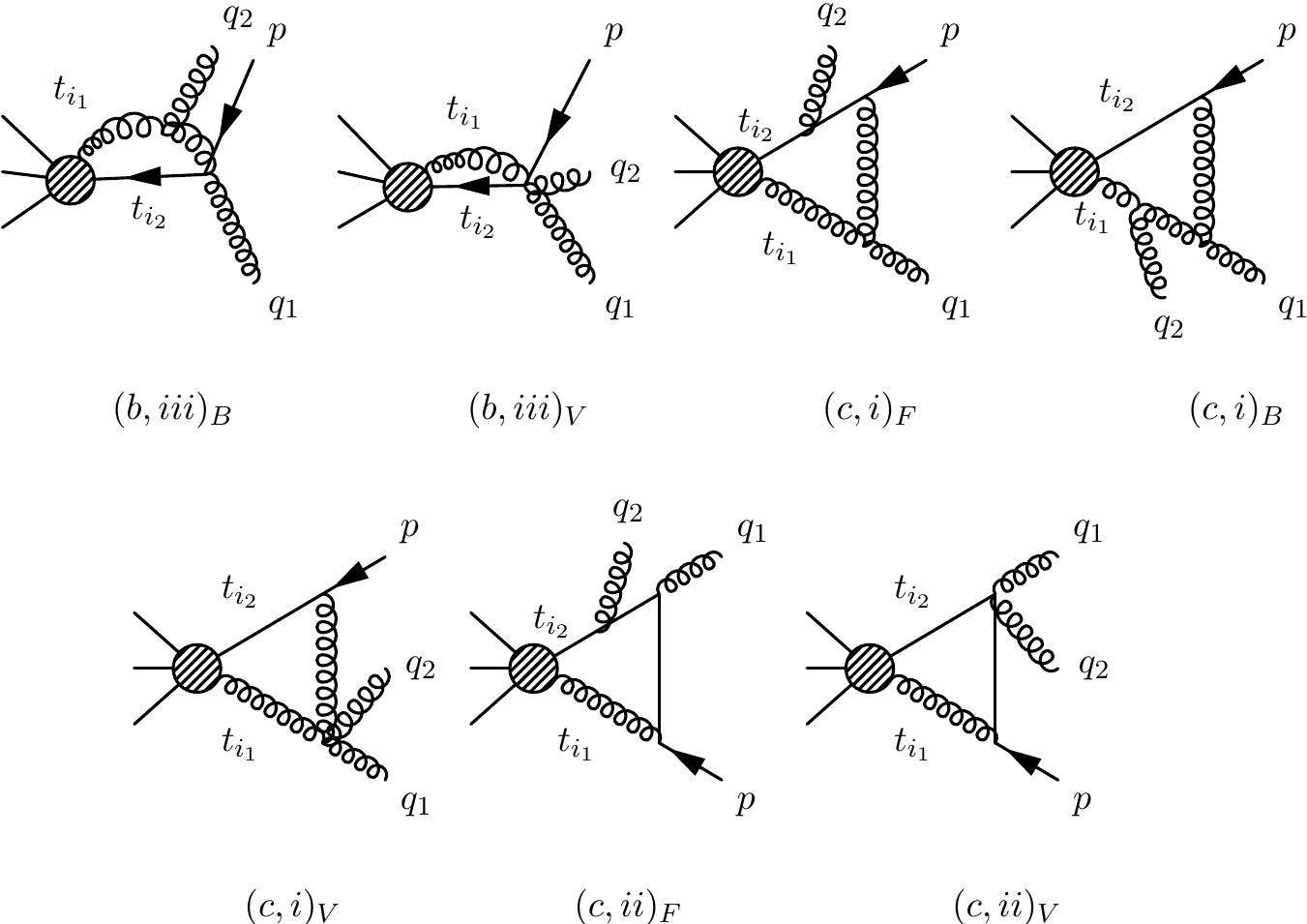}
\end{center}
\caption{\label{fig:qgg}
1PI diagrams contributing to the mixing $J^{B2}_{{\cal A}\partial\chi}\to J^{C2}_{{\cal AA}\chi}$ of a $B$- into a $C$-type current in the collinear
sector with fermion number $F_i=1$.
}
\end{figure}

The remaining diagrams can be classified as follows: one-particle irreducible 
(1PI) diagrams are derived from the diagrams of type $(b)$ and $(c)$ in 
Fig.~\ref{fig:qg} with an additional gluon emitted off either an internal 
fermion (quark) line (subscript $F$), an internal boson (gluon) line ($B$), 
a vertex ($V$), or directly from the operator ($J$). 
In addition, there are 1PR diagrams (called type $(d)$ loops), that we will 
discuss further below. 
The relevant 1PI diagrams are shown in Fig.~\ref{fig:qgg}.
Diagrams that differ only by permutation of the gluon lines are not included. In addition, when generating the diagrams according to the
procedure described above, it is possible to obtain the same diagram several times. Accordingly, we omitted equivalent diagrams.
For example a potential contribution $(b, iii)_J$, 
for which the gluon with momentum $q_2$ is attached to the operator, is already taken into account by $(b, i)_V$ when permuting the gluon lines.
In addition, diagrams for which one of the external gluon lines is attached directly to the Wilson line contained within the
fermionic building block $\chi$ are not shown, because they vanish for external $\perp$ polarization.

Several of the displayed diagrams are zero due to our choice of external 
momenta and polarization vectors: 
\begin{itemize}
\item In diagram $(b, i)_J$ at least one of the external gluons is attached to a Wilson line,
and it therefore vanishes. 
\item In diagram $(b, i)_V$, since the external gluon line with momentum 
$q_1$ has $\perp$ polarization, the internal gluon line picks
up a factor $\np$ from the Feynman rule for the gluon building block. When 
multiplying with the vertex (\ref{eq:Vxixicc}), one obtains zero.
\item Similarly, in diagram $(b, ii)_J$ both internal gluons come with 
factors of $\np$. The three-gluon vertex \eqref{eq:threegluon} contracted 
as $\np^\rho\np^\lambda Q_{\rho\lambda\sigma}\epsilon_{1\perp}^{*\sigma} =0$ 
vanishes. 
\item In diagram $(b, ii)_V$ the internal gluon attached to the fermionic building 
block involves a factor $\np$, and the one to the gluonic building
block either $\np$ or $\perp$, such that there are two possible
contractions of the four-gluon vertex \eqref{eq:fourgluon},
$\np^\rho\np^\lambda Q_{\rho\lambda\sigma\kappa}\epsilon_{1\perp}^{*\sigma}
\epsilon_{2\perp}^{*\kappa} =0$,  $\np^\rho g_\perp^{\mu\lambda} 
Q_{\rho\lambda\sigma\kappa}\epsilon_{1\perp}^{*\sigma}\epsilon_{2\perp}^{
*\kappa} =0$ that both vanish. 
\item Diagram $(b, iii)_V$ involves a vertex with two collinear quarks and 
three collinear gluons. From the collinear SCET Lagrangian 
\eqref{eq:SCETLagrangian} one sees that at most two gluon fields can be 
transverse, hence in the above vertex at least one gluon comes from a 
Wilson line and therefore picks up a factor $\np$. This has to be the 
internal line, since the two external line have $\perp$ polarization. Then 
the $\np$ multiplied with the Feynman rule for the gluon building block 
vanishes.
\end{itemize}

The left-over diagrams are $(b, i)_{F/B}$, $(b, ii)_B$, $(b, iii)_{F/B}$, 
$(c, i/ii)_{F/V}$, $(c, i)_B$. In the limit $\epsilon\to 0$ they yield a 
single $1/\epsilon$ pole and are non-diagonal in momentum fractions and 
therefore contribute to the anomalous dimension. Some of them feature a 
simple pole singularity in collinear momentum fractions for particular 
configurations. We checked that these poles either cancel when adding up all 
diagrams, or lie outside of the support of Heaviside functions multiplying 
them. For example, $(c, i)_F$ has a single pole for $\bar x\to y_2$, that 
cancels with the corresponding pole of a diagram related
to $(b, iii)_B $ by interchanging the external gluon lines. Further, $(c, ii)_F$  has single poles for $\bar x\to y_2$ and $x\to y_3$.
The singularity $\propto 1/(y_3-x)$ cancels with $(b, iii)_F $, and the singularity $\propto 1/(\bar x-y_2)$ with $(c,ii)_V$.
Diagram $(c,ii)_V$ has a further singularity $\propto 1/(\bar x-y_1)$ that cancels with the contribution
analogous to $(c,ii)_F$ with interchanged external gluon lines. Note that the diagram $(c, ii)_V$ remains unchanged when interchanging external
gluons, and therefore one should \emph{not} add a diagram with permuted external lines in this case.

\begin{figure}
\begin{center}
  \includegraphics[width=0.85\textwidth]{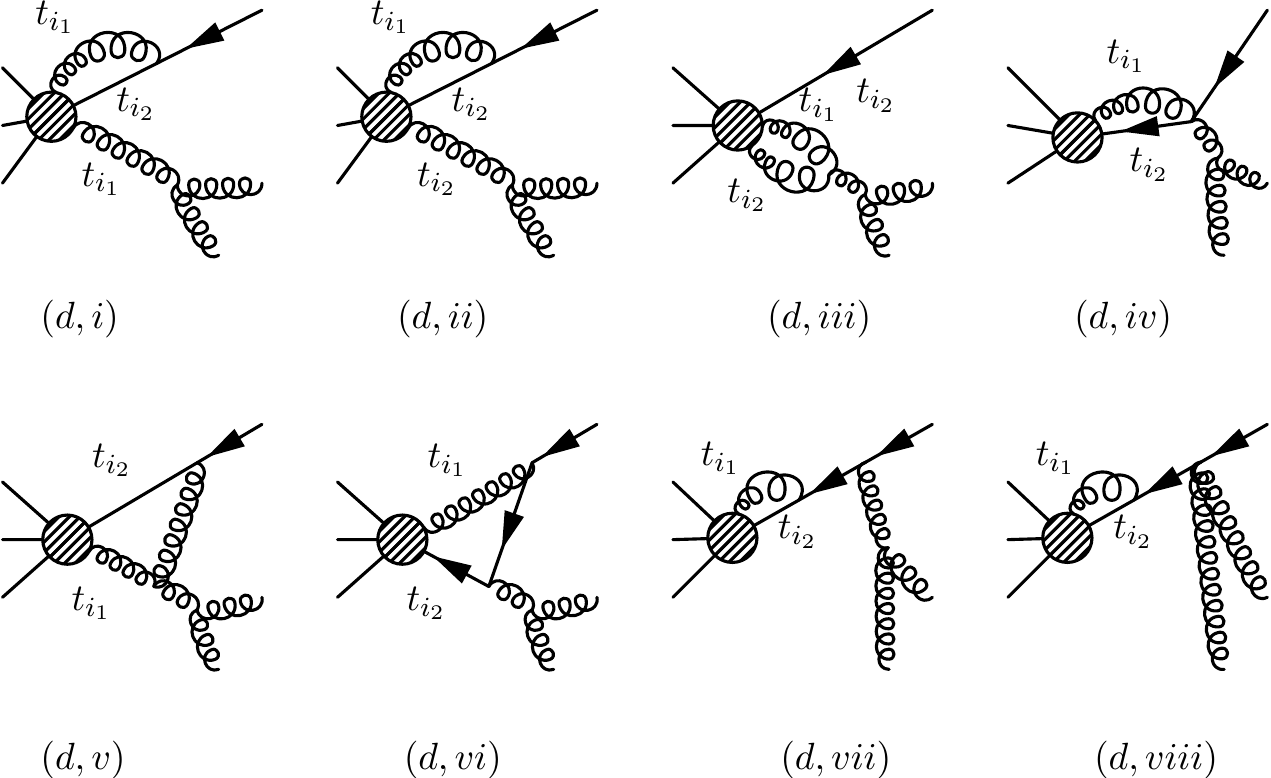}
\end{center}
\caption{\label{fig:qgg_1PR}
1PR diagrams contributing to the mixing $J^{B2}_{{\cal A}\partial\chi}\to J^{C2}_{{\cal AA}\chi}$ of a $B$- into a $C$-type current in the collinear
sector with fermion number $F_i=1$.
}
\end{figure}

In addition, as in the previous section, 1PR diagrams for which the 1PR 
propagator is cancelled have to be included. The corresponding diagrams are 
shown in Fig.~\ref{fig:qgg_1PR}. 
Diagrams where external gluons are radiated 
off the external fermion line vanish for external momenta without $\perp$ 
component and pure $\perp$ polarization, due to the structure of the SCET 
vertex \eqref{eq:Vxixic}. The only non-zero contributions involving a
1PR fermion propagator are the last two.

All diagrams except the last one involve a three-gluon vertex. For these 
diagrams it is possible to first compute the corresponding
diagram without the gluon splitting, and a single external gluon (using a polarization vector $\epsilon^{*\rho}$ and adjoint colour index $a$),
which we denote by ${\cal M}_{\rho a}\epsilon^{*\rho}$. We keep all possible polarizations for $\epsilon^{*\rho}$,
including also the longitudinal component (i.e. $\epsilon^*\cdot q\not= 0$).
The diagram with gluon splitting is then obtained by the replacement
\be
  {\cal M}_{\rho a}\epsilon^{*\rho} \to {\cal M}_{\rho a}\frac{-ig_s}{q^2} f^{dea}\left( (q_2-q_1)^\rho \epsilon_1^*\cdot\epsilon_2^* -2\epsilon_2^{*\rho} \epsilon_1^*\cdot q_2  
  +2\epsilon_1^{*\rho} \epsilon_2^*\cdot q_1\right)\,,
\ee
where $d(e)$ are the adjoint gluon colour indices for the two external gluons 
with momenta $q_1(q_2)$ and polarization vectors $\epsilon_1^*(\epsilon_2^*)$, 
and we used $\epsilon_i^*\cdot q_i=0$.  Furthermore, $q\equiv q_1+q_2$ denotes 
the momentum of the 1PR propagator, and $p$ the momentum of the external 
outgoing antiquark.

The contribution of the 1PR diagrams to $B$-to-$C$ mixing
corresponds to the divergent part which is not already
accounted for by the time-ordered product of the 1PI subdiagram
in on-shell kinematics with the three-gluon interaction.
Consistency requires that this contribution must be local, that
is, the $1/q^2$ from the 1PR gluon propagator must be
cancelled. To extract this contribution, we have to temporarily
restore $\perp$ components for the external momenta $q_i$, such
that $q^2$ is independent from the small regulating
offshellnesses $q_i^2\to 0$, $p^2\to 0$.\footnote{Otherwise one 
could express $q^2=\np q\nm (q_1+q_2)=
\np q\,(q_1^2/\np q_1+q_2^2/\np q_2)$ in terms of $q_i^2$, such that the
limit $q_i^2\to 0$ could not be taken while keeping $q^2$ finite.}
This allows us to independently take the on-shell limit
$q_i^2, p^2\to 0$ at finite $q^2$ and then $q^2\to 0$. In this
limit the relevant contribution from the 1PR diagrams becomes
independent of the transverse momenta by power counting due
to the homogeneous $\lambda$ scaling of all expressions. It
is therefore possible and convenient to perform the calculation
for the special configuration $q_{1\perp}=-q_{2\perp}$ such that
$q=q_1+q_2$ has no $\perp$ component. 
The divergent contribution from the time-ordered product of the 1PI subdiagram
in on-shell kinematics with the three-gluon interaction vanishes
in this case. The reason is that the 1PI subdiagram in on-shell 
kinematics must be a $B2$ operator, i.e. is proportional to a linear
combination of $q_\perp$ or $p_\perp$. Therefore, it vanishes 
for the kinematic configuration considered here, analogously
to Eq.\,\eqref{eq:BtoCatype}. This reduces our task to evaluating
the 1PR diagrams in Fig.~\ref{fig:qgg_1PR}. The matrix element
${\cal M}_{\rho a}={\cal M}_{\rho a}(q,p)$ can be decomposed as
\be
{\cal M}_{\rho a} = {\cal M}^-_{a}{\nm}_\rho +   {\cal M}^+_{a}{\np}_\rho +
{\cal M}^\perp_{\rho a}\,.
\ee
With $q_\rho=\frac12\np\, q {\nm}_\rho+\frac12\nm \,q{\np}_\rho$
for  $q_\perp=0$, we can
always re-express the matrix element in the form
\be
\label{eq:Mdecomp}
{\cal M}_{\rho a} = {\cal M}^L_{a}q_\rho +   \tilde {\cal M}^+_{a}{\np}_\rho 
+ {\cal M}^\perp_{\rho a}\,,
\ee
where
\be
\label{eq:Mtilde}
\tilde {\cal M}^+_{a} = {\cal M}^+_{a} - 
{\cal M}^-_{a}\frac{q^2}{(\np q)^2}\;.
\ee
Then, using that $\epsilon_i=\epsilon_{i\perp}$ are assumed to be polarized 
in the $\perp$ direction, gives for the gluon splitting
\bea
{\cal M}_{\rho a}\epsilon^{*\rho} &\to& \frac{-ig_s}{q^2} f^{dea}
\,\Bigg(\!\left({\cal M}^L_a(q_2^2-q_1^2)+\tilde {\cal M}^+_{a}\np(q_2-q_1)
+{\cal M}^\perp_a\cdot(q_2-q_1)_\perp\right) \epsilon_1^*\cdot\epsilon_2^* 
\nn\\
&& {} -2\epsilon_2^*\cdot {\cal M}^\perp_a \epsilon_1^*\cdot q_{2\perp}    
+2\epsilon_1^*\cdot {\cal M}^\perp_a  \epsilon_2^*\cdot q_{1\perp}\Bigg)\,.
\label{eq:1PRgluon}
\eea
For ${\cal M}^+_{a}$, we find that the divergent part of the loop 
amplitude can be expanded for small $q^2$ and $p^2$ in the form 
\be\label{eq:Mplusdecomp}
  {\cal M}^+_{a}={\cal M}^{+(p)}_{a}p^2+{\cal M}^{+(q)}_{a}q^2\,.
\ee
At this point we can take the on-shell limit $q_i^2, p^2\to 0$ with $q^2$ 
finite, such that the first term in the bracket on the right-hand side 
of Eq.~\eqref{eq:1PRgluon} vanishes, and ${\cal M}^{+(p)}_{a}$ in the 
previous equation can be dropped. After that, we can safely perform the 
limit $q_{i\perp}\to 0$, such that we finally arrive at the following 
rule, 
\bea
\label{eq:split}
{\cal M}_{\rho a}\epsilon^{*\rho} &\to& -ig_s f^{dea} \left({\cal M}^{+(q)}_{a} 
  - \frac{{\cal M}^-_{a}}{(\np q)^2}\right) \np(q_2-q_1) 
\epsilon_1^*\cdot\epsilon_2^* \,.
\eea
The $1/q^2$ factor is manifestly cancelled in this expression, which 
therefore contributes to the mixing into a $C$-type operator.

In summary, we need to compute the matrix elements ${\cal M}^\pm_{a}$ for 
quark-gluon final states, for external momenta with vanishing $\perp$ 
components, and gluon polarization in the $\pm$ directions. This is different 
from the quark/gluon matrix elements computed in Sec.~\ref{sec:BB}, and 
therefore we recomputed these diagrams for the required configuration of 
momenta and polarization vectors.

We find that the diagrams $(d, i)$ and $(d, ii)$ are proportional to $p^2/q^2$ 
(i.e. only ${\cal M}^{+(p)}_{a}$ is non-zero), 
and therefore vanish for $p^2\to 0$. For $(d,iii)$ only ${\cal M}^{+(q)}_{a}$ is non-zero, i.e. it gives a contribution to the
anomalous dimension. Diagram $(d,iv)$ gives ${\cal M}_{\rho a}\propto (p+q)^2$ which
can be brought in the form \eqref{eq:Mplusdecomp} using $(p+q)^2=\np(p+q)\nm(p+q)=\frac{p^2}{y_3}+\frac{q^2}{y_1+y_2}$.
For diagrams $(d,v)$ and $(d,vi)$ both ${\cal M}^-_{a}$ and ${\cal M}^{+(q)}_{a}$ yield non-zero contributions.
A singularity $\propto \gamma_\perp^\nu\gamma_\perp^\mu/(y_3-x)$ cancels in the sum of $(d,vi)$ and $(d,iv)$.

For diagram $(d, vii)$ only ${\cal M}^-_{a}$ is non-zero. Still, the loop gives an 
additional factor $(p+q)^2$, that however cancels with the 1PR fermion 
propagator. Therefore this diagram also contributes. 
Finally, the diagram $(d, viii)$ is special because it does not contain a three-gluon vertex. A direct computation
shows that the 1PR fermion propagator cancels with a factor $(p+q)^2$ obtained from the loop integral, similar as for $(d, vii)$.

As discussed before, counterterm diagrams involving $A$- and $B$-type 
operators necessarily involve some powers of external $\perp$ momenta, and 
therefore vanish. The only non-zero counterterm diagram is therefore the one 
involving $J^{C2}_{{\cal A}^\mu{\cal A}^\nu\chi}$,
\bea
  \langle g_{d}(q_1)g_{e}(q_2)\bar q(p) | J^{C2}_{{\cal A}^{\mu b}{\cal A}^{\nu c}\chi} |  0\rangle_{\rm tree} &=& g_s^2 e^{i(t_{i_1}\np q_1+t_{i_2}\np q_2+t_{i_3}\np p)}
  \epsilon_1^{*\mu}\epsilon_2^{*\nu} \delta_{bd}\delta_{ce} v_c \nn\\
  && {} + (q_1,d,\epsilon_1^* \leftrightarrow q_2,e,\epsilon_2^*)\,,
\eea
where we made explicit the (adjoint) colour indices for external gluons and 
gluon building blocks. After Fourier transformation with respect to the 
$t_{i_j}$ (with $y_3=1-y_1-y_2$),
\bea
  \lefteqn{\langle g_{d}(q_1)g_{e}(q_2)\bar q(p) | J^{C2}_{{\cal A}^{\mu b}{\cal A}^{\nu c}\chi}(y_1,y_2) |  0\rangle_{\rm tree} } \nn\\
  &=& P_i^3 g_s^2 \delta(P_iy_1-\np q_1)\delta(P_iy_2-\np q_2)\delta(P_iy_3-\np p)   \epsilon_1^{*\mu}\epsilon_2^{*\nu} \delta_{bd}\delta_{ce} v_c \nn\\
  && {}+ (q_1,d,\epsilon_1^* \leftrightarrow q_2,e,\epsilon_2^*) \nn\\
  &=& P_i g_s^2 \delta(y_1-\hat y_1)\delta(y_2-\hat y_2)\delta(P_i-\np (q_1+q_2+p))   \epsilon_1^{*\mu}\epsilon_2^{*\nu} \delta_{bd}\delta_{ce} v_c \nn\\
  && {}+ (q_1,d,\epsilon_1^* \leftrightarrow q_2,e,\epsilon_2^*) \,.
\eea
Here we defined the momentum
fractions $\hat y_{1(2)}\equiv\np q_{1(2)}/\np(p+q_1+q_2)$ via the external momenta.

On the other hand, the divergent part of the one-loop matrix element of 
$J^{B2}_{{\cal A}^{\mu a}\partial^\nu\xi}$ can be written in the form
\bea
\lefteqn{ \langle g_{d}(q_1)g_{e}(q_2)\bar q(p) | J^{B2}_{{\cal A}^{\mu a}
\partial^\nu\chi} | 0\rangle_{\rm 1-loop}^{\rm div} } 
\nn\\
&=& 
\frac{g_s^4}{16\pi^2\epsilon}\int_0^1 dx' e^{i(t_{i_1}x'+t_{i_2}\bar x')
\np(q_1+q_2+p)} I_{ade}^{\mu\nu\sigma\lambda}(x',\hat y_1,\hat y_2)
\epsilon_1^{*\sigma}\epsilon_2^{*\lambda} v_c\,,
\eea
which defines the function $I_{ade}^{\mu\nu\sigma\lambda}(x',\hat y_1,
\hat y_2)$. After Fourier transformation,
\bea
\lefteqn{ \langle g_{d}(q_1)g_{e}(q_2)\bar q(p)| J^{B2}_{{\cal A}^{\mu a}
\partial^\nu\chi}(x) | 0 \rangle_{\rm 1-loop}^{\rm div} } 
\nn\\
&=& 
P_i^2\frac{g_s^4}{16\pi^2\epsilon}\int_0^1 dx' \delta(P_ix-x'\np(q_1+q_2+p))
\delta(P_i\bar x-\bar x'\np(q_1+q_2+p))\nn\\
&&\times\, I_{ade}^{\mu\nu\sigma\lambda}(x',\hat y_1,\hat y_2)
\epsilon_1^{*\sigma}\epsilon_2^{*\lambda} v_c 
\nn\\
&=& P_i\frac{g_s^4}{16\pi^2\epsilon}\delta(P_i-\np(q_1+q_2+p))
I_{ade}^{\mu\nu\sigma\lambda}(x,\hat y_1,\hat y_2)\epsilon_1^{*\sigma}
\epsilon_2^{*\lambda} v_c
\nn\\
&=& \frac{g_s^2}{16\pi^2\epsilon} \frac{1}{2} \int dy_1 dy_2 \,
I_{abc}^{\mu\nu\sigma\lambda}(x,y_1,y_2)\,
\langle  g_{d}(q_1)g_{e}(q_2)\bar q(p)| J^{C2}_{{\cal A}^{\sigma b}{\cal A}^{
\lambda c}\chi}(y_1,y_2) | 0 \rangle_{\rm tree}\,,\qquad
\eea
where in the last step we used that $I_{ade}^{\mu\nu\sigma\lambda}(x,y_1,y_2)=I_{aed}^{\mu\nu\lambda\sigma}(x,y_2,y_1)$
due to symmetry under exchange of the two external gluon lines, leading to the additional factor $1/2$.
From the last relation we can read off the anomalous dimension,
\be\label{eq:gamma_Adelchi_AAchi}
  \gamma^i_{{\cal A}^{\mu a}\partial^\nu\xi,{\cal A}^{\sigma d}{\cal A}^{\lambda e}\xi}(x,y_1,y_2) = - \frac{\alpha_s}{8\pi}\,I_{ade}^{\mu\nu\sigma\lambda}(x,y_1,y_2)\,.
\ee
From the explicit one-loop results one can read off $I_{ade}^{\mu\nu\sigma\lambda}(x,y_1,y_2)$.
The results are provided in App.~\ref{app:gamma_Adelchi_AAchi}.

\subsubsection{Mixing $J^{B2}_{{\cal A}\partial\chi}(x)\to J^{C2}_{\chi\bar\chi\chi}(y_1,y_2)$ }

For this case we consider the matrix element of the current with three 
fermions, two outgoing antiquarks and one outgoing quark, all 
with external momenta that have vanishing $\perp$ components. For simplicity 
we assume that the third fermionic building block
has a different flavour from the first two, and comment on the generalization 
below.

\begin{figure}
\begin{center}
  \includegraphics[width=0.65\textwidth]{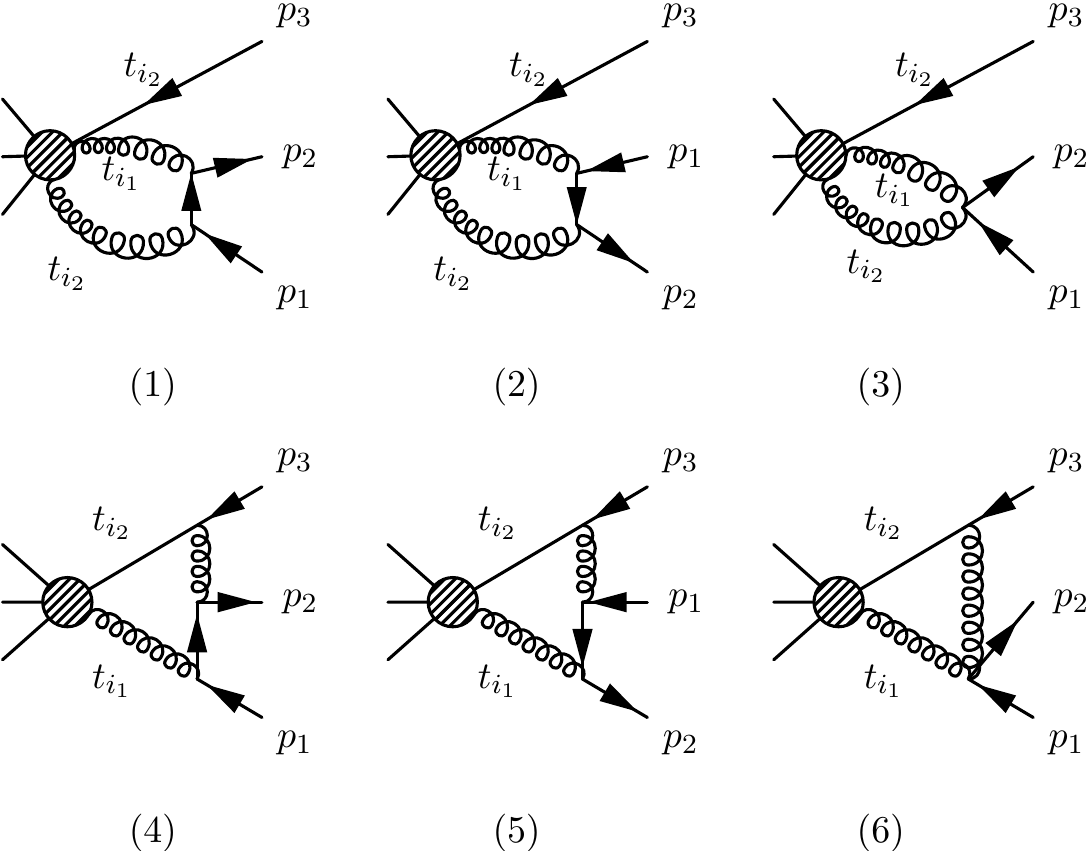}
\end{center}
\caption{\label{fig:qqbarq}
1PI diagrams contributing to the mixing $J^{B2}_{{\cal A}\partial\chi}\to 
J^{C2}_{\chi\bar\chi\chi}$ of a $B$- into a $C$-type current in the collinear
sector with fermion number $F_i=1$.
}
\end{figure}

The relevant 1PI diagrams are shown in Fig.~\ref{fig:qqbarq}.
Diagram (3) vanishes, because the gluon line attached to the fermionic 
building block (label $t_{i_2}$) picks
up a factor $\np$ from the Wilson line, and this gives zero when multiplied with the two-fermion vertex.
Diagram (1) would become singular for $x\to \bar y_3\equiv y_1+y_2$, but one can check that the Heaviside functions
obtained from the collinear loop vanish in the domain of the pole for $0<y_i<1$.
Furthermore, a potential singularity for $x\to y_1$ cancels in the sum of (4) and (6), and for $x\to y_2$ in the sum of (5) and (6).

The relevant 1PR diagrams are similar to the 1PR diagrams shown in 
Fig.~\ref{fig:qgg_1PR} for the $qgg$ final state. They can be obtained by 
replacing the three-gluon vertex attached to 1PR gluon propagator by
a fermion-fermion-gluon vertex for diagram $(d, i)$ to $(d,vii)$. 
Diagram $(d,viii)$ does not exist for the $q\bar q q$ final state.
As before, we can infer the contribution from the matrix element 
${\cal M}_{\rho a}\epsilon^{*\rho}$ of quark-gluon final states obtained by 
cutting the 1PR gluon propagator. For vanishing external $\perp$ momenta, 
taking the $g\to \bar q q$ splitting into account amounts to (assuming the 
same-flavour antiquark and quark attached to the 1PR propagator have momenta 
$p_1$ and $p_2$, respectively)
\be
  {\cal M}_{\rho a}\epsilon^{*\rho} \to {\cal M}_{\rho a}\frac{-i}{q^2}\bar u_c(p_2) ig_st^a \left(\nm^\rho + \frac{\slashed{p}_{2\perp}\gamma_\perp^\rho}{\np p_2} 
  + \frac{\gamma_\perp^\rho\slashed{p}_{1\perp}}{\np p_1}  - \np^\rho\frac{\slashed{p}_{2\perp}\slashed{p}_{1\perp}}{\np p_1\np p_2} \right)\frac{\nps}{2} v_c(p_1) \,,
\ee
where now $q=p_1+p_2$ is the momentum through the 1PR propagator.
In order to be able to take the limit $p_1^2\to 0$ and $p_2^2\to 0$ with $q^2=(p_1+p_2)^2$ finite,
we introduce again for a moment a non-zero $\perp$ momentum such that $p_{1\perp}=-p_{2\perp}$.
Then, as before, the matrix element ${\cal M}_{\rho a}={\cal M}_{\rho a}(q,p_3)$ is independent of the $\perp$
momenta. Using the decomposition \eqref{eq:Mdecomp} we get
\bea
  {\cal M}_{\rho a}\epsilon^{*\rho} &\to& \frac{-i}{q^2}\bar u_c(p_2) ig_st^a 
\nn\\
&& 
\hspace*{-1cm}\times\,\left(2\tilde{\cal M}^+_a + \frac{\slashed{p}_{2\perp}\slashed{\cal M}_a^\perp}{\np p_2} 
  + \frac{\slashed{\cal M}_a^\perp\slashed{p}_{1\perp}}{\np p_1}  + {\cal M}_a^L \left(\frac{p_1^2}{\np p_1}+\frac{p_2^2}{\np p_2}\right) \right)\frac{\nps}{2} v_c(p_1)\,. 
\eea
At this point we can take the limit $p_i^2\to 0$, and afterwards let the 
$\perp$ momenta go to zero again, obtaining
\bea
  {\cal M}_{\rho a}\epsilon^{*\rho} &\to& \frac{g_s}{q^2}\left({\cal M}^{+(q^2)}_{a}q^2 - {\cal M}^-_{a}\frac{q^2}{(\np q)^2}\right) \bar u_c(p_2) t^a \nps v_c(p_1) \,,
\eea
where we used Eqs.~\eqref{eq:Mtilde} and \eqref{eq:Mplusdecomp} with 
$p\mapsto p_3$. As for the $qgg$ case, the 1PR momentum $q^2$ cancels and we 
obtain a finite result. The result for all 1PR diagrams $(d, i)$ to $(d,vii)$ 
can therefore be obtained from the corresponding $qgg$ final state by 
replacing 
\be\label{eq:rescaling}
  -if^{dea}(y_2-y_1)\epsilon_1^*\cdot\epsilon_2^* \mapsto \frac{1}{P_i}\bar u_c(p_2) t^a \nps v_c(p_1)\,.
\ee

The tree-level contribution from the $C$-type operator is (displaying colour 
and Dirac indices and accounting for signs from anticommutations)
\be
  \langle \bar q(p_1) q(p_2)\bar q(p_3) | J^{C2}_{\xi^\beta_j\bar\xi^\gamma_k\xi^\delta_l} |  0 \rangle_{\rm tree} = (-1)e^{i(t_{i_1}\np p_1+t_{i_2}\np p_2+t_{i_3}\np p_3)}
  v_{cj}^\beta(p_1) \bar u_{ck}^\gamma(p_2) v_{cl}^\delta(p_3) \,.
\ee
The loop amplitude for the sum of the 1PI and 1PR diagrams can be written as 
(we extract a factor $1/P_i$ of total collinear momentum for later 
convenience, for dimensional reasons, and 
$\hat y_i\equiv \np p_i/\np(p_1+p_2+p_3)$)
\bea
 \lefteqn{ \langle \bar q(p_1) q(p_2)\bar q(p_3)| J^{B2}_{{\cal A}^{\mu a}\partial^\nu\xi^\alpha_i} |   0 \rangle_{\rm 1-loop}^{\rm div} } \nn\\
  &=& 
  \frac{g_s^4}{16\pi^2\epsilon} \frac{1}{P_i} \int_0^1 dx' \,e^{i(t_{i_1}x'+t_{i_2}\bar x')\np(p_1+p_2+p_3)} I_{aijkl}^{\mu\nu\alpha\beta\gamma\delta}(x',\hat y_1,\hat y_2)v_{cj}^\beta(p_1) \bar u_{ck}^\gamma(p_2) v_{cl}^\delta(p_3)\,, \qquad
\eea
which defines the kernel  $I_{aijkl}^{\mu\nu\alpha\beta\gamma\delta}(x,\hat y_1,\hat y_2)$.
After Fourier transformation,
\bea\label{eq:qqqmixing}
\lefteqn{ \langle \bar q(p_1) q(p_2)\bar q(p_3)| J^{B2}_{{\cal A}^{\mu a}
\partial^\nu\xi^\alpha_i}(x) |  0 \rangle_{\rm 1-loop}^{\rm div} } 
\nn\\
&=& 
{} - \frac{g_s^4}{16\pi^2\epsilon} \frac{1}{P_i} \int dy_1 dy_2 \,
I_{aijkl}^{\mu\nu\alpha\beta\gamma\delta}(x,y_1,y_2)\,
\langle \bar q(p_1) q(p_2)\bar q(p_3)| J^{C2}_{\xi^\beta_j\bar\xi^\gamma_k
\xi^\delta_l}(y_1,y_2) |  0 \rangle_{\rm tree}\,. 
\nn\\[-0.2cm]
\eea
From this relation, the anomalous dimension can be read off,
\be\label{eq:gamma_Adelchi_chichibarchi}
  \gamma^i_{{\cal A}^{\mu a}\partial^\nu\chi^\alpha_i,\chi^\beta_j\bar\chi^\gamma_k\chi^\delta_l}(x,y_1,y_2) =  \frac{\alpha_s }{8\pi} g_s^2  \frac{1}{P_i} I_{aijkl}^{\mu\nu\alpha\beta\gamma\delta}(x,y_1,y_2)\,.
\ee
The factor $g_s^2$ is due to our convention for the normalization of the collinear building blocks, and the factor $1/P_i$ of the total collinear
momentum arises for dimensional reasons. The results for $I_{aijkl}^{\mu\nu\alpha\beta\gamma\delta}(x,y_1,y_2)$ are collected in 
App.~\ref{app:gamma_Adelchi_chichibarchi}.

If all fermion building blocks are of the the same flavour, the expression on the right-hand side needs to be
anti-symmetrized with respect to interchanging the first and the last building block, i.e.
\be
  I_{aijkl}^{\mu\nu\alpha\beta\gamma\delta}(x,y_1,y_2) \to \frac12\left[ I_{aijkl}^{\mu\nu\alpha\beta\gamma\delta}(x,y_1,y_2)-I_{ailkj}^{\mu\nu\alpha\delta\gamma\beta}(x,y_3,y_2)\right]\,,
\ee
where, as before, $y_3=1-y_1-y_2$.

\subsection{Mixing of $C$-type currents into $C$-type 
currents}\label{sec:CC}

As discussed in  Ref.~\cite{Beneke:2017ztn}, at the one-loop order, this 
mixing arises from diagrams for which only two out of the three building 
blocks of the $C$-type current are attached to lines belonging to the loop.
Therefore, the anomalous dimension can be obtained from the one of the 
corresponding $B$-type currents at ${\cal O}(\lambda)$. We denote the 
independent momentum fractions of the first and second collinear building block
by $x_1\equiv x_{i_1}$ and  $x_2\equiv x_{i_2}$, and set $x_3=1-x_1-x_2$ for 
the third one. For example, for $J^{C2}_{{\cal AA}\chi}$ $x_3$ denotes the 
fraction of collinear momentum carried by the fermion. For the anomalous 
dimension that corresponds to 
$J^{C2}_{{\cal AA}\chi}(x_1,x_2)\to J^{C2}_{{\cal AA}\chi}(y_1,y_2)$ the momentum fractions $y_{1,2,3}$ of the second operator are defined
analogously, and we find
\bea
\label{eq:gamma_AAchi_AAchi}
 \gamma^i_{{\cal A}^\mu{\cal A}^\nu\chi_\alpha,{\cal A}^\rho{\cal A}^\sigma\chi_\beta}(x_1,x_2,y_1,y_2) 
&=& 
\Bigg[\frac{1}{1-y_2}\delta(x_2-y_2)g_\perp^{\nu\sigma} \gamma^i_{{\cal A}^\mu\chi_\alpha,{\cal A}^\rho\chi_\beta}\left(\frac{x_1}{1-x_2},\frac{y_1}{1-y_2}\right) \nn\\
 && 
\hspace*{-3cm}
+ \,\frac{1}{1-y_1}\delta(x_1-y_1)g_\perp^{\mu\rho} \gamma^i_{{\cal A}^\nu\chi_\alpha,{\cal A}^\sigma\chi_\beta}\left(\frac{x_2}{1-x_1},\frac{y_2}{1-y_1}\right)\nn\\
 && \hspace*{-3cm}
+ \,\frac{1}{1-y_3}\delta(x_3-y_3) \delta_{\alpha\beta}\gamma^i_{{\cal A}^\mu{\cal A}^\nu,{\cal A}^\rho{\cal A}^\sigma}\left(\frac{x_1}{1-x_3},\frac{y_1}{1-y_3}\right) \Bigg]_{\rm sym}\,.
\eea
Here the square bracket refers to symmetrization with respect to 
$(y_1,\rho,b_1)\leftrightarrow(y_2,\sigma,b_2)$ where $b_1$ ($b_2$) denotes 
the adjoint colour index carried by the gluon building block ${\cal A}^\rho$ 
(${\cal A}^\sigma$). These indices are left implicit in the equation above, 
including a Kronecker symbol for the colour indices
of the two gluon building blocks not contained in 
$\gamma^i_{{\cal A}^\mu\chi_\alpha,{\cal A}^\rho\chi_\beta}$ or 
$\gamma^i_{{\cal A}^\nu\chi_\alpha,{\cal A}^\sigma\chi_\beta}$ in the first
and second line, respectively. A similar statement refers to the 
quark fields in the third line and the equations below in this section.
Symmetrization refers here to the \emph{average} over the expression given
in the square bracket, and the corresponding expression obtained when replacing $(y_1,\rho,b_1)\to(y_2,\sigma,b_2)$, i.e. includes
a normalization factor $1/2$. The anomalous dimension $\gamma^i_{{\cal A}^\mu{\cal A}^\nu,{\cal A}^\rho{\cal A}^\sigma}$ will
be provided in a future work dedicated to the case of fermion number $F_i=0$.

The mixing $J^{C2}_{{\cal AA}\chi}(x_1,x_2)\to J^{C2}_{\chi\bar\chi\chi}(y_1,y_2)$ vanishes if we assume that all fermions in the latter
operator carry a different flavour quantum number. If the fermions $\chi\bar\chi$ are of the same flavour, and the fermion
in the last building block has a different flavour, we find
\be
\label{eq:gamma_AAchi_chichibarchi}
\gamma^i_{{\cal A}^\mu{\cal A}^\nu\chi_\alpha,\chi_\beta\bar\chi_\gamma
\chi_\delta}(x_1,x_2,y_1,y_2)
= -\frac{1}{1-y_3}\delta(x_3-y_3)\delta_{\alpha\delta}\, 
\gamma^i_{{\cal A}^\mu{\cal A}^\nu,\bar\chi_\gamma\chi_\beta}
\left(\frac{x_1}{1-x_3},\frac{y_2}{1-y_3}\right)\,.
\ee
Note the minus sign due to the interchange of fermion indices. For 
$\gamma^i_{{\cal A}^\mu{\cal A}^\nu,\bar\chi_\gamma\chi_\beta}$ we also refer
to future work on the $F_i=0$ case. If all fermions are of the same flavour, 
the anomalous dimension can be obtained by \emph{anti}symmetrizing
Eq.~\eqref{eq:gamma_AAchi_chichibarchi} with respect to 
$(y_1,\beta,c_1)\leftrightarrow(y_3,\delta,c_3)$, where $c_k$ denote the 
fundamental colour indices of the first and third fermion building block, 
respectively. As before, antisymmetrization is understood to include a 
normalization factor $1/2$.

For the contribution corresponding to $J^{C2}_{\chi\bar\chi\chi}(x_1,x_2)\to J^{C2}_{\chi\bar\chi\chi}(y_1,y_2)$, we find for the case where
the first and last building blocks carry distinct flavour,
\bea
\label{eq:gamma_chichibarchi_chichibarchi}
\gamma^i_{\chi_\alpha\bar\chi_\beta\chi_\gamma,\chi_{\alpha'}\bar\chi_{\beta'}\chi_{\gamma'}}(x_1,x_2,y_1,y_2) 
 &=& \frac{1}{1-y_2}\delta(x_2-y_2)\delta_{\beta\beta'} \gamma^i_{\chi_\alpha\chi_\gamma,\chi_{\alpha'}\chi_{\gamma'}}\left(\frac{x_1}{1-x_2},\frac{y_1}{1-y_2}\right)  \nn\\
 &&  \hspace*{-3cm}
+ \,\frac{1}{1-y_1}\delta(x_1-y_1)\delta_{\alpha\alpha'} \gamma^i_{\bar\chi_\beta\chi_\gamma,\bar\chi_{\beta'}\chi_{\gamma'}}\left(\frac{x_2}{1-x_1},\frac{y_2}{1-y_1}\right) \nn\\
 &&  \hspace*{-3cm}
+ \,\frac{1}{1-y_3}\delta(x_3-y_3)\delta_{\gamma\gamma'}  \gamma^i_{\bar\chi_\beta\chi_\alpha,\bar\chi_{\beta'}\chi_{\alpha'}}\left(\frac{x_2}{1-x_3},\frac{y_2}{1-y_3}\right)  \,.
\eea
Note that the last line requires \emph{two} fermion permutations leading to 
the positive sign, and that even for the case of different flavour quantum 
numbers three distinct loop contributions exist that lead to the three terms 
on the right-hand side. For $\gamma^i_{\chi_\alpha\chi_\gamma,
\chi_{\alpha'}\chi_{\gamma'}}$ we refer to Ref.~\cite{Beneke:2017ztn},
and for $\gamma^i_{\bar\chi_\beta\chi_\alpha,\bar\chi_{\beta'}\chi_{\alpha'}}$ to future work on the $F_i=0$ case. If the flavour of the fermions in the
first and last building block are identical, one needs to antisymmetrize the right-hand side with respect to $(y_1,\beta,c_1)\leftrightarrow(y_3,\delta,c_3)$
as before.

For $J^{C2}_{\chi\bar\chi\chi}(x_1,x_2)\to J^{C2}_{{\cal AA}\chi}(y_1,y_2)$ we first provide the result obtained if all fermions have
identical flavour,
\bea\label{eq:gamma_chichibarchi_AAchi}
 \gamma^i_{\chi_\alpha\bar\chi_\beta\chi_\gamma,{\cal A}^\mu{\cal A}^\nu\chi_\delta}(x_1,x_2,y_1,y_2) &=& 
  \frac{1}{1-y_3}\delta(x_1-y_3)\delta_{\alpha\delta} \gamma^i_{\bar\chi_\beta\chi_\gamma,{\cal A}^\mu{\cal A}^\nu}\left(\frac{x_2}{1-x_1},\frac{y_1}{1-y_3}\right)  \nn\\
 &&  \hspace*{-3cm}
-\, \frac{1}{1-y_3}\delta(x_3-y_3)\delta_{\gamma\delta} \gamma^i_{\bar\chi_\beta\chi_\alpha,{\cal A}^\mu{\cal A}^\nu}\left(\frac{x_2}{1-x_3},\frac{y_1}{1-y_3}\right) \,. 
\eea
If the first fermion $\chi_\alpha$ has a different flavour from the other two, only the first line contributes on the right-hand side.
If, on the other hand, the third fermion $\chi_\gamma$ has a different flavour, only the second line contributes.
In this case we do not need to explicitly symmetrize with respect to interchanging the gluonic building blocks, 
because this symmetrization is already taken care of in the anomalous dimension $\gamma^i_{\bar\chi_\alpha\chi_\beta,{\cal A}^\mu {\cal A}^\nu}$.

Finally, we discuss the case $F_i=3$, where the only the mixing  
$J^{C2}_{\chi\chi\chi}(x_1,x_2)\to J^{C2}_{\chi\chi\chi}(y_1,y_2)$ is 
possible. For three fermions with mutually
distinct flavour quantum numbers, the result has the expected form
\bea\label{eq:gamma_chichichi_chichichi}
\gamma^i_{\chi_\alpha\chi_\beta\chi_\gamma,\chi_{\alpha'}\chi_{\beta'}\chi_{\gamma'}}(x_1,x_2,y_1,y_2) 
 &=& \frac{1}{1-y_2}\delta(x_2-y_2)\delta_{\beta\beta'} \gamma^i_{\chi_\alpha\chi_\gamma,\chi_{\alpha'}\chi_{\gamma'}}\left(\frac{x_1}{1-x_2},\frac{y_1}{1-y_2}\right) \nn\\
 &&  \hspace*{-3cm} 
+\, \frac{1}{1-y_1}\delta(x_1-y_1)\delta_{\alpha\alpha'} \gamma^i_{\chi_\beta\chi_\gamma,\chi_{\beta'}\chi_{\gamma'}}\left(\frac{x_2}{1-x_1},\frac{y_2}{1-y_1}\right) \nn\\
 && \hspace*{-3cm}
 +\, \frac{1}{1-y_3}\delta(x_3-y_3)\delta_{\gamma\gamma'} \gamma^i_{\chi_\alpha\chi_\beta,\chi_{\alpha'}\chi_{\beta'}}\left(\frac{x_1}{1-x_3},\frac{y_1}{1-y_3}\right) \,.
\eea
If, for example, the first and last fermion have identical flavour the 
right-hand side needs to be antisymmetrized with respect to the interchange 
of the corresponding momentum fractions and Dirac as well as colour indices $(y_1,\alpha',c_1)\leftrightarrow(y_3,\gamma',c_3)$.
If all three fermions have identical flavour, the right-hand side needs to be fully antisymmetrized with respect to all possible $3!$ permutations,
including a normalization factor $1/6$ and a minus sign for odd permutations, due to fermion anticommutation.

\section{Summary}
\label{sec:result}

In this work we extended the computation of the one-loop anomalous dimension 
matrix of subleading-power $N$-jet operators started in 
Ref.~\cite{Beneke:2017ztn}. The operator basis can be characterized by the 
number and type of collinear building blocks for each of the $N$ collinear 
directions. In addition, homogeneous power counting in $\lambda$ of the 
anomalous dimension requires to take into account 
time-ordered products of $N$-jet currents with insertions of
the power-suppressed terms ${\cal L}^{(n)}$ of the SCET Lagrangian. The 
general structure of the anomalous dimension matrix \eqref{eq:GammaPQ} 
encompasses universal contributions that are diagonal with respect to 
collinear momentum and the type of operators, as well as off-diagonal 
contributions. The latter can be divided into a contribution $\gamma^i$ that 
describes current-current mixing and arises from collinear loops along the $i$ 
direction, and $\gamma^{ij}$ that captures mixing of time-ordered products 
into currents. It originates from soft loops connecting directions $i$ and $j$, 
and represents a qualitatively new feature compared to 
Ref.~\cite{Beneke:2017ztn}. In this work we provide complete results 
for $\gamma^i$ for currents with 
fermion number $|F_i|=1,3$ in direction $i$, and for $\gamma^{ij}$ for 
$|F_i|=|F_j|=1$.\footnote{Ref.\,\cite{Beneke:2017ztn} covers the case 
$|F_i|=2$ for $\gamma^i$, while $\gamma^{ij}$ vanishes for $|F_i|>1$ or 
$|F_j|>1$.} In addition, we find several general properties of $\gamma^{ij}$:
\begin{itemize}
\item Time-ordered products containing a single insertion
of ${\cal L}^{(1)}$ or ${\cal L}^{(2)}$, or double insertions along a single 
collinear direction, do not mix into currents. This implies in particular
that $\gamma^{ij}$ vanishes at order $\lambda$. 
\item Time-ordered products involving power-suppressed 
interactions of massless soft quarks, given by ${\cal L}^{(1)}_{\xi q}$, 
also do not mix into currents. As a consequence, fermion number is conserved 
separately for every collinear direction in the massless theory.  
\item For $|F_i|=|F_j|=1$, operators containing a product of two time-ordered 
products $J^{T1}J^{T1}$ in directions $i$ and $j$ can only mix into
a product of two ${\cal O}(\lambda)$ currents $J^{X1}J^{Y1}$ with $X,Y=A,B$, 
but not into $J^{X2}J^{A0}$ with $X=A,B,C$. Thus we observe that the
level of power suppression is also ``conserved'' along each collinear direction. 
\end{itemize}
Altogether, non-zero contributions to $\gamma^{ij}$ can arise only from 
time-ordered products containing an insertion of ${\cal L}^{(1)}_{\xi}$ or
 ${\cal L}^{(1)}_{\rm YM}$ along direction $i$, and another one along 
direction $j$. For the case $F_i=F_j=1$ the structure of $\gamma^{ij}$ is 
therefore given by\\[0.1cm]
\bea\label{eq:gammasoft}
 \gamma^{ij}_{PQ} \quad = \quad 
 \begin{array}{c||cccc|cc|} 
  & J^{A1}_{\partial\chi} J^{A1}_{\partial\chi}  &  J^{B1}_{\calA\chi} J^{A1}_{\partial \chi} &   J^{A1}_{\partial \chi} J^{B1}_{\calA\chi}& J^{B1}_{\calA\chi} J^{B1}_{\calA\chi} & J^{X2} J^{A0}  &  J^{A0} J^{X2}   \\ \hline\hline
  J^{T1}_{\chi,\xi}\;J^{T1}_{\chi,\xi} & \eqref{eq:T1xiT1xi_A1A1} & \eqref{eq:T1xiT1xi_B1A1}  &\eqref{eq:T1xiT1xi_B1A1}&\eqref{eq:T1xiT1xi_B1B1} &0 &0  \\ 
  J^{T1}_{\chi,\rm YM}\;J^{T1}_{\chi,\xi} & 0  & \eqref{eq:T1YMT1xi_B1A1} &0& \eqref{eq:T1YMT1xi_B1B1} &0 &0 \\ 
  J^{T1}_{\chi,\xi}\;J^{T1}_{\chi,\rm YM} & 0  & 0 &\eqref{eq:T1YMT1xi_B1A1}& \eqref{eq:T1YMT1xi_B1B1} &0 &0 \\ 
  J^{T1}_{\chi,\rm YM}\;J^{T1}_{\chi, \rm YM} & 0  & 0 &0& \eqref{eq:T1YMT1YM_B1B1} & 0 & 0   \\ \hline
  J^{T2}\;J^{A0} & 0  & 0 & 0 & 0 & 0 &0\\ 
  J^{T1}\;J^{Y1} & 0  & 0 & 0 & 0 & 0 &0\\ 
  J^{T1}_{\chi,\xi q}\;J^{T1}& 0  & 0 & 0 & 0 &0 &0\\[0.2cm] 
\end{array}\quad
\eea
where $X=A,B,C$ and $Y=A,B$. The non-zero entries refer to the equation numbers 
in which the result is given or to which it is related up to interchanging 
$i\leftrightarrow j$. For $F_i=-1$ or $F_j=-1$ the anomalous dimension is 
obtained by hermitian conjugation (see App.~\ref{app:cc} for details).
For $F_i=-F_j$ mixing into operators with $F_i=F_j=0$ via soft quark exchange 
vanishes due to the conservation of fermion number along each collinear 
direction as observed above.

Apart from the soft contributions to the anomalous dimension, we provide 
results for the collinear part $\gamma^i$. For this part we find that it is 
sufficient to consider current-current mixing. Mixing of time-ordered products
into currents vanishes in the collinear sector, while mixing of time-ordered 
products into themselves is identical to the
corresponding current-current mixing. Furthermore, collinear loops involve 
only a single collinear direction, denoted by $i$.
The operator basis for $F_i=1$ contains two operators at ${\cal O}(\lambda)$ 
(one $A$- and one $B$-type), and five at ${\cal O}(\lambda^2)$ 
(one $A$-, two $B$- and two $C$-type). The corresponding
$2\times 2$ and $5\times 5$ matrices $\gamma^i_{PQ}$ are given in 
Eq.~\eqref{eq:gammacoll_lambda} and in Eq.~\eqref{eq:gammacoll}, respectively.
The latter contains non-zero mixings of the form $J^{B2}\to J^{B2},J^{C2}$ and 
$J^{C2}\to J^{C2}$. Operators with fermion number $F_i=3$ start at 
${\cal O}(\lambda^2)$, see Eq.~\eqref{eq:gamma_chichichi_chichichi}.

To complete the one-loop renormalization programme of ${\cal O}(\lambda^2)$ 
SCET $N$-jet operators, the calculation of the anomalous dimension in the 
$F=0$ sector is required. This includes the case of gluon jets at leading 
power and the mixing of two-gluon into quark-antiquark $B1$-type operators 
at the power-suppressed level. Work on this is in progress. It should then 
be feasible to consider next-to-leading logarithmic resummation of power 
corrections to jet processes of the SCET$_{\rm I}$ type.

\subsubsection*{Acknowledgements} 
We thank A.~Broggio and S.~Jaskiewicz for useful discussions. 
This work has been supported by the Bundesministerium f\"ur Bildung und 
Forschung (BMBF) grant nos. 05H15WOCAA and 05H18WOCA1. 

\begin{appendix}

\section{SCET Feynman rules}
\label{sec:feynman}

\subsection{Preliminaries}

In this appendix we give explicit expressions for the Feynman rules in the
position-space formulation of SCET~\cite{Beneke:2002ph} up to 
${\cal O}(\lambda^2)$, derived from the multipole-expanded Lagrangian given 
in Ref.~\cite{Beneke:2002ni}. The field content consists of collinear 
quarks ($\xi$) with scaling $\xi\sim \lambda$,  collinear gluons ($A_c$) with 
scaling $(\np A_c, A_{c\perp},\nm A_c)\sim (1,\lambda,\lambda^2)$, soft 
quarks ($q\sim \lambda^3$) 
and soft gluons ($A_s\sim \lambda^2 $).\footnote{The soft fields here 
were called ultrasoft in Ref.~\cite{Beneke:2002ni}.} 
The Lagrangian can be split into a purely bosonic part ${\cal L}_{\rm YM}$ 
and a part involving fermions (denoted by ${\cal L}$). Each part can be 
expanded in powers of $\lambda$ \cite{Beneke:2002ni} 
\bea
&& {\cal L} = \bar\xi\left(i\nm D+i\slashed{D}_{\perp }
\frac{1}{i\np D}i\slashed{D}_{\perp }\right)\frac{\nps}{2}\xi
   +\bar q i\slashed{D}_s q
   + {\cal L}_\xi^{(1)}
   + {\cal L}_\xi^{(2)}
   + {\cal L}_{\xi q}^{(1)}
   + {\cal L}_{\xi q}^{(2)} \,,
\nn\\
&&  {\cal L}_{\rm YM} = -\frac12{\rm tr}\left(F_c^{\mu\nu} 
F^c_{\mu\nu}\right) -\frac12{\rm tr}\left(F_s^{\mu\nu}
F^s_{\mu\nu}\right)
   + {\cal L}_{\rm YM}^{(1)}
   + {\cal L}_{\rm YM}^{(2)} \,,
\label{eq:SCETLagrangian}
\eea
where $g_s F_c^{\mu\nu}=i \,[D^\mu,D^\nu]$, 
$g_s F_s^{\mu\nu}=i \,[D_s^\mu,D_s^\nu]$ and
\bea
D^\mu &=& \partial^\mu -ig_sA_c^\mu(x)-ig_s\nm A_s(x_-)\frac{\np^\mu}{2} \,,
\nn\\
D_s^\mu &=& \partial^\mu -ig_sA_s^\mu(x)\,.
\eea
The subleading-power interactions at order $\lambda^n$ are contained in 
${\cal L}^{(n)}$, which can be split into interactions involving collinear 
quarks (${\cal L}_\xi^{(n)}$), collinear and soft quarks 
(${\cal L}_{\xi q}^{(n)}$), and soft and collinear gluons only 
(${\cal L}_{\rm YM}^{(n)}$). For completeness we reprint the 
power-suppressed SCET Lagrangian up to ${\cal O}(\lambda^2)$ from  
Ref.~\cite{Beneke:2002ni}:
\begin{eqnarray}
\label{eq:Lxi1}
{\cal L}^{(1)}_{\xi} &=&
\bar{\xi} \left( x_\perp^\mu n_-^\nu \, W_c \,g_s F_{\mu\nu}^s W_c^\dagger 
\right) \frac{\slash n_+}{2} \xi,
\\[0.0cm]
{\cal L}^{(2)}_{\xi} &=&
  \frac{1}{2} \, \bar \xi \left(
  (n_-x) \, n_+^\mu n_-^\nu \, W_c \,g_sF_{\mu\nu}^{s}  W_c^\dagger
  + x_\perp^\mu x_{\perp\rho} n_-^\nu W_c \big[D^\rho_{s}, 
   g_s F_{\mu\nu}^{s}\big] W_c^\dagger  \right) 
   \frac{\slash  n_+}{2} \xi
\nn \\
&& + \,\frac{1}{2} \, \bar \xi \left(
    i \Slash D_{\perp }  \,
 \frac{1}{i n_+ D} \, x_\perp^\mu \gamma_\perp^\nu \,
 W_c \,g_s F_{\mu\nu}^{s}W_c^\dagger  +   x_\perp^\mu \gamma_\perp^\nu \,
 W_c \,g_s F_{\mu\nu}^{s}W_c^\dagger  \,
 \frac{1}{i n_+ D} \, i \Slash D_{\perp }
 \right) \frac{\slash n_+}{2}\xi,
\nonumber\\[-0.3cm]
&&\\
{\cal L}^{(1)}_{\xi q} &=& 
    \bar q \,W_{c}^\dagger i\Slash{D}_{\perp } \,\xi - 
    \bar{\xi} \,i\overleftarrow{\Slash D}_{\perp } W_{c} q , 
\\[0.0cm]
{\cal L}^{(2)}_{\xi q} &=&  
\bar{q} \,\Wc^\dagger \left( i n_- D
   + i \Slash{D}_{\perp }\,
     \left(i n_+ D\right)^{-1} i \Slash{D}_{\perp }\right)
     \frac{\slash{n}_+}{2} \,\xi
+ \, \bar q \,
  \overleftarrow{D}^\mu_{s} x_{\perp\mu}
   W_{c}^\dagger \,i \Slash D_{\perp } \xi 
\nn \\
&& - \, \bar{\xi}\,\frac{\slash{n}_+}{2} \left( i n_- \overleftarrow{D}
   + i \overleftarrow{\Slash D}_{\perp }\,
     \left(i n_+ \overleftarrow{D}\right)^{-1}
    i \overleftarrow{\Slash D}_{\perp }\right) \Wc\, q
- \, \bar \xi \, i \overleftarrow{\Slash D}_{\perp } W_{c} \,x_{\perp\mu}
  D^\mu_{s} q.
\label{eq:Lxiq2}
\\
\label{eq:LYM1}
{\cal L}^{(1)}_{\rm YM} &=& 
\mbox{tr}\left(n_+^\mu F^c_{\mu\nu_\perp} \! W_c \,i\Big[x_\perp^\rho 
n_-^\sigma F^{s}_{\rho\sigma},W_c^\dagger [iD^{\nu_\perp} W_c]
\Big] W_c^\dagger\right)
- \mbox{tr}\left(n_{+\mu} F_c^{\mu\nu_\perp} W_c n_-^\rho F^{s}_{\rho
    \nu_\perp} W_c^\dagger \right),
\cr &&
\\[0.0cm]
{\cal L}^{(2)}_{\rm YM} &=& 
\frac{1}{2} \,\mbox{tr}\left(n_+^\mu F^c_{\mu\nu_\perp} W_c 
\,i\Big[\nm x\,n_+^\rho n_-^\sigma F^{s}_{\rho\sigma} + 
x_\perp^\rho x_{\perp\omega} n_-^\sigma 
\big[D_{s}^\omega,F^{s}_{\rho\sigma}\big],
W_c^\dagger [iD^{\nu_\perp} W_c]
\Big] W_c^\dagger\right)
\nonumber\\[0.1cm]
&&-\,\frac{1}{2} \,\mbox{tr}\left(n_{+\mu} F_c^{\mu\nu_\perp} W_c 
\,i\Big[x_\perp^\rho F^{s}_{\rho\nu_\perp},
W_c^\dagger i\nm D W_c - i\nm D_{s}
\Big] W_c^\dagger\right)
\nonumber\\[0.1cm]
&&+ \,\mbox{tr}\left(F_c^{\mu_\perp\nu_\perp} W_c 
\, i\Big[x_\perp^\rho  F^{s}_{\rho\mu_\perp},
W_c^\dagger [iD_{  \nu_\perp} W_c]\Big] W_c^\dagger\right)
\nonumber\\[0.1cm]
&&+\,\frac{1}{2} \,\mbox{tr}\left(n_+^\mu n_-^\nu F^c_{\mu\nu}
W_c n_+^\rho n_-^\sigma F^{s}_{\rho\sigma}W_c^\dagger \right) - 
\mbox{tr}\left(F_c^{\mu_\perp\nu_\perp} W_c F^{s}_{\mu_\perp\nu_\perp}
W_c^\dagger\right)
\nonumber\\[0.1cm]
&&-\,\mbox{tr}\left(n_{+\mu} F_c^{\mu\nu_\perp} W_c n_-^\rho
  x_{\perp\sigma} \big[D_{s}^\sigma,F^{s}_{\rho
    \nu_\perp}\big] W_c^\dagger \right).
\label{eq:LYM2}
\end{eqnarray}  
These Lagrangians are exact, i.e. its coefficients are not modified by
radiative corrections, neither do radiative corrections induce 
new operators \cite{Beneke:2002ph}. We note that interactions among 
collinear fields, without a soft field, exist only at leading power, 
while all subleading-power interactions {\em always} contain at least 
one soft field. The leading-power soft Lagrangian (second terms in
${\cal L}$ and ${\cal L}_{\rm YM}$, respectively) coincides with
the standard QCD Lagrangian for the soft fields. The leading-power 
collinear Lagrangian (first terms in ${\cal L}$ and ${\cal L}_{\rm YM}$, 
respectively) contains the soft field $\nm A_s(x_-)$, evaluated at 
position
\be
  x_-^\mu=\np x\frac{\nm^\mu}{2}\,,
\ee
only via the $\nm$ projection  $\nm D$ of the covariant derivative.
Soft fields that enter in ${\cal L}^{(n)}$ are also understood to be 
evaluated at $x_-$. 
The soft field entering in the leading-power collinear Lagrangian 
via $F_c^{\mu\nu}$ or $\nm D$ is evaluated at $x_-$ \emph{before} taking
derivatives, such that e.g. $\partial_\perp A_s(x_-)=\np\partial A_s(x_-)=0$ 
vanishes identically. In momentum space this corresponds
to setting $k_\perp=\np k=0$ for the soft field.
On the contrary, due to the multipole expansion, soft fields entering 
in ${\cal L}^{(n)}$ should be evaluated at $x_-$ \emph{after}
taking derivatives, e.g. $F_s^{\mu\nu}(x_-) = (\partial^\mu A_s^\nu)(x_-) 
- (\partial^\nu A_s^\mu)(x_-)+ \dots$ or $(D_sq)(x_-)=(\partial q)(x_-)+\dots$.
In momentum space this means that derivatives acting on soft fields yield 
a factor proportional to the \emph{full} soft momentum $k^\mu$ 
including the $\perp$ and $+$ components. Evaluating the soft expressions 
at $x_-$ then implies that $k_\perp$ and $\np k$ should be set to zero 
only inside of the momentum-conserving Dirac delta-function at the 
interaction vertex.

In addition to the Lagrangian presented in Ref.~\cite{Beneke:2002ni} we 
specify the gauge-fixing Lagrangian  
\be
\label{eq:gf}
{\cal L}_{\rm gf} = -\frac{1}{\alpha_c}\mbox{tr}\left(\frac12 (\np\partial) 
(\nm A_c) + \frac12 (\nm D_{s})(\np A_c) + \partial_\perp A_c^\perp 
\right)^{\!2}
-\frac{1}{\alpha_{s}}\mbox{tr}\left(\partial A_{s}\right)^2\,,
\ee
with $D_{s}^\mu A_c^\nu=\partial^\mu A_c^\nu-ig_s [ A_{s}^\mu(x_-), A_c^\nu]$. 
Using soft background field gauge for the collinear field in this form 
ensures that the gauge-fixing term for the collinear gauge symmetry 
preserves the soft gauge symmetry $A_c \to U_{s}(x_-)A_cU_{s}^\dag(x_-)$. 
One may use different gauge-fixing 
parameters $\alpha_s$ and $\alpha_c$ for the soft and collinear gauge 
symmetry, respectively. We present Feynman rules for this general choice, 
but use $\alpha_c=\alpha_s=1$ in our computations.
The corresponding ghost sector reads
\bea
{\cal L}_{\rm FP} &=& 
2\,{\rm tr} \left[ \bar c_{s} \left(-\partial_\mu D^{\mu}_{s}(x) 
\right) c_{s} \right] 
\nn\\
&& {} + 2\,{\rm tr}\left[ \bar c_{c} \left(-\frac12 (\np \partial) (\nm D) 
- \frac12 (\nm D_{s}(x_-)) (\np D) - \partial_{\perp\mu} 
D_{\perp}^\mu\right)  c_{c}\right]
\nn\\
&=& \bar c_{s}^a \left(-\partial^2 \delta^{ac} - g_s f^{abc} 
\partial A_{s}^b(x) \right) c_{s}^c
\nn\\
&& {} + \bar c_{c}^a \bigg(-\partial^2 \delta^{ac} - g_s f^{abc} 
\Big[ \partial A_{c}^b +   (\nm A_{s}^b(x_-))(\np \partial)\Big] 
\nn\\
&&  -\frac12 g_s^2 f^{ade}f^{ebc} (\nm A_{s}^d(x_-)) (\np A_c^b) \bigg) c_{c}^c\,.
\eea
By construction, the gauge-fixing term contributes only at leading power, 
and this property is inherited by the ghost interactions.

At subleading power, the multipole expansion produces terms in the 
Lagrangian proportional to powers of $x^\mu$, which leads to derivatives 
in momentum space. We explicitly include the momentum-conservation 
Dirac delta-functions for $x$-dependent vertices, using the notation
\bea
X^\mu &\equiv& \partial^\mu\!\left[ (2\pi)^4\delta^{(4)}\left(\sum p_{in} 
- \sum p_{out}\right)\right]\,, 
\nn\\
X^\mu X^\nu &\equiv& \partial^\mu \partial^\nu 
\!\left[ (2\pi)^4\delta^{(4)}
\left(\sum p_{in} - \sum p_{out}\right)\right]\,,
\eea
where the derivative $\partial=\partial/\partial p_{in}$ acts on one 
(arbitrarily chosen) incoming momentum in the argument of the delta-function, 
or equivalently on one outgoing momentum, $\partial=-\partial/\partial p_{out}$.
Note that a factor $x_\perp^\mu$ in the interaction term gives a factor 
$i X_\perp^\mu$ in the Feynman rule, where a projection on the perpendicular 
component is taken. Following the discussion above, for soft fields 
that are evaluated at position $x_-$ in the Lagrangian, the momentum 
components $\np k$ and $k_\perp$ must be set to zero inside 
the momentum-conservation delta-function \emph{after} the derivatives 
are taken. 
Spatial derivatives in the Lagrangian translate as $\partial_\mu\to -ip_\mu$ 
for incoming momentum, and $\partial_\mu\to ip_\mu$
for outgoing momentum, as usual. 

The gluon propagators take the standard form of general covariant 
gauge, $-i(g_{\mu\nu}-(1-\alpha )k_\mu k_\nu/k^2))/(k^2+i\varepsilon)$, with 
$\alpha=\alpha_c\,(\alpha_s)$ for collinear (soft) gluons, the soft quark 
propagator is also standard, $i\slashed{k}/(k^2+i\varepsilon)$, and the 
collinear quark propagator is 
\be
  \frac{i\np k}{k^2+i\varepsilon} \frac{\nms}{2}\,.
\ee

\subsection{Derivative operators and Wilson lines}

To derive Feynman rules, one can use the following expansion of the 
inverse collinear derivative operator
\bea
\frac{1}{i\np D} &=& \frac{1}{i\np \partial + g_s\np A_c} 
\nn \\
&=& \frac{1}{i\np \partial} - \frac{1}{i\np \partial}g_s\np A_c
\frac{1}{i\np \partial} + \frac{1}{i\np \partial}g_s\np A_c
\frac{1}{i\np \partial}g_s\np A_c\frac{1}{i\np \partial} -\dots \,.
\qquad
\label{eq:Dinv}
\eea
Inverse collinear derivative operators are always understood with a 
$i\np \partial \to i\np \partial+i\varepsilon$ prescription,
such that in momentum space
\be
  \frac{1}{\np p} \equiv \frac{1}{\np p+i\varepsilon}\,. 
\ee

To expand the collinear Wilson lines we use the identities
\bea
\frac{1}{i\np \partial}\phi(x) &=& (-i)\int_{-\infty}^0 ds\; \phi(x+s \np)\,, 
\nn \\
\frac{1}{i\np \partial}\phi(x) \frac{1}{i\np \partial}\phi(x) &=& 
(-i)^2  \,\frac12 \,P \int_{-\infty}^0 \!\!ds_1 \int_{-\infty}^0 \!\!ds_2\, 
\phi(x+s_1 \np)\phi(x+s_2 \np)\,, \quad
\label{eq:phi2}
\eea
where $P$ denotes path ordering with respect to the $s_i$.
This gives the following expansion of the collinear Wilson line, which we 
use to derive Feynman rules
\bea
W_c(x) &=& P \exp\left[ ig_s  \int_{-\infty}^0 ds\; \np A_c(x+s \np)
\right]
\nn\\
&=& 1 - \left[ \frac{1}{i\np \partial} g_s\np A_c\right] + 
\left[ \frac{1}{i\np \partial} g_s\np A_c \frac{1}{i\np \partial} 
g_s\np A_c\right]-\dots \,,
\label{eq:W}\\
W_c^\dag(x) &=& 1 + \left[ \frac{1}{i\np \partial} g_s\np A_c\right] 
+ \left[ \frac{1}{i\np \partial}\left[\frac{1}{i\np \partial} 
g_s\np A_c\right] g_s\np A_c \right]+\dots \,,
\nn
\eea
where we used Eq.~(\ref{eq:phi2}) for the second-order term.
Derivative operators act only inside square brackets. After inserting 
these expansions into the SCET Lagrangian, the Feynman rules can be read 
off in the standard way. For example, for one (two) incoming collinear 
gluon line(s) with momentum $k\,(q)$, Lorentz index $\mu(\nu)$ and colour 
$a\,(b)$,
\be
\label{eq:W2}
W_c \to \left\{\begin{array}{ll} 
\displaystyle -\frac{g_st^a{\np}_\mu}{\np k} & 
{\rm one\ gluon} \\[0.2cm]
g_s^2W^{ab}_{\mu\nu}(k,q) \equiv {\displaystyle 
g_s^2\frac{{\np}_\mu{\np}_\nu}{\np(k+q)}\left(\frac{t^at^b}{\np q} 
+\frac{t^bt^a}{\np k}\right)}\hspace*{0.4cm}
& {\rm two\ gluons}\end{array}
\right.
\ee
and
\be
\label{eq:W2hat}
W_c^\dag \to \left\{\begin{array}{ll} 
\displaystyle \frac{g_st^a{\np}_\mu}{\np k} & 
{\rm one\ gluon} \\[0.2cm]
g_s^2\hat W^{ab}_{\mu\nu}(k,q) \equiv g_s^2
 W^{ba}_{\mu\nu}(k,q)\hspace*{0.4cm}
& {\rm two\ gluons}\end{array}
\right.
\ee

\subsection{Notation for Yang-Mills Feynman rules}

A single collinear gluon with incoming momentum $k$, Lorentz index $\mu$ and 
colour index $a$ produces the following terms in the Feynman rules when 
attached to the operators shown on the left,
\bea
i (n_\pm)_\rho F_c^{\rho\nu_\perp} &\to& t^a f_\mu^{\pm\nu_\perp}(k) \,,
\nn\\
i (n_+)_\rho (n_-)_\nu F_c^{\rho\nu} &\to& t^a f_\mu^{+-}(k) \,,
\nn\\
i F_c^{\rho_\perp\nu_\perp} &\to& t^a f_\mu^{\rho_\perp\nu_\perp}(k) \,,
\eea
with the definitions
\bea
f_\mu^{\pm\nu_\perp}(k) &=&
n_\pm^\kappa g_\perp^{\nu\sigma} (k_\kappa g_{\mu\sigma} - g_{\mu\kappa} 
k_\sigma)  
= (n_\pm k) \delta_{\perp\mu}^\nu - {n_\pm}_\mu k_\perp^\nu \,,
\nn\\
f_\mu^{+-}(k) &=& 
\np^\kappa \nm^\sigma (k_\kappa g_{\mu\sigma} - g_{\mu\kappa} k_\sigma) 
= (\np k){\nm}_\mu - (\nm k){\np}_\mu \,,
\nn\\
f_\mu^{\rho_\perp\nu_\perp}(k) &=& 
g_\perp^{\rho\kappa} g_\perp^{\nu\sigma}(k_\kappa g_{\mu\sigma} 
- g_{\mu\kappa} k_\sigma) 
= k_{\perp}^\rho \delta_{\perp\mu}^\nu - k_{\perp}^\nu 
\delta_{\perp\mu}^\rho \,.
\eea
The three expressions can be written in a compact form by introducing 
the ``projectors'' $P_\mu^+\equiv{\np}_\mu$,  $P_\mu^-\equiv{\nm}_\mu$, 
$P_\mu^{\lambda_\perp}\equiv \delta_{\perp\mu}^{\lambda}\equiv 
g_\perp^{\lambda\nu}g_{\nu\mu}$ on the light-cone basis. Using the notation 
$F_c^{AB}\equiv P^A_\mu P^B_\nu F_c^{\mu\nu}$ as well as
\be
f_\mu^{AB}(k) = P^A_\kappa P^B_\sigma (k^\kappa \delta_\mu^\sigma 
- k^\sigma \delta_\mu^\kappa )\,,
\ee
for $A,B\in\{+,-,\lambda_\perp\}$, the rules from above can be summarized as
\bea
  i F^{AB} &\to& t^a f_\mu^{AB}(k) \,.
\eea
Similarly, defining $iD^A\equiv P_\mu^A iD^\mu$,
\bea
 \left[ i D^{A} W_c \right] &\to& g_st^a \frac{1}{\np k} f_\mu^{+A}(k) \,.
\eea
To lower $\perp$ indices we use the convention $f_{\nu\lambda_\perp}^\pm\equiv f_\nu^{\pm\kappa_\perp}g_{\kappa\lambda}$, 
$f_{\nu\rho_\perp\lambda_\perp}\equiv f_\nu^{\sigma_\perp\kappa_\perp}g_{\sigma\rho}g_{\kappa\lambda}$.

Two collinear gluons with incoming momenta, labeled by $k\mu a$ and 
$q \nu b$, respectively, Lagrangian terms map into Feynman rules as 
follows:
\bea
i F_c^{AB} &\to& g_s[t^a,t^b] f_{\mu\nu}^{AB} \,,
\nn\\
\left[ i D^{A} W_c \right] &\to& g_s^2\left(P_\kappa^A(k+q)^\kappa 
W_{\mu\nu}^{ab}(k,q) - t^at^b \delta_\mu^A\frac{\np^\nu}{\np q} - 
t^bt^a \delta_\nu^A\frac{\np^\mu}{\np q}\right)\,,
\nn\\[0.1cm]
i [D^\omega,  F_c^{AB}] &\to& g_s[t^a,t^b] \big((k+q)^\omega f_{\mu\nu}^{AB} 
+ \delta^\omega_\mu f_\nu^{AB}(q) - \delta^\omega_\nu f_\mu^{AB}(k)
\big)\,,
\eea
where
\be
f_{\mu\delta}^{AB} \equiv P^A_\kappa P^B_\sigma (\delta_\mu^\kappa 
\delta_\delta^\sigma - \delta_\delta^\kappa \delta_\mu^\sigma)\,.
\ee

\subsection{Fermionic Feynman rules}

Note: for vertices not containing a momentum-derivative, the standard 
momentum conserving delta-function $(2\pi)^4\delta^{(4)}\left(\sum p_{in} 
- \sum p_{out}\right)$ is not written explicitly. Otherwise we write 
$X^\mu$ as defined above.

\subsubsection{Purely collinear or purely soft vertices}

\vskip0.2cm
\be\label{eq:Vxixic}
  \raisebox{-13mm}{\includegraphics[width=0.2\textwidth]{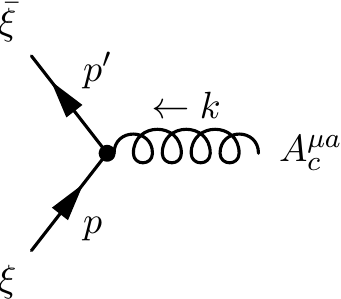}} \quad
  i g_s t^a \left\{ \begin{array}{ll}
     C_\mu(p',p)\, \displaystyle \npshalf   & {\cal O}(\lambda^0) \\
     0 & {\cal O}(\lambda) \\[0.1cm]
     0 & {\cal O}(\lambda^2)
  \end{array}
  \right.
\ee
\vskip0.2cm
\be\label{eq:Vxixicc}
  \raisebox{-17mm}{\includegraphics[width=0.2\textwidth]{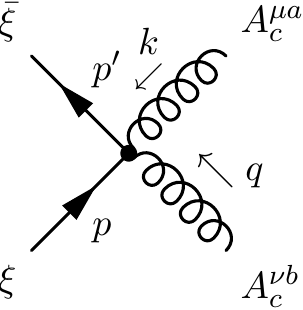}} \quad
  i g_s^2 \left\{ \begin{array}{ll}
     C_{\mu\nu}^{ab}(p',p,k,q) \displaystyle \npshalf  
  & {\cal O}(\lambda^0) \\
     0 & {\cal O}(\lambda) \\[0.1cm]
     0 & {\cal O}(\lambda^2)
  \end{array}
  \right.
\ee
where
\bea
C^\mu(p',p) &\equiv& 
\nm^\mu + \frac{\slashed{p}'_\perp}{\np p'} \gamma_\perp^\mu 
+ \gamma_\perp^\mu \frac{\slashed{p}_\perp}{\np p}  
- \frac{\slashed{p}'_\perp}{\np p'} \np^\mu \frac{\slashed{p}_\perp}{\np p} 
\,, 
\nn \\
C_{\mu\nu}^{ab}(p',p,k,q) &\equiv & \Gamma_\mu(p')\frac{t^at^b}{\np (p+q)} 
\Gamma_\nu(p) + \Gamma_\nu(p')\frac{t^bt^a}{\np (p+k)} \Gamma_\mu(p) \,,
\nn\\
\Gamma^\mu(p) &\equiv& \gamma_\perp^\mu - \frac{\slashed{p}_\perp}{\np p}
\np^\mu\,.
\eea

\be
  \raisebox{-13mm}{\includegraphics[width=0.2\textwidth]{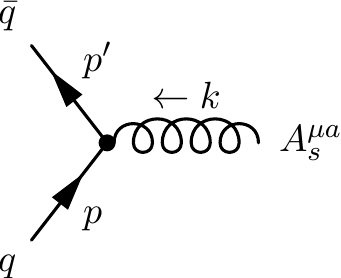}} \quad
  i g_s t^a \left\{ \begin{array}{ll}
     \gamma_\mu  & {\cal O}(\lambda^0) \\
     0 & {\cal O}(\lambda) \\
     0 & {\cal O}(\lambda^2)
  \end{array}
  \right.
\ee
We recall that there are no sub-leading power vertices of this type 
to any order in the $\lambda$ expansion.

\subsubsection{Soft-collinear interaction vertices} 

\vskip0.2cm
\be\label{eq:Vccs}
\raisebox{-13mm}{\includegraphics[width=0.2\textwidth]{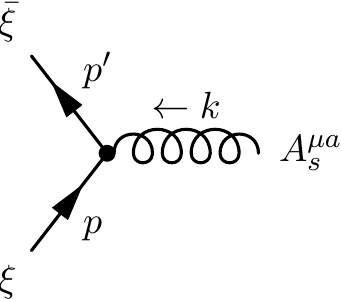}} 
\quad
i g_s t^a \left\{ \begin{array}{ll}
     \displaystyle \npshalf {\nm}_\mu  & {\cal O}(\lambda^0) \\[0.3cm]
     \displaystyle  \npshalf X_\perp^\rho \nm^\nu 
     (k_\rho g_{\nu\mu}-k_\nu g_{\rho\mu}) & {\cal O}(\lambda) \\
     S^{\rho\nu}(k,p,p')  \displaystyle \npshalf 
     (k_\rho g_{\nu\mu}-k_\nu g_{\rho\mu})& {\cal O}(\lambda^2)
\end{array}\right.
\ee
where
\be\label{eq:Srnonu}
S^{\rho\nu}(k,p,p') \equiv \frac12 \left[ (\nm X) \np^\rho \nm^\nu + 
(k X_\perp) X_\perp^\rho \nm^\nu
+X_\perp^\rho \left( \frac{\slashed{p}'_\perp}{\np p'} \gamma_\perp^\nu 
+ \gamma_\perp^\nu \frac{\slashed{p}_\perp}{\np p} \right) \right].
\ee
{\em After} the derivative in $X_\perp^\rho$ is taken, 
$p^\prime_\perp$ can be set to $p_\perp$. ($\np p'=\np p$ may be 
set from the start.)

\be
\raisebox{-17mm}{\includegraphics[width=0.2\textwidth]{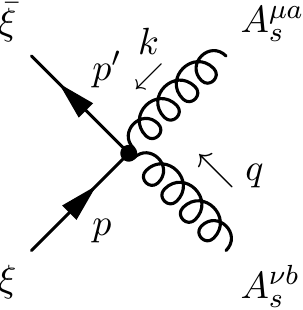}} 
\quad
i g_s^2 [t^a,t^b] \left\{ \begin{array}{ll}
0  & {\cal O}(\lambda^0) \\[0.2cm]
\displaystyle \npshalf X_\perp^\rho \nm^\sigma (g_{\rho\mu} g_{\sigma\nu}-g_{\rho\nu} g_{\sigma\mu}) & {\cal O}(\lambda) \\
\displaystyle S^{\rho\sigma}(k+q,p,p') \displaystyle 
\npshalf (g_{\rho\mu} g_{\sigma\nu}-g_{\rho\nu} g_{\sigma\mu}) \\
\displaystyle + \frac12 \npshalf X_\perp^\rho X_\perp^\sigma \nm^\lambda 
\Big[g_{\rho\mu}(q_\sigma g_{\lambda\nu}-q_\lambda g_{\sigma\nu}) \\[0.15cm]
- g_{\rho\nu}(k_\sigma g_{\lambda\mu}-k_\lambda g_{\sigma\mu})\Big] & 
{\cal O}(\lambda^2)
\end{array}
\right.
\ee

\be
\raisebox{-17mm}{\includegraphics[width=0.2\textwidth]{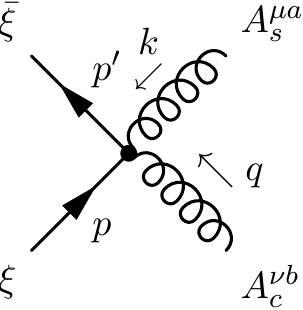}} \quad
i g_s^2 \left\{ \begin{array}{ll}
0 & {\cal O}(\lambda^0) \\[0.2cm]
\displaystyle \npshalf X_\perp^\rho \nm^\sigma (k_\rho g_{\sigma\mu} - 
k_\sigma g_{\rho\mu}) \frac{{\np}_\nu}{\np q} [t^a,t^b]  & 
{\cal O}(\lambda) \\[0.2cm]
\displaystyle \Big[ \frac12 X_\perp^\rho \left(\Gamma_\nu(p')
\frac{\gamma_\perp^\sigma}{\np (p'-q)}t^bt^a + \frac{\gamma_\perp^\sigma}
{\np (p+q)}\Gamma_\nu(p)t^at^b\right) \\
\displaystyle + S^{\rho\sigma}(k,p,p')  \frac{{\np}_\nu}{\np q} [t^a,t^b]   
\Big] \npshalf (k_\rho g_{\sigma\mu} - k_\sigma g_{\rho\mu})  
& {\cal O}(\lambda^2)
  \end{array}
  \right.
\ee

\be
\raisebox{-13mm}{\includegraphics[width=0.2\textwidth]{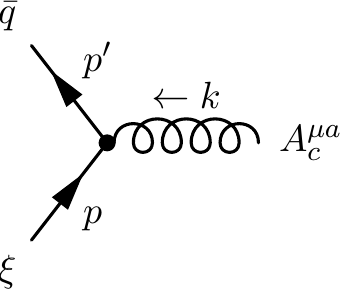}} \quad
  i g_s t^a \left\{ \begin{array}{ll}
     0  & {\cal O}(\lambda^0) \\[0.1cm]
      \Gamma_\mu(p) & {\cal O}(\lambda) \\
     \displaystyle \Big[{\nm}_\mu       + \gamma_{\perp\mu} \frac{\slashed{p}_\perp}{\np p} +\frac{{\np}_\mu}{\np k}\frac{p^2}{\np p}\Big]\frac{\nps}{2} 
      - (p' X_\perp)\Gamma_\mu(p)\;\; & {\cal O}(\lambda^2)
  \end{array}
  \right.
\ee

\be
\raisebox{-13mm}{\includegraphics[width=0.2\textwidth]{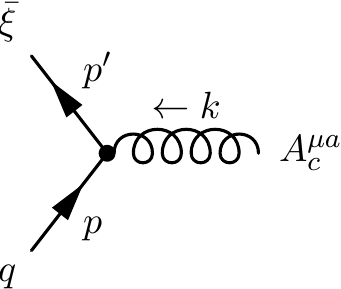}} \quad
i g_s t^a \left\{ \begin{array}{ll}
0  & {\cal O}(\lambda^0) \\
\Gamma_\mu(p') & {\cal O}(\lambda) \\
\displaystyle \frac{\nps}{2}\big[{\nm}_\mu +\frac{\slashed{p}'_\perp}
{\np p'} \gamma_{\perp\mu} -\frac{{\np}_\mu}{\np k}\frac{(p')^2}{\np p'}
\big] + (p X_\perp)\Gamma_\mu(p')\;\; & {\cal O}(\lambda^2)
  \end{array}
  \right.
\ee

\be
\label{eq:qxicc}
\raisebox{-17mm}{\includegraphics[width=0.2\textwidth]{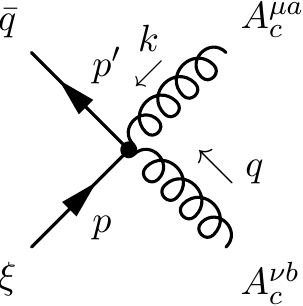}} \quad
i g_s^2 \left\{ \begin{array}{ll}
0 & {\cal O}(\lambda^0) \\[0.1cm]
     \hat W_{\mu\nu}^{ab}(k,q) \slashed{p}_\perp + D^{ab}_{\mu\nu}(k,q) & {\cal O}(\lambda) \\[0.1cm]
      \displaystyle V^{ab}_{\mu\nu}(p',p,k,q) \frac{\nps}{2} 
      - (p' X_\perp) \Big[ \hat W_{\mu\nu}^{ab}(k,q) \slashed{p}_\perp + D^{ab}_{\mu\nu}(k,q) \Big]\;\; & {\cal O}(\lambda^2)
  \end{array}
  \right.
\ee

\be\label{eq:xiqcc}
  \raisebox{-17mm}{\includegraphics[width=0.2\textwidth]{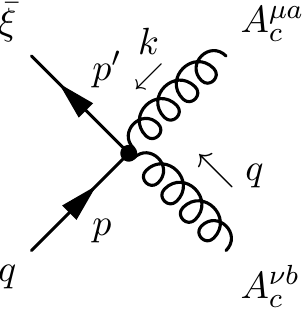}} \quad
  i g_s^2 \left\{ \begin{array}{ll}
     0 & {\cal O}(\lambda^0) \\
     \slashed{p}'_\perp W_{\mu\nu}^{ab}(k,q)  -  D^{ba}_{\mu\nu}(k,q) & {\cal O}(\lambda) \\[0.1cm]
     \displaystyle \frac{\nps}{2} \bar V^{ab}_{\mu\nu}(p',p,k,q)  
      + (p X_\perp) \Big[ \slashed{p}'_\perp W_{\mu\nu}^{ab}(k,q)  -  D^{ba}_{\mu\nu}(k,q)  \Big]\;\; & {\cal O}(\lambda^2)
  \end{array}
  \right.
\ee
\vskip0.2cm\noindent
where the second-order contributions $W_{\mu\nu}^{ab}(k,q)$ and $\hat W_{\mu\nu}^{ab}(k,q)$ from the Wilson 
lines were defined in Eq.~\eqref{eq:W2} and Eq.~\eqref{eq:W2hat}, respectively, and
\bea
D^{ab}_{\mu\nu}(k,q) &\equiv& \frac{{\np}_\mu}{\np k}\gamma_{\perp\nu} 
t^at^b + \frac{{\np}_\nu}{\np q}\gamma_{\perp\mu} t^bt^a \,,
\nn\\
V^{ab}_{\mu\nu}(p',p,k,q) &\equiv& t^at^b \Bigg[ - \frac{{\np}_\mu{\np}_\nu}
{\np p\np k}\frac{p^2}{\np p} - \frac{\gamma_{\perp\mu}}{\np k}\Gamma_\nu(p)
    + \frac{{\np}_\mu}{\np k}C_\nu(-k,p) \Bigg] \nn\\
  && + \,(k\mu a \leftrightarrow q\nu b)\,, \nn\\
\bar V^{ab}_{\mu\nu}(p',p,k,q) &\equiv& t^at^b \Bigg[ \frac{{\np}_\mu{\np}_\nu}{\np p'\np q}\frac{{p'}^2}{\np p'} + \Gamma_\mu(p')\frac{\gamma_{\perp\nu}}{\np q}
    - \frac{{\np}_\nu}{\np q}C_\mu(p',q) \Bigg] \nn\\
  && + \,(k\mu a \leftrightarrow q\nu b)\,.
\eea

\be
\raisebox{-17mm}{\includegraphics[width=0.2\textwidth]{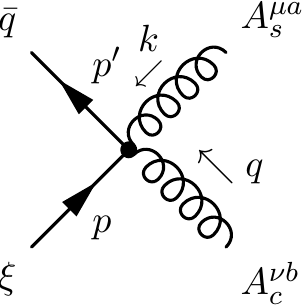}} \quad
i g_s^2 \left\{ \begin{array}{ll}
     0 & {\cal O}(\lambda^0) \\
     0 & {\cal O}(\lambda) \\
     \displaystyle t^bt^a  \,\frac{{\nm}_\mu {\np}_\nu}{\np q}\frac{\nps}{2}
     - t^at^b \,X_{\perp\mu} \Gamma_\nu(p)\;\; & {\cal O}(\lambda^2)
  \end{array}
  \right.
\ee

\be
\hskip2cm
  \raisebox{-17mm}{\includegraphics[width=0.2\textwidth]{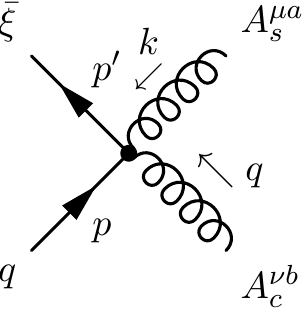}} \quad
  i g_s^2 \left\{ \begin{array}{ll}
     0 & {\cal O}(\lambda^0) \\
     0 & {\cal O}(\lambda) \\
\displaystyle - t^at^b \,\frac{ {\nm}_\mu {\np}_\nu}{\np q} \frac{\nps}{2} 
+  t^bt^a \, \Gamma_\nu(p')  X_{\perp\mu}\;\; & {\cal O}(\lambda^2)
\end{array}
\right.
\ee

\subsection{Three gluon vertices}

\subsubsection{Purely collinear or purely soft vertices}

We use the abbreviation
\be\label{eq:threegluon}
  Q^{\mu\nu\rho}(k,q,p) \equiv g^{\mu\nu}(k-q)^\rho+g^{\nu\rho}(q-p)^\mu+g^{\rho\mu}(p-k)^\nu\,,
\ee
for the structure of the standard QCD vertex. The leading-power SCET
vertices involving three collinear gluons or three soft gluons, 
respectively, are identical to the QCD vertex:

\be
  \raisebox{-13mm}{\includegraphics[width=0.2\textwidth]{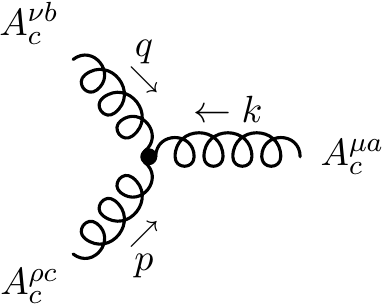}} \quad
  g_s f^{abc} \left\{ \begin{array}{ll}
     Q_{\mu\nu\rho}(k,q,p) \;\; & {\cal O}(\lambda^0) \\
     0 & {\cal O}(\lambda) \\
     0 & {\cal O}(\lambda^2)
  \end{array}
  \right.
\ee

\be
  \raisebox{-13mm}{\includegraphics[width=0.2\textwidth]{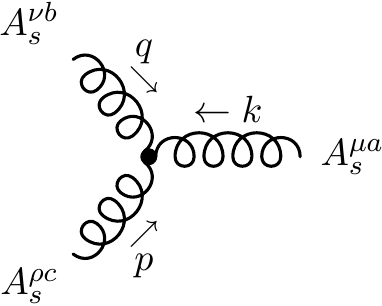}} \quad
  g_s f^{abc} \left\{ \begin{array}{ll}
     Q_{\mu\nu\rho}(k,q,p) \;\; & {\cal O}(\lambda^0) \\
     0 & {\cal O}(\lambda) \\
     0 & {\cal O}(\lambda^2)
  \end{array}
  \right.
\ee

\subsubsection{Soft-collinear interaction vertices} 

\vskip0.2cm
\be\label{eq:Vuscc}
  \raisebox{-13mm}{\includegraphics[width=0.2\textwidth]{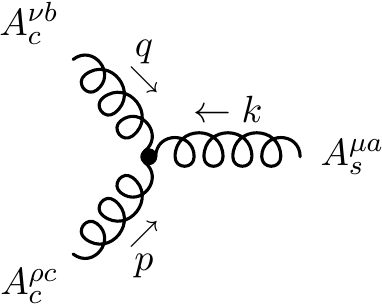}} \quad
  \frac12 g_s f^{abc} \left\{ \begin{array}{ll}
     V^{(0)}_{\mu\nu\rho}(k,q,p)  & {\cal O}(\lambda^0) \\[0.1cm]
      V^{(1)}_{\mu\nu\rho}(k,q,p) & {\cal O}(\lambda) \\[0.1cm]
     \sum_{i=1}^6 V^{(2),i}_{\mu\nu\rho}(k,q,p)\;\;
     & {\cal O}(\lambda^2)
  \end{array}
  \right.
\ee
where
\bea
V^{(0)}_{\mu\nu\rho}(k,q,p) &\equiv&  
\left[ -2g_{\nu\rho}(\np p)+\left(1- \frac{1}{\alpha_c}\right)
({\np}_\nu p_\rho - {\np}_\rho q_\nu) \right] {\nm}_\mu \,.
\eea
At ${\cal O}(\lambda)$ we find
\be\label{eq:V1}
V^{(1)}_{\mu\nu\rho}(k,q,p) \equiv V^{(1),\lambda_\perp}_{\nu\rho}(q,p)
f_{\mu\lambda_\perp}^{-}(k) \,,
\ee
with
\bea
\label{eq:V1b}
V^{(1),\lambda_\perp}_{\nu\rho}(q,p) &\equiv&  f_\nu^{+\sigma_\perp}(q)
f_{\rho\sigma_\perp}^+(p)\left(\frac{1}{\np q}-\frac{1}{\np p}\right)
X_{\perp}^{\lambda} + H_{\nu\rho}^{+\lambda_\perp}(q,p)  
\nn\\
&\to & 2\left((\np p)g_{\perp\nu\rho}+{\np}_\nu q_{\perp\rho}-{\np}_\rho 
p_{\perp\nu}-{\np}_\nu{\np}_\rho\frac{q_\perp p_\perp}{\np p}\right)
X_{\perp}^{\lambda} \,,
\eea
and 
\be
  H_{\nu\rho}^{+\lambda_\perp}(q,p) \equiv f_\nu^{+\lambda_\perp}(q)
\frac{{\np}_\rho}{\np p} - f_\rho^{+\lambda_\perp}(p)\frac{{\np}_\nu}{\np q} -
\np^\sigma(g_{\sigma\nu}g_{\kappa\rho}-g_{\sigma\rho}g_{\kappa\nu})
g_\perp^{\kappa\lambda} \,.
\ee
The two terms on the right-hand side of the first line of 
Eq.~\eqref{eq:V1b} correspond to the two terms of 
${\cal L}_{\rm YM}^{(1)}$ in Eq.~\eqref{eq:LYM1}. The last line has been 
obtained using $\np p=-\np q$, while $p_\perp=-q_\perp$ has been used only 
for the term that does \emph{not} involve $X_\perp$ (which can then be 
shown to vanish). 
At ${\cal O}(\lambda^2)$ we find six vertex factors that correspond to the 
six terms in ${\cal L}_{\rm YM}^{(2)}$ in Eq.~\eqref{eq:LYM2},
\bea\label{eq:V2}
 V^{(2),1}_{\mu\nu\rho}(k,q,p) &\equiv& V^{(2),1}_{\nu\rho}(q,p) \left(-\nm X f_\mu^{+-}(k) + kX_\perp X_{\perp}^\lambda f_{\mu \lambda_\perp}^{-}(k)\right)\,, \nn\\[0.05cm]
  V^{(2),2+3}_{\mu\nu\rho}(k,q,p) &\equiv& V^{(2),2+3,\lambda_\perp}_{\nu\rho}(q,p) X_{\perp}^{\kappa} f_{\mu\kappa_\perp\lambda_\perp}(k)\,, \nn\\[0.05cm]
  V^{(2),4}_{\mu\nu\rho}(k,q,p) &\equiv& V^{(2),4}_{\nu\rho}(q,p) f_\mu^{+-}(k) \,, \nn\\[0.05cm]
  V^{(2),5}_{\mu\nu\rho}(k,q,p) &\equiv& V^{(2),5,\lambda_\perp\sigma_\perp}_{\nu\rho}(q,p) f_{\mu\lambda_\perp\sigma_\perp}(k)\,,   \nn\\[0.05cm]
  V^{(2),6}_{\mu\nu\rho}(k,q,p) &\equiv& V^{(2),6,\lambda_\perp}_{\nu\rho}(q,p) kX_\perp  f_{\mu\lambda_\perp}^{-}(k)  \,,
\eea
where
\bea\label{eq:V2coeff}
  V^{(2),1}_{\nu\rho}(q,p) &\equiv& \frac12 f_\nu^{+\sigma_\perp}(q)f_{\rho\sigma_\perp}^+(p)\left(\frac{1}{\np q} -\frac{1}{\np p}\right)  \nn\\
  &\to & (\np p)g_{\perp\nu\rho}+{\np}_\nu q_{\perp\rho}-{\np}_\rho p_{\perp\nu}-{\np}_\nu{\np}_\rho\frac{q_\perp p_\perp}{\np p} \,,  \nn\\
  V^{(2),2+3,\lambda_\perp}_{\nu\rho}(q,p)  &\equiv&  (q-p)_\perp^\lambda g_{\nu\rho} + \frac{pq}{\np p}({\np}_\rho \delta_{\perp\nu}^\lambda + {\np}_\nu \delta_{\perp\rho}^\lambda) 
     -q_\rho \delta_{\perp\nu}^\lambda     +p_\nu \delta_{\perp\rho}^\lambda \nn \\
&& {} - \frac{q_\perp^\lambda p_\nu {\np}_\rho + p_\perp^\lambda q_\rho {\np}_\nu}{\np p}  \,, \nn\\
  V^{(2),4}_{\nu\rho}(q,p) &\equiv&  \frac12 \frac{{\np}_\nu{\np}_\rho }{\np p} \, \nm (p+q) \,,\nn\\
  V^{(2),5,\lambda_\perp\sigma_\perp}_{\nu\rho}(q,p) &\equiv&  \left(p_\perp^\lambda \frac{{\np}_\nu \delta_{\perp\rho}^\sigma  -{\np}_\rho \delta_{\perp\nu}^\sigma }{\np p} +  \delta_{\perp\rho}^\lambda\delta_{\perp\nu}^\sigma\right)  - (\lambda\leftrightarrow\sigma) \,,\nn\\
   V^{(2),6,\lambda_\perp}_{\nu\rho}(q,p) &\equiv& -\frac{{\np}_\nu{\np}_\rho}{\np p}(p+q)_\perp^\lambda  \;.
\eea
Except for the first line we used $\np (p+q)=0$ and, for terms without 
any $X_\perp$, $(p+q)_\perp=0$. Note that in Eq.~(\ref{eq:V2}), one should
 use the expression for $V^{(2),1}_{\nu\rho}(q,p)$
from the first line above in the contribution involving $\nm X$.

\subsection{Four gluon vertices}

\subsubsection{Purely collinear or purely soft vertices}

The standard QCD four-gluon vertex is proportional to
\bea
Q^{abcd}_{\mu\nu\rho\delta} &=& f^{eab}f^{ecd}(g_{\mu\rho}g_{\nu\delta}
-g_{\mu\delta}g_{\nu\rho})
    + f^{eac}f^{ebd}(g_{\mu\nu}g_{\rho\delta}-g_{\mu\delta}g_{\rho\nu}) 
\nn \\
&& + f^{ead}f^{ebc}(g_{\mu\nu}g_{\delta\rho}-g_{\mu\rho}g_{\delta\nu})\,.
\label{eq:fourgluon}
\eea
The leading-power SCET 
vertices involving four collinear gluons or four soft gluons 
are identical to the QCD vertex.

\be
  \raisebox{-17mm}{\includegraphics[width=0.2\textwidth]{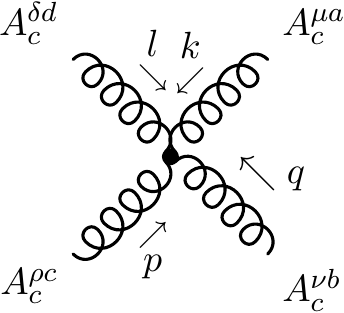}} \quad
  - i g_s^2 \left\{ \begin{array}{ll}
     Q^{abcd}_{\mu\nu\rho\delta} \;\;& {\cal O}(\lambda^0) \\
     0 & {\cal O}(\lambda) \\
     0 & {\cal O}(\lambda^2)
  \end{array}
  \right.
\ee

\be
\raisebox{-17mm}{\includegraphics[width=0.2\textwidth]{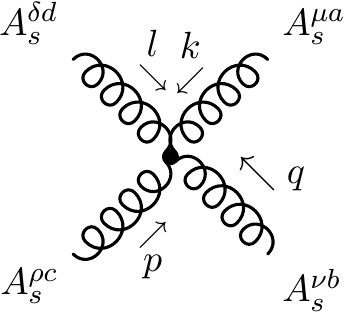}} 
\quad
  - i g_s^2 \left\{ \begin{array}{ll}
     Q^{abcd}_{\mu\nu\rho\delta} \;\;& {\cal O}(\lambda^0) \\
     0 & {\cal O}(\lambda) \\
     0 & {\cal O}(\lambda^2)
  \end{array}
  \right.
\ee

\subsubsection{Soft-collinear interaction vertices} 

The two collinear--two soft gluon vertex reads
\be
\raisebox{-14mm}{\includegraphics[width=0.2\textwidth]{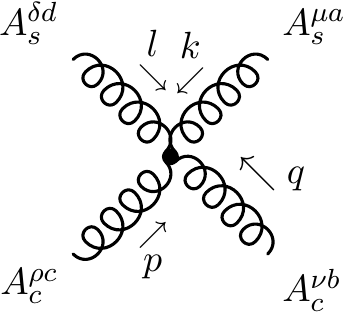}} \quad
  \!- i g_s^2 \left\{ \begin{array}{ll}
  \displaystyle \frac14 \, {\nm}_\mu {\np}_\nu {\np}_\rho {\nm}_\delta 
  \left(1-\frac{1}{\alpha_c}\right)\left(f^{eab}f^{ecd}+f^{eac}f^{ebd}
  \right) \;\;& {\cal O}(\lambda^0) \\[0.2cm]
   \displaystyle -\frac12 \,f^{ead}f^{ebc} V^{(1)}_{\nu\rho,\lambda_\perp}(q,p) f_{\mu\delta}^{-\lambda_\perp}  & {\cal O}(\lambda) \\[0.2cm]
    \displaystyle  -\frac12 \,f^{ead}f^{ebc}  \sum_{i=1}^6 V^{(2),i}_{\mu\nu\rho\delta}(k,q,p,l) + \Delta V^{(2),2,abcd}_{\mu\nu\rho\delta}(k,q,p,l)\;\;
 & {\cal O}(\lambda^2)
  \end{array}
  \right.
\ee
where
\bea\label{eq:V2ccusus}
V^{(2),1}_{\mu\nu\rho\delta}(k,q,p,l) &\equiv& V^{(2),1}_{\nu\rho}(q,p) 
\Big[-\nm X f_{\mu\delta}^{+-} + X_{\perp\lambda}  \big( (k+l)X_\perp f_{\mu\delta}^{- \lambda_\perp} \nn\\
  && {} + X_{\perp\mu}f_\delta^{-\lambda_\perp}(l) -  X_{\perp\delta}f_\mu^{-\lambda_\perp}(k)\big)\Big] \,,
\nn\\[0.05cm]
  V^{(2),2+3}_{\mu\nu\rho\delta}(k,q,p,l) &\equiv& V^{(2),2+3}_{\nu\rho,\lambda_\perp}(q,p) X_{\perp\kappa} f_{\mu\delta}^{\kappa_\perp\lambda_\perp} \,,
\nn\\[0.05cm]
  V^{(2),4}_{\mu\nu\rho\delta}(k,q,p,l) &\equiv& V^{(2),4}_{\nu\rho}(q,p) f_{\mu\delta}^{+-} +\frac12 \frac{{\np}_\nu{\np}_\rho}{\np p}\left({\nm}_\mu f_\delta^{+-}(l)-{\nm}_\delta f_\mu^{+-}(k) \right) \,,
\nn\\[0.05cm]
  V^{(2),5}_{\mu\nu\rho\delta}(k,q,p,l) &\equiv& V^{(2),5}_{\nu\rho,\lambda_\perp\sigma_\perp}(q,p) f_{\mu\delta}^{\lambda_\perp\sigma_\perp} \,,  \nn\\
  V^{(2),6}_{\mu\nu\rho\delta}(k,q,p,l) &\equiv& V^{(2),6}_{\nu\rho,\lambda_\perp}(q,p) \left( (k+l)X_\perp  f_{\mu\delta}^{-\lambda_\perp} + X_{\perp\mu}f_\delta^{-\lambda_\perp}(l) -  X_{\perp\delta}f_\mu^{-\lambda_\perp}(k)\right) \,,
\nn\\
\eea
with $V^{(2),i}(q,p)$ on the right-hand side defined already for the 
three-gluon vertices in Eq.~\eqref{eq:V2coeff}. The extra term 
$\Delta V^{(2),2}$ arises from attaching a soft gluon to 
$W_c^\dag \nm D W_c - \nm D_{s}$ in the second term  in 
${\cal L}_{\rm YM}^{(2)}$ in Eq.~\eqref{eq:LYM2},
\bea
\Delta V^{(2),2,abcd}_{\mu\nu\rho\delta}(k,q,p,l) &\equiv& 
\frac14  \,\Big(f_\nu^{+\lambda_\perp}(q) X_{\perp\kappa} 
f_{\mu\lambda_\perp}^{\kappa_\perp}(k) \frac{{\np}_\rho}{\np p} 
{\nm}_\delta 
\nn\\
&& + f_\rho^{+\lambda_\perp}(p) X_{\perp\kappa} 
f_{\delta\lambda_\perp}^{\kappa_\perp}(l) \frac{{\np}_\nu}{\np q} 
{\nm}_\mu \Big)\, f^{eab}f^{ecd} + (k\mu a \leftrightarrow l\delta d) \;. 
\nn\\[-0.3cm]
\eea
As before, one should use the expression for $V^{(2),1}_{\nu\rho}(q,p)$
from the first line of Eq.~\eqref{eq:V2coeff} in the contribution involving 
$\nm X$, before the collinear momentum conservation is imposed.

The three collinear--one soft gluon vertex reads
\be
\raisebox{-17mm}{\includegraphics[width=0.2\textwidth]{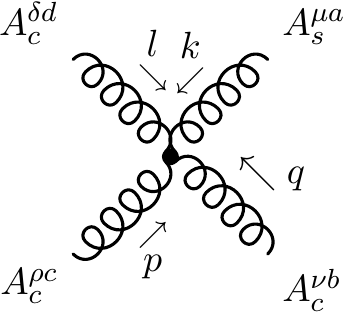}} \quad
  - i g_s^2 \left\{ \begin{array}{ll}
  \displaystyle \np^\sigma Q^{abcd}_{\sigma\nu\rho\delta}
   \frac{{\nm}_\mu}{2} & {\cal O}(\lambda^0) \\[0.3cm]
     \sum_{i=1}^2 V^{(1),i,abcd}_{\mu\nu\rho\delta}(k,q,p,l) \ + 5\ {\rm  permutations}\;\; & {\cal O}(\lambda) \\[0.3cm]
     \sum_{i=1}^6 V^{(2),i,abcd}_{\mu\nu\rho\delta}(k,q,p,l) \ + 5\ {\rm  permutations} & {\cal O}(\lambda^2)
  \end{array}
  \right.
\ee
In the subleading power vertices one needs to add the terms that are 
obtained from permutations of the three collinear gluons. The auxiliary 
functions are defined by
\bea
V^{(1),1,abcd}_{\mu\nu\rho\delta}(k,q,p,l) &\equiv& 
H^{abcd,+\perp\perp}_{\nu\rho\delta}(q,p,l)  X_{\perp\kappa} 
f_\mu^{-\kappa_\perp}(k)\,,
\nn\\
V^{(1),2,abcd}_{\mu\nu\rho\delta}(k,q,p,l) &\equiv& 
H^{abcd,+\lambda_\perp}_{\nu\rho\delta}(q,p,l) 
f^{-}_{\mu\lambda_\perp}(k)\,,
\eea
at ${\cal O}(\lambda)$, and at ${\cal O}(\lambda^2)$
\bea
V^{(2),1,abcd}_{\mu\nu\rho\delta}(k,q,p,l) &=& \frac12 H^{abcd,+\perp\perp}_{\nu\rho\delta}(q,p,l) \left[-\nm X f_\mu^{+-}(k) + kX_\perp X_{\perp\kappa} f_\mu^{-\kappa_\perp}(k)\right]\,,
\nn\\
  V^{(2),2,abcd}_{\mu\nu\rho\delta}(k,q,p,l) &=&  \frac12 H^{abcd,+\lambda_\perp-}_{\nu\rho\delta}(q,p,l) X_{\perp}^{\kappa} f_{\mu\kappa_\perp\lambda_\perp}(k) \,,
\nn\\
  V^{(2),3,abcd}_{\mu\nu\rho\delta}(k,q,p,l) &=& -H^{abcd,\lambda_\perp\perp\perp}_{\nu\rho\delta}(q,p,l)  X_{\perp}^{\kappa} f_{\mu\kappa_\perp\lambda_\perp}(k) \,,
\nn\\
  V^{(2),4,abcd}_{\mu\nu\rho\delta}(k,q,p,l) &=& -\frac12 H^{abcd,+-}_{\nu\rho\delta}(q,p,l) f^{+-}_{\mu}(k)\,,
\nn\\
  V^{(2),5,abcd}_{\mu\nu\rho\delta}(k,q,p,l) &=& H^{abcd,\kappa_\perp\lambda_\perp}_{\nu\rho\delta}(q,p,l) f_{\mu\kappa_\perp\lambda_\perp}(k)\,,
\nn\\
  V^{(2),6,abcd}_{\mu\nu\rho\delta}(k,q,p,l) &=& H^{abcd,+\lambda_\perp}_{\nu\rho\delta}(q,p,l) kX_\perp f^{-}_{\mu\lambda_\perp}(k)\,,
\eea
together with
\bea
H^{abcd,ABC}_{\nu\rho\delta}(q,p,l) &\equiv& f_\nu^{AB}(q)\frac{{\np}_\delta}{\np l}\bigg\{ - \left(P^C_\rho + P^C_\lambda(p+l)^{\lambda}\frac{{\np}_\rho}{\np q}\right)\mbox{tr}[t^b[t^a,t^ct^d]] 
\nn\\
  && {} + \frac{f_{\rho}^{+C}(p)}{\np p}\left(\frac12 f^{eac}f^{ebd}+\mbox{tr}[t^b[t^a,t^dt^c]]]\right)\bigg\} 
-\frac12 f^{eac}f^{ebd}\frac{f_{\nu\delta}^{AB}}{2}\frac{f_{\rho}^{+C}(p)}{\np p}  \,,\nn\\
  H^{abcd,A\perp\perp}_{\nu\rho\delta}(q,p,l) &\equiv& H^{abcd,A\lambda_\perp\sigma_\perp}_{\nu\rho\delta}(q,p,l) g^\perp_{\lambda\sigma} \,,\nn\\
H^{abcd,AB}_{\nu\rho\delta}(q,p,l) &\equiv&  
\frac{{\np}_\delta}{\np l}\Bigg[ f_\nu^{AB}(q){\np}_\rho 
\mbox{tr}\left[\frac{t^bt^ct^at^d}{\np p} + \frac{t^bt^ct^dt^a+t^bt^at^dt^c}{\np q} 
\right] 
+\frac12 f^{ebc}f^{ead} \frac{f_{\nu\rho}^{AB}}{2}\Bigg] \,.\quad\nn\\
\eea
The traces can be evaluated using
\bea
\label{eq:tr3}
\mbox{tr}[t^at^bt^c] &=& \frac14(if^{abc}+d^{abc}) \,,
\nn\\
 \mbox{tr}[t^b[t^a,t^ct^d]] &=& 
\frac14(f^{eab}f^{ecd}+if^{eac}d^{ebd}+if^{ead}d^{ebc})\,,
\eea
and \eqref{eq:tatb}. 
The additional factor $1/2$ in the terms involving $f_{\nu\delta}^{AB}$ and 
$f_{\nu\rho}^{AB}$ accounts for an overcounting that would occur when adding 
the terms where the collinear gluons are permuted.

\subsection{Ghost vertices}

\vskip0.2cm
\be
  \raisebox{-13mm}{\includegraphics[width=0.2\textwidth]{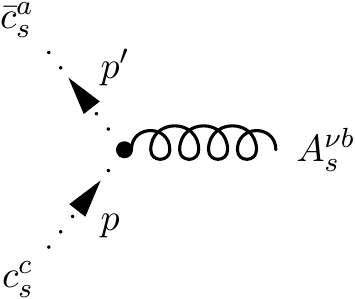}} \quad
  g_s f^{abc} \left\{ \begin{array}{ll}
     -p'_{\nu}  \;\;& {\cal O}(\lambda^0) \\
     0 & {\cal O}(\lambda) \\
     0 & {\cal O}(\lambda^2)
  \end{array}
  \right.
\ee

\be
  \raisebox{-13mm}{\includegraphics[width=0.2\textwidth]{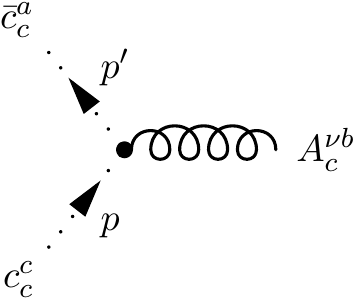}} \quad
  g_s f^{abc} \left\{ \begin{array}{ll}
     -p'_{\nu}  \;\;& {\cal O}(\lambda^0) \\
     0 & {\cal O}(\lambda) \\
     0 & {\cal O}(\lambda^2)
  \end{array}
  \right.
\ee

\be
  \raisebox{-13mm}{\includegraphics[width=0.2\textwidth]{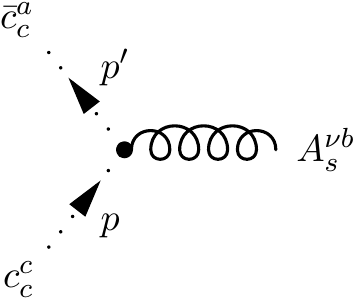}} \quad
  g_s f^{abc} \left\{ \begin{array}{ll}
     - (\np p)  {\nm}_\nu \;\;& {\cal O}(\lambda^0) \\
     0 & {\cal O}(\lambda) \\
     0 & {\cal O}(\lambda^2)
  \end{array}
  \right.
\ee

\be
  \raisebox{-17mm}{\includegraphics[width=0.2\textwidth]{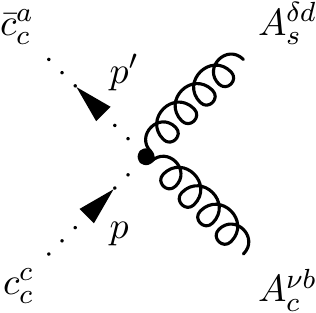}} \quad
  - i g_s^2f^{ade}f^{ebc} \left\{ \begin{array}{ll}
  \displaystyle   \frac12   {\nm}_\delta {\np}_\nu \;\;& {\cal O}(\lambda^0) \\[0.1cm]
     0 & {\cal O}(\lambda) \\
     0 & {\cal O}(\lambda^2)
  \end{array}
  \right.
\ee
We recall that the ghost Lagrangian contains only leading-power 
interactions.

\subsection{Collinear building blocks}

In addition to the Feynman rules derived from the SCET Lagrangian, we also 
give Feynman rules for insertions of the collinear building blocks 
${\cal A}_\perp^{\rho b}t^b = [W_c^\dag iD_\perp^\rho W_c]$ and 
$\chi_\beta=W_c^\dag \xi_\beta$.
\bea
  \raisebox{-4mm}{\includegraphics[width=0.16\textwidth]{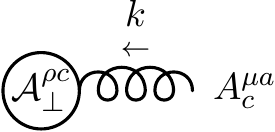}} &&
  g_s \delta^{ac}\frac{1}{\np k}f_\mu^{+\rho_\perp}(k) = g_s\delta^{ac} \left(\delta_{\perp\mu}^\rho - \frac{k_\perp^\rho{\np}_\mu }{\np k}\right) 
\\
  \raisebox{-14mm}{\includegraphics[width=0.12\textwidth]{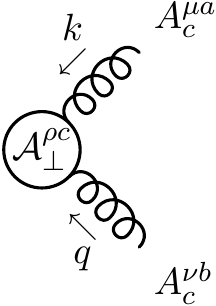}} &&
  ig_s^2f^{abc}\Bigg(\frac{{\np}_\mu{\np}_\nu (\np k k_\perp^\rho-\np q q_\perp^\rho)}{\np(k+q)\np k\np q}
  +\frac{{\np}_\mu \delta_{\perp\nu}^\rho}{\np k} - \frac{{\np}_\nu \delta_{\perp\mu}^\rho}{\np q} \Bigg) 
\\
  \raisebox{-2mm}{\includegraphics[width=0.14\textwidth]{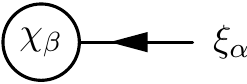}} &&
  \delta_{\alpha\beta}
\\
  \raisebox{-3mm}{\includegraphics[width=0.14\textwidth]{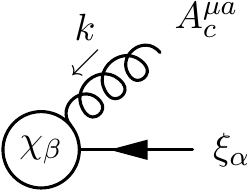}} &&
  \frac{g_s t^a {\np}_\mu}{\np k}\delta_{\alpha\beta}
\\
  \raisebox{-13mm}{\includegraphics[width=0.14\textwidth]{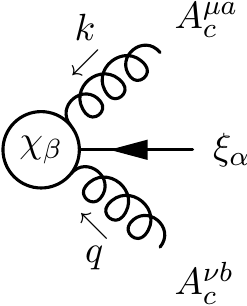}} &&
  g_s^2 \hat W^{ab}_{\mu\nu}(k,q)\delta_{\alpha\beta}
\eea
$\hat W^{ab}_{\mu\nu}(k,q)$ has been defined in Eq.~\eqref{eq:W2hat}.

\section{Soft master integral}
\label{sec:integrals}

The integral repeatedly referred to in Sec.~\ref{sec:soft} is 
given by 
\bea
I &=& \frac{-i\tilde\mu^{4-d}}{(2\pi)^d}\int d^dl \, 
\frac{1}{l^2+i\varepsilon}\frac{1}{A_1+B_1\nnm{1}l+i\varepsilon}
\frac{1}{A_2+B_2\nnm{2}l+i\varepsilon} 
\nn\\
  &=& -\frac{e^{\epsilon\gamma_E}\Gamma(\epsilon)}{16\pi^2}\frac{\pi}{\sin(\pi\epsilon)}\frac{2}{B_1B_2(\nnm{1}\nnm{2})}\left(\frac{B_1B_2(\nnm{1}\nnm{2})\mu^2}{2(A_1+i\varepsilon)(A_2+i\varepsilon)}\right)^\epsilon 
\nn\\
&=& -F_\epsilon \times 
\frac{1}{B_1^{1-\epsilon}(A_1+i\varepsilon)^\epsilon} \times 
\frac{1}{B_2^{1-\epsilon}(A_2+i\varepsilon)^\epsilon}
\nn\\
&=& -\frac{1}{16\pi^2}\frac{2}{B_1B_2(\nnm{1}\nnm{2})}\left(\frac{1}{\epsilon^2}+\frac{1}{\epsilon}L+\frac{L^2}{2}+\frac{\pi^2}{4}+{\cal O}(\epsilon)\right)\,,
\eea
with
\bea\label{eq:Fepsilon}
F_\epsilon &\equiv& \mu^{2\epsilon} \,
\frac{e^{\epsilon\gamma_E}\Gamma(\epsilon)}{16\pi^2}
\frac{\pi}{\sin(\pi\epsilon)}\left(\frac{\nnm{1}\nnm{2}}{2}
\right)^{-1+\epsilon} \,,
\nn\\
L &\equiv& \ln\left(\frac{B_1B_2(\nnm{1}\nnm{2})\mu^2}
{2(A_1+i\varepsilon)(A_2+i\varepsilon)}\right)\,.
\eea
The general version is
\bea\label{eq:master_scalar}
I(a_1,a_2,b_1,b_2) &=& 
\frac{-i\tilde\mu^{4-d}}{(2\pi)^d}\int d^dl \, \frac{1}{l^2+i\varepsilon}
\frac{(\nnm{1} l)^{b_1}}{(A_1+B_1\nnm{1}l+i\varepsilon)^{a_1}}
\frac{(\nnm{2} l)^{b_2}}{(A_2+B_2\nnm{2}l+i\varepsilon)^{a_2}} 
\nn\\
&=& \frac{1}{(A_1+i\varepsilon)^{a_1-b_1-1}B_1^{b_1}}
\frac{1}{(A_2+i\varepsilon)^{a_2-b_2-1}B_2^{b_2}} \,D(a_1,b_1,\epsilon) 
D(a_2,b_2,\epsilon) \times I\quad
\nn\\
&=& - F_\epsilon \times \frac{D(a_1,b_1,\epsilon)}
{(A_1+i\varepsilon)^{a_1-b_1-1+\epsilon}B_1^{b_1+1-\epsilon}}
\times \frac{D(a_2,b_2,\epsilon)}
{(A_2+i\varepsilon)^{a_2-b_2-1+\epsilon}B_2^{b_2+1-\epsilon}}\,,
\eea
where
\bea\label{eq:softcoeff}
D(a,b,\epsilon) &\equiv& \sum_{k=0}^{\min(b,a-1)}\left(-1\right)^{b-k}\,
\left(b\atop k\right)\frac{\Gamma(a-k+\epsilon-1)}
{\Gamma(a-k)\Gamma(\epsilon)} \nn\\ 
&=& (-1)^b  \frac{\Gamma (a-b+\epsilon-1)}{\Gamma (a) \Gamma (\epsilon-b)} \,. 
\eea

\section{Auxiliary functions entering the collinear anomalous dimension}
\label{app:coll}

\subsection{$B$-to-$B$ mixing}

Here we provide results for the coefficient functions entering the 
collinear anomalous dimensions $\gamma^{i}_{{\cal A}^\mu\partial^\nu\chi, {\cal A}^\rho\partial^\sigma\chi}(x,y)$
and $\gamma^{i}_{{\cal A}^\mu\partial^\nu\chi, \partial^\sigma({\cal A}^\rho\chi)}(x,y)$
discussed in Sec.\,\ref{sec:BB}.
The coefficients that are obtained from diagram $(c, i)$ in Fig.\,\ref{fig:qgg}
are given by
\bea
  M^{\mu\nu,\rho\sigma}(x,y) &\equiv& M_{\perp}^{\mu\nu,\rho\sigma}(x,y) - M_{-}^{\mu\nu,\rho\sigma}(x,y)\,,\nn\\
  \hat M^{\mu\nu,\rho\sigma}(x,y) &\equiv& \hat M_{\perp}^{\mu\nu,\rho\sigma}(x,y) + M_{-}^{\mu\nu,\rho\sigma}(x,y)\,,
\eea
where $M_{\perp}$ and $\hat M_{\perp}$ arise from contributions proportional 
to the $\perp$ component of the polarization vector $\epsilon$ of the 
external gluon, while $M_{-}$ captures the contributions obtained from 
$\nm\epsilon$.
\bea
  M_\perp^{\mu\nu,\rho\sigma}(x,y) &\equiv& 2a(x,y)\Bigg[\frac{2\bar x}{x}g_\perp^{\rho\sigma}g_\perp^{\mu\nu}-2g_\perp^{\mu\sigma}\gamma_\perp^\nu\gamma_\perp^\rho
  +g_\perp^{\rho\sigma}\gamma_\perp^\nu\gamma_\perp^\mu
  +\frac{2\bar xy}{x\bar y}g_\perp^{\mu\nu}\gamma_\perp^\rho\gamma_\perp^\sigma \nn\\
 && {} + \frac{2}{\bar y}g_\perp^{\mu\rho}\gamma_\perp^\nu\gamma_\perp^\sigma
  -\frac{2\bar x}{\bar y}g_\perp^{\nu\rho}\gamma_\perp^\mu\gamma_\perp^\sigma\Bigg]
 + 2e(x,y)\bar x g_\perp^{\nu\sigma}g_\perp^{\mu\rho} \nn\\
 && {} + 2a(x,y)^2\Bigg[4\frac{2x-1}{x}g_\perp^{\nu\sigma}g_\perp^{\mu\rho}-\frac{4\bar x}{x}\left(g_\perp^{\rho\sigma}g_\perp^{\mu\nu}+g_\perp^{\mu\sigma}g_\perp^{\nu\rho}\right)\nn\\
 && {} -2\left(g_\perp^{\nu\sigma}\gamma_\perp^\rho\gamma_\perp^\mu+g_\perp^{\rho\sigma}\gamma_\perp^\nu\gamma_\perp^\mu+g_\perp^{\nu\rho}\gamma_\perp^\sigma\gamma_\perp^\mu \right) \nn\\
 && {} +\frac{2y}{x}\left(g_\perp^{\nu\sigma}\gamma_\perp^\mu\gamma_\perp^\rho+g_\perp^{\mu\sigma}\gamma_\perp^\nu\gamma_\perp^\rho
       + g_\perp^{\mu\nu}\gamma_\perp^\sigma\gamma_\perp^\rho\right)\Bigg]\,,
\eea
\bea
  \hat M_\perp^{\mu\nu,\rho\sigma}(x,y) &\equiv& 2a(x,y)\Bigg[\frac{2\bar x}{x} g_\perp^{\rho\sigma}g_\perp^{\mu\nu}
  + \frac{4\bar x}{x} g_\perp^{\mu\sigma}g_\perp^{\nu\rho}
  + \frac{2(x-y)}{x}g_\perp^{\mu\sigma}\gamma_\perp^\nu\gamma_\perp^\rho \nn\\
  && {} + g_\perp^{\rho\sigma}\gamma_\perp^\nu\gamma_\perp^\mu -2g_\perp^{\mu\rho}\gamma_\perp^\nu\gamma_\perp^\sigma\Bigg] 
  - 2d(x,y)\bar x g_\perp^{\nu\sigma}g_\perp^{\mu\rho} \nn\\
  && +2c(x,y)\Bigg(-\frac{\bar x}{x}g_\perp^{\rho\sigma}g_\perp^{\mu\nu}-\frac{\bar x}{x}g_\perp^{\mu\sigma}g_\perp^{\nu\rho}
  +\frac{y}{2x}\left(g_\perp^{\mu\nu}\gamma_\perp^\sigma+g_\perp^{\mu\sigma}\gamma_\perp^\nu+g_\perp^{\nu\sigma}\gamma_\perp^\mu\right)\gamma_\perp^\rho\nn\\
  && {} -\frac12 \left(g_\perp^{\rho\nu}\gamma_\perp^\sigma+g_\perp^{\rho\sigma}\gamma_\perp^\nu+g_\perp^{\nu\sigma}\gamma_\perp^\rho\right)\gamma_\perp^\mu 
  + g_\perp^{\nu\sigma}g_\perp^{\mu\rho}\Bigg)\,,
\eea
\bea
   M_{-}^{\mu\nu,\rho\sigma}(x,y) &\equiv& \frac{4\bar x}{y} a(x,y)\left(-\frac{y+2x}{x}g_\perp^{\mu\nu}+\frac12\frac{y-2x}{\bar x}\gamma_\perp^\nu\gamma_\perp^\mu\right) g_\perp^{\rho\sigma}\,,
\eea
where we define
\bea
  a(x,y) &\equiv& \frac{\bar x}{2\bar y}\theta(x-y)+\frac{ x}{2 y}\theta(y-x)\,,\nn\\
  c(x,y) &\equiv& \frac{\bar x^2}{\bar y}\theta(x-y)+\frac{ x(\bar xy+y-x)}{ y^2}\theta(y-x)\,,\nn\\
  d(x,y) &\equiv& \frac{\bar x}{x\bar y}\theta(x-y)+\frac{ 1}{ y}\theta(y-x)\,,\nn\\
  e(x,y) &\equiv& -\frac{\bar x}{\bar y^2}\theta(x-y)+\frac{ x}{ y^2}\theta(y-x)\,.
\eea

The coefficients that are obtained from diagram $(c, ii)$ and part of $(b, iii)$ in Fig.\,\ref{fig:qgg} (see main text for details)
are given by
\bea
  N^{\mu\nu,\rho\sigma}(x,y) &\equiv& N_{\perp}^{\mu\nu,\rho\sigma}(x,y) - N_{-}^{\mu\nu,\rho\sigma}(x,y)\,,\nn\\
  \hat N^{\mu\nu,\rho\sigma}(x,y) &\equiv& \hat N_{\perp}^{\mu\nu,\rho\sigma}(x,y) + N_{-}^{\mu\nu,\rho\sigma}(x,y)\,,
\eea
with contributions from $\epsilon_\perp$ and $\nm\epsilon$ analogous to above,
\bea
  N_{\perp}^{\mu\nu,\rho\sigma}(x,y) &\equiv& 2a(x,\bar y)\Bigg[
  -\frac{2\bar x^2}{xy}\left(g_\perp^{\mu\rho}g_\perp^{\nu\sigma}+g_\perp^{\mu\sigma}g_\perp^{\nu\rho}\right)
  - \frac{\bar x}{y}\left(g_\perp^{\nu\sigma}\gamma_\perp^\rho+g_\perp^{\nu\rho}\gamma_\perp^\sigma+g_\perp^{\rho\sigma}\gamma_\perp^\nu\right) \gamma_\perp^\mu
   \nn\\
  && {} + 2g_\perp^{\mu\sigma}\gamma_\perp^\nu\gamma_\perp^\rho
  +\frac{\bar x}{x}\left(g_\perp^{\nu\sigma}\gamma_\perp^\mu\gamma_\perp^\rho+g_\perp^{\mu\sigma}\gamma_\perp^\nu\gamma_\perp^\rho
  +g_\perp^{\mu\nu}\gamma_\perp^\rho\gamma_\perp^\sigma\right) 
  -\frac{2x}{\bar y}g_\perp^{\nu\rho}\gamma_\perp^\mu\gamma_\perp^\sigma \nn\\
  && {}
  +\frac{1}{\bar y}\gamma^\rho\gamma^\nu\gamma^\mu\gamma^\sigma
  -\frac{2\bar x(\bar x-y)}{xy}g_\perp^{\rho\sigma}g_\perp^{\mu\nu}\Bigg] 
  +\frac{\theta(\bar y-x)}{2\bar y^2 }\Bigg[\frac{2x}{y}(\bar x+y)g_\perp^{\nu\sigma}\gamma_\perp^\rho\gamma_\perp^\mu \nn\\
  && {} +(\bar x-y)\Bigg(\frac{2x}{y}\left(g_\perp^{\nu\rho}\gamma_\perp^\sigma + g_\perp^{\rho\sigma}\gamma_\perp^\nu\right)\gamma_\perp^\mu 
  + \frac{4\bar x}{y}\left(g_\perp^{\mu\rho}g_\perp^{\nu\sigma}+g_\perp^{\mu\sigma}g_\perp^{\nu\rho}\right) \nn\\
  && {}
  -2\left(\gamma_\perp^\mu g_\perp^{\nu\sigma}+\gamma_\perp^\nu g_\perp^{\mu\sigma}\right)\gamma_\perp^\rho
  + \frac{4\bar x}{y}g_\perp^{\mu\nu}g_\perp^{\rho\sigma}  -2g_\perp^{\mu\nu}\gamma_\perp^\sigma\gamma_\perp^\rho\Bigg)\Bigg]\,,
\eea
\bea
  \hat N_{\perp}^{\mu\nu,\rho\sigma}(x,y) &\equiv& 2a(x,\bar y)\Bigg[\frac{\bar x-y-xy}{xy}\left(2g_\perp^{\mu\sigma}g_\perp^{\nu\rho}
  -xg_\perp^{\nu\rho}\gamma_\perp^\mu\gamma_\perp^\sigma-yg_\perp^{\mu\sigma}\gamma_\perp^\nu\gamma_\perp^\rho\right) \nn\\
  && {} - \frac{\bar x\bar y}{x}g_\perp^{\mu\nu}\gamma_\perp^\rho\gamma_\perp^\sigma
  +\frac{\bar x(1+y)}{xy}\Big( g_\perp^{\nu\sigma}\left(2g_\perp^{\mu\rho}-(x+y)\gamma_\perp^\mu\gamma_\perp^\rho\right) \nn\\
  && {} + g_\perp^{\rho\sigma}\left(2(\bar x-y)g_\perp^{\mu\nu}+x\gamma_\perp^\nu\gamma_\perp^\mu\right)\Big)
  + \gamma_\perp^\rho\gamma_\perp^\nu\gamma_\perp^\sigma\gamma_\perp^\mu\Bigg]\nn\\
  && -\frac{1}{y\bar y}\theta(\bar y-x)\Bigg[x\left((\bar x-y)g_\perp^{\nu\rho}\gamma_\perp^\sigma
  +\left(\bar x-y+\frac{2y}{x}\right)g_\perp^{\nu\sigma}\gamma_\perp^\rho\right)\gamma_\perp^\mu
  \nn\\
  && {} +(\bar x-y)  \Bigg(g_\perp^{\mu\sigma}\left(2\bar xg_\perp^{\nu\rho}-y\gamma_\perp^\nu\gamma_\perp^\rho\right)
  + g_\perp^{\nu\sigma}\left(2\bar xg_\perp^{\mu\rho}-y\gamma_\perp^\mu\gamma_\perp^\rho\right)
  + y g_\perp^{\mu\nu}\gamma_\perp^\rho\gamma_\perp^\sigma \nn\\
  && {} + g_\perp^{\rho\sigma}\left(2(\bar x-y)g_\perp^{\mu\nu}+x\gamma_\perp^\nu\gamma_\perp^\mu\right)\Bigg)\Bigg]\,,
\eea
\bea
   N_{-}^{\mu\nu,\rho\sigma}(x,y) &\equiv& -\frac{4\bar x}{y} a(x,\bar y) \left(\frac{2\bar y}{x}g_\perp^{\mu\nu}-\gamma_\perp^\mu\gamma_\perp^\nu\right)g_\perp^{\rho\sigma}\,.
\eea

\subsection{$B$-to-$C$ mixing}

Here we report the explicit results for the contributions to the anomalous 
dimensions discussed in Sec.\,\ref{sec:BC}.

\subsubsection{$J^{B2}_{{\cal A}\partial\chi}(x)\to J^{C2}_{{\cal AA}\chi}(y_1,y_2)$}\label{app:gamma_Adelchi_AAchi}

We list the non-zero results for the kernels $I_{ade}^{\mu\nu\sigma\lambda}$ 
that enter in the anomalous dimension \eqref{eq:gamma_Adelchi_AAchi} from the 
various diagrams shown in Fig.~\ref{fig:qgg} and Fig.~\ref{fig:qgg_1PR} 
(plus the ones with interchanged external gluon lines, if applicable).
Notation $y=y_1+y_2$, $\bar y_k=1-y_k$, $y_1+y_2+y_3=1$. We leave the Dirac 
indices implicit.

\bea
  \lefteqn{ I_{ade}^{\mu\nu\sigma\lambda}(x,y_1,y_2)|_{(b,i)_F} } \nn\\
  &=& g_\perp^{\mu\sigma} \theta(x-y_1)\frac{\bar x(\bar x-y_2)}{(x-y_1)(y_2+y_3)}\left(\frac{\gamma_\perp^\nu\gamma_\perp^\lambda}{\bar x}+\frac{\gamma_\perp^\lambda\gamma_\perp^\nu}{\bar x-y_2}\right) \nn\\
  && \left(\theta(x-\bar y_2)\frac{\bar x}{y_2}+ \theta(\bar y_2-x)\frac{x-y_1}{y_3}\right) if^{bda} t^et^b + (y_1 d \sigma \leftrightarrow y_2 e \lambda)\,,
\eea
\bea
    \lefteqn{ I_{ade}^{\mu\nu\sigma\lambda}(x,y_1,y_2)|_{(b,i)_B} } \nn\\
  &=&  -\frac12 g_\perp^{\mu\sigma} \theta(x-y_1) \frac{\bar x}{(x-y_1)\bar y_1}\left(\frac{\gamma_\perp^\nu\gamma_\perp^\lambda}{\bar x}(y_1+2y_2-x)-4g_\perp^{\nu\lambda}\right) \nn\\
  && \left(\theta(x-\bar y_3)\frac{\bar x}{y_3}+ \theta(\bar y_3-x)\frac{x-y_1}{y_2}\right) f^{cbe}f^{bda} t^c + (y_1 d \sigma \leftrightarrow y_2 e \lambda)\,,
\eea
\bea
    \lefteqn{ I_{ade}^{\mu\nu\sigma\lambda}(x,y_1,y_2)|_{(b,ii)_B} } \nn\\
  &=& -\frac12\theta(y-x) \frac{1}{y(y-x)}\Big(2(y_2-y_1-x)\left(g_\perp^{\mu\sigma}g_\perp^{\nu\lambda}-g_\perp^{\mu\nu}g_\perp^{\sigma\lambda}\frac{y_2}{x}\right) \nn\\
  && {} -g_\perp^{\nu\sigma}g_\perp^{\mu\lambda}(3y_2+x-y_1)\Big) 
 \left(\theta(x-y_2)\frac{\bar x-y_3}{y_1}+ \theta(y_2-x)\frac{x}{y_2}\right) f^{ace}f^{bcd} t^b \nn\\
&& {} + (y_1 d \sigma \leftrightarrow y_2 e \lambda)\,,
\eea
\bea
    \lefteqn{ I_{ade}^{\mu\nu\sigma\lambda}(x,y_1,y_2)|_{(b,iii)_F} } \nn\\
  &=&  \frac12 \bar x(\bar x-y_2)\left(\frac{\gamma_\perp^\nu\gamma_\perp^\lambda}{\bar x}+\frac{\gamma_\perp^\lambda\gamma_\perp^\nu}{\bar x-y_2}\right)  t^e 
\left(\frac{\gamma_\perp^\mu\gamma_\perp^\sigma t^at^d}{y_1+y_3}+\frac{\gamma_\perp^\sigma\gamma_\perp^\mu t^dt^a}{y_3-x}\right) \nn \\
  &&
  \left(\theta(x-\bar y_2)\frac{\bar x}{y_2}+ \theta(\bar y_2-x)\frac{x}{\bar y_2}\right) + (y_1 d \sigma \leftrightarrow y_2 e \lambda) \,,
\eea
\bea
    \lefteqn{ I_{ade}^{\mu\nu\sigma\lambda}(x,y_1,y_2)|_{(b,iii)_B} } \nn\\
  &=&  \frac{if^{abe}}{2}\bar x \left(\theta(x- y_2)\frac{\bar x}{\bar y_2}+ \theta( y_2-x)\frac{x}{ y_2}\right)\nn\\
  && \Bigg\{ \left(2g_\perp^{\nu\lambda}\gamma_\perp^\mu\gamma_\perp^\sigma-2\frac{y_2}{x}g_\perp^{\mu\nu}\gamma_\perp^\lambda\gamma_\perp^\sigma-\frac{1+y_2}{\bar x}g_\perp^{\mu\lambda}\gamma_\perp^\nu\gamma_\perp^\sigma\right)\frac{t^bt^d}{y_1+y_3} \nn\\
  && {} + \left(2g_\perp^{\nu\lambda}\gamma_\perp^\sigma\gamma_\perp^\mu-2\frac{y_2}{x}g_\perp^{\mu\nu}\gamma_\perp^\sigma\gamma_\perp^\lambda-g_\perp^{\mu\lambda}\gamma_\perp^\sigma\gamma_\perp^\nu\right)\frac{t^dt^b}{y_2+y_3-x}\Bigg\}+ (y_1 d \sigma \leftrightarrow y_2 e \lambda)\,,\nn\\
\eea
\bea
    \lefteqn{ I_{ade}^{\mu\nu\sigma\lambda}(x,y_1,y_2)|_{(c,i)_F} } \nn\\
  &=&  \bar x (\bar x-y_2)\Bigg\{ \frac14 \Bigg[ - \theta(\bar x-y_2)\theta(\bar y_1-\bar x)\frac{\bar x^2\bar y_2+ x^2\bar y_1-\bar y_2\bar y_1}{\bar y_2 y_3\bar y_1} \nn\\
  && {} + \theta(y_2-\bar x)\theta(\bar x)\frac{\bar x^2}{y_2\bar y_1} + \theta(x)\theta(\bar x-\bar y_1)\frac{x^2}{\bar y_2 y_1}\Bigg] 
  \Bigg[  \left(\frac{\gamma_\perp^\nu\gamma_\perp^\lambda}{\bar x}+\frac{\gamma_\perp^\lambda\gamma_\perp^\nu}{\bar x-y_2}\right)\Bigg( \frac{\gamma_\perp^\sigma\gamma_\perp^\mu}{\bar x-y_2}\nn\\
  && {} +2g_\perp^{\mu\sigma}\frac{\bar x-y_2-x}{x(\bar x-y_2)} -\frac{y_1}{x(\bar x-y_2)}\gamma_\perp^\mu\gamma_\perp^\sigma\Bigg) 
  + \left(\frac{\gamma_\perp^\sigma\gamma_\perp^\lambda}{\bar x}+\frac{\gamma_\perp^\lambda\gamma_\perp^\sigma}{\bar x-y_2}\right)\Bigg(\frac{\gamma_\perp^\nu\gamma_\perp^\mu}{\bar x-y_2}+\frac{2g_\perp^{\mu\nu}}{x}\Bigg) \nn\\
  && + \left(g_\perp^{\nu\sigma}\gamma_\perp^\lambda\gamma_\perp^\mu- \frac{y_1}{x}\gamma_\perp^\lambda\gamma_\perp^\sigma g_\perp^{\mu\nu}\right)\frac{2}{(\bar x-y_2)^2}
  + \left(\frac{\gamma_\perp^\mu\gamma_\perp^\lambda}{\bar x}+\frac{\gamma_\perp^\lambda\gamma_\perp^\mu}{\bar x-y_2}\right)
    \left(\frac{2g_\perp^{\nu\sigma}}{x}-\frac{y_1\gamma_\perp^\nu\gamma_\perp^\sigma}{x(\bar x-y_2)}\right)\Bigg] \nn\\
  && +\frac12 \Bigg[ \theta(\bar x-y_2)\theta(\bar y_1-\bar x)\frac{x \bar y_1 - \bar x \bar y_2}{\bar y_2 y_3\bar y_1}  
   + \theta(y_2-\bar x)\theta(\bar x)\frac{\bar x}{y_2\bar y_1} - \theta(x)\theta(\bar x-\bar y_1)\frac{x}{\bar y_2 y_1}\Bigg] \nn\\
  &&  g_\perp^{\mu\sigma} \left(\frac{\gamma_\perp^\nu\gamma_\perp^\lambda}{\bar x}+\frac{\gamma_\perp^\lambda\gamma_\perp^\nu}{\bar x-y_2}\right) \Bigg\} \,(-if^{abd})t^et^b  + (y_1 d \sigma \leftrightarrow y_2 e \lambda)\,,
\eea
\bea
    \lefteqn{ I_{ade}^{\mu\nu\sigma\lambda}(x,y_1,y_2)|_{(c,i)_B} } \nn\\
  &=&  \bar x \Bigg\{ \Bigg[ - \theta(x-y_2)\theta(\bar y_3-x)\frac{x^2\bar y_2+\bar x^2\bar y_3-\bar y_2\bar y_3}{\bar y_2 y_1\bar y_3} \nn\\
  && {} + \theta(y_2-x)\theta(x)\frac{x^2}{y_2\bar y_3} + \theta(\bar x)\theta(x-\bar y_3)\frac{\bar x^2}{\bar y_2 y_3}\Bigg] 
  \Bigg[\frac12 g_\perp^{\nu\lambda}\left(\frac{\gamma_\perp^\sigma\gamma_\perp^\mu}{\bar x}+\frac{2g_\perp^{\mu\sigma}}{x\bar x}(\bar x-x)\right) \nn\\
  && {} + \frac12 g_\perp^{\nu\sigma}\left(\frac{\gamma_\perp^\lambda\gamma_\perp^\mu}{\bar x}+\frac{g_\perp^{\mu\lambda}(2\bar x-x)}{x\bar x}\right) 
  + \frac12 g_\perp^{\lambda\sigma}\left(\frac{\gamma_\perp^\nu\gamma_\perp^\mu}{\bar x}+\frac{2g_\perp^{\mu\nu}(\bar x+y_2)}{x\bar x}\right) 
  +\frac{x-y_1}{2x\bar x}g_\perp^{\mu\lambda}\gamma_\perp^\nu\gamma_\perp^\sigma \nn\\
  && -\frac{y_1}{2x\bar x}\left(g_\perp^{\nu\lambda}\gamma_\perp^\mu\gamma_\perp^\sigma+g_\perp^{\mu\nu}\gamma_\perp^\lambda\gamma_\perp^\sigma\right)
 -\frac{y_2}{2x\bar x}\left(g_\perp^{\mu\nu}\gamma_\perp^\sigma\gamma_\perp^\lambda+g_\perp^{\nu\sigma}\gamma_\perp^\mu\gamma_\perp^\lambda+g_\perp^{\mu\sigma}\gamma_\perp^\nu\gamma_\perp^\lambda\right)  \Bigg] \nn\\
  && +\frac12 \Bigg[ \theta(x-y_2)\theta(\bar y_3-x)\frac{\bar x \bar y_3 - x \bar y_2}{\bar y_2 y_1\bar y_3}  
   + \theta(y_2-x)\theta(x)\frac{x}{y_2\bar y_3} - \theta(\bar x)\theta(x-\bar y_3)\frac{\bar x}{\bar y_2 y_3}\Bigg] \nn\\
  &&  \Bigg[ 
  -2g_\perp^{\nu\lambda}g_\perp^{\mu\sigma}+\frac{2y_2}{x}g_\perp^{\mu\nu}g_\perp^{\lambda\sigma}
  -g_\perp^{\mu\lambda}g_\perp^{\nu\sigma}-g_\perp^{\mu\lambda}\gamma_\perp^\nu\gamma_\perp^\sigma\frac{x-y_1}{\bar x}
  \Bigg] \Bigg\} f^{abe}f^{bcd} t^c + (y_1 d \sigma \leftrightarrow y_2 e \lambda)\,,\nn\\
\eea
\bea
    \lefteqn{ I_{ade}^{\mu\nu\sigma\lambda}(x,y_1,y_2)|_{(c,i)_V} } \nn\\
  &=& \frac12 \bar x \Bigg[\left(f^{cab}f^{cde}-f^{cae}f^{cbd}\right)\frac{g_\perp^{\mu\sigma}\gamma_\perp^\nu\gamma_\perp^\lambda}{\bar x} \nn\\
  && {} + \left(f^{cab}f^{ced}-f^{cad}f^{cbe}\right)\frac{g_\perp^{\mu\lambda}\gamma_\perp^\nu\gamma_\perp^\sigma}{\bar x} 
  + \left(f^{cad}f^{cbe}+f^{cae}f^{cbd}\right)g_\perp^{\lambda\sigma}\left(\frac{2g_\perp^{\mu\nu}}{x}+\frac{\gamma_\perp^\nu\gamma_\perp^\mu}{\bar x}\right)\Bigg]  t^b  \nn\\
  && {}  \left(\theta(x-y)\frac{\bar x}{\bar y}+\theta(y-x)\frac{x}{y}\right)\,,
\eea
\bea
    \lefteqn{ I_{ade}^{\mu\nu\sigma\lambda}(x,y_1,y_2)|_{(c,ii)_F} } \nn\\
  &=&  -\frac18 \bar x(\bar x-y_2)(y_3-x) \Bigg\{ \Bigg[ - \theta(x-y_3)\theta(\bar y_2-x)\nn\\
  && {}  \frac{x^2\bar y_3+\bar x^2\bar y_2-\bar y_3\bar y_2}{\bar y_3 y_1\bar y_2} 
   + \theta(y_3-x)\theta(x)\frac{x^2}{y_3\bar y_2} + \theta(\bar x)\theta(x-\bar y_2)\frac{\bar x^2}{\bar y_3 y_2}\Bigg] \nn\\
  && {}  \Bigg[ \frac{2}{x}\left(\frac{\gamma_\perp^\nu\gamma_\perp^\lambda}{\bar x}+\frac{\gamma_\perp^\lambda\gamma_\perp^\nu}{\bar x-y_2}\right)
  \left(\frac{\gamma_\perp^\mu\gamma_\perp^\sigma}{\bar x-y_2} + \frac{\gamma_\perp^\sigma\gamma_\perp^\mu y_3}{(y_3-x)^2}\right) \nn\\
  && + 2\left(\frac{\gamma_\perp^\lambda\gamma_\perp^\sigma}{(\bar x-y_2)^2} + \frac{\gamma_\perp^\sigma\gamma_\perp^\lambda}{\bar x (y_3-x)}\right)
  \left(\frac{2g_\perp^{\mu\nu}}{x}+\frac{\gamma_\perp^\nu\gamma_\perp^\mu}{y_3-x}\right) \nn\\
  && {} + \left(\frac{\gamma_\perp^\rho\gamma_\perp^\lambda}{\bar x}+\frac{\gamma_\perp^\lambda\gamma_\perp^\rho}{\bar x-y_2}\right) 
  \left(\frac{\gamma_\perp^\nu\gamma_\perp^\sigma}{\bar x-y_2} + \frac{\gamma_\perp^\sigma\gamma_\perp^\nu }{y_3-x}\right)
  \left(\frac{2g_\perp^{\mu\rho}}{x}+\frac{\gamma_\perp^\rho\gamma_\perp^\mu}{y_3-x}\right) \Bigg]  \Bigg\}\,  t^et^dt^a\nn\\
&& {}  + (y_1 d \sigma \leftrightarrow y_2 e \lambda)\,,
\eea
\bea
I_{ade}^{\mu\nu\sigma\lambda}(x,y_1,y_2)|_{(c,ii)_V} 
  &=& \frac12 \bar x(y_3-x)   
  \left(\frac{\gamma_\perp^\sigma\gamma_\perp^\lambda t^dt^e}{\bar x-y_1}+\frac{\gamma_\perp^\lambda\gamma_\perp^\sigma t^et^d}{\bar x-y_2}\right)
  \left(\frac{2g_\perp^{\mu\nu}}{x}+\frac{\gamma_\perp^\nu\gamma_\perp^\mu}{y_3-x}\right) t^a  \nn\\
  && 
\left(\theta(x-y_3)\frac{\bar x}{\bar y_3}+\theta(y_3-x)\frac{x}{y_3}\right)\,,
\eea
\bea
     I_{ade}^{\mu\nu\sigma\lambda}(x,y_1,y_2)|_{(d,iii)} 
  &=&  \frac12\theta(y-x)\frac{y-x}{y^3} f^{cba}f^{dec} t^b g_\perp^{\mu\nu} \, (y_1-y_2)  g_\perp^{\lambda\sigma}\,,
\eea
\bea
    I_{ade}^{\mu\nu\sigma\lambda}(x,y_1,y_2)|_{(d,iv)} 
  &=&   \frac12\frac{x\bar x}{\bar x-y}\gamma_\perp^\nu\gamma_\perp^\mu if^{deb} t^bt^a \frac{y_1-y_2}{y}g_\perp^{\lambda\sigma}\,,
\eea
\bea
    \lefteqn{ I_{ade}^{\mu\nu\sigma\lambda}(x,y_1,y_2)|_{(d,v)} } \nn\\
  &=& \bar x \, \Bigg\{  \frac{\gamma_\perp^\nu\gamma_\perp^\mu}{2\bar x y^2}\Bigg(\theta(x-y)\frac{\bar x(x-y+xy)}{\bar y} \nn\\
  && {}  +\theta(y-x)\frac{x(2x-2y+xy)}{y}\Bigg) + g_\perp^{\mu\nu}\Bigg(\frac{(x+y+xy)\bar x}{xy^2\bar y}\theta(x-y) \nn\\
  && +\frac{x(3-4x)+y(2x\bar x+1)}{2\bar xy^3}\theta(y-x)\Bigg)\Bigg\} f^{bac}f^{deb} t^c \, (y_1-y_2) g_\perp^{\lambda\sigma}\,,
\eea
\bea
  \lefteqn{ I_{ade}^{\mu\nu\sigma\lambda}(x,y_1,y_2)|_{(d,vi)} } \nn\\
  &=& \frac12 \left(\frac{2}{x}g_\perp^{\mu\nu}+\frac{1}{\bar x-y}\gamma_\perp^\nu\gamma_\perp^\mu\right) \nn\\
  && \Bigg(\theta(x-\bar y)\frac{\bar x^2}{y^3}(2\bar x-(2+x) y)  +\theta(\bar y-x)\frac{\bar xx(\bar x-y-xy)}{\bar y y^2}\Bigg) if^{deb} t^bt^a\, (y_1-y_2) g_\perp^{\lambda\sigma} \,,\nn\\
\eea
\bea
I_{ade}^{\mu\nu\sigma\lambda}(x,y_1,y_2)|_{(d,vii)} 
  &=&  \frac12 x\bar x^2\left(\frac{2}{x}g_\perp^{\mu\nu}+\frac{1}{\bar x}\gamma_\perp^\nu\gamma_\perp^\mu\right)  (-if^{deb}) t^at^b\, \frac{y_1-y_2}{y^2} g_\perp^{\lambda\sigma} \,, \qquad
\eea
\bea
  \lefteqn{ I_{ade}^{\mu\nu\sigma\lambda}(x,y_1,y_2)|_{(d,viii)} } \nn\\
  &=&  -\frac12 x\bar x^2\left(\frac{2}{x}g_\perp^{\mu\nu}+\frac{1}{\bar x}\gamma_\perp^\nu\gamma_\perp^\mu\right)    
 t^a\,\left(\frac{\gamma_\perp^\sigma\gamma_\perp^\lambda t^dt^e}{y_3+y_2}+\frac{\gamma_\perp^\lambda\gamma_\perp^\sigma t^et^d}{y_3+y_1}\right) \,.
\eea

\subsubsection{$J^{B2}_{{\cal A}\partial\chi}(x)\to J^{C2}_{\chi\bar\chi\chi}(y_1,y_2)$}\label{app:gamma_Adelchi_chichibarchi}

We list the non-zero results for the kernels $I_{aijkl}^{\mu\nu\alpha\beta\gamma\delta}(x,y_1,y_2)$ that enter in the anomalous dimension
\eqref{eq:gamma_Adelchi_chichibarchi} from the various diagrams shown in Fig.\,\ref{fig:qqbarq}
and for the 1PR diagrams corresponding to all but the last diagrams in Fig.\,\ref{fig:qgg_1PR}, with the three-gluon vertex replaced by
a fermion-fermion-gluon vertex, as discussed in the main text.
We assume that the fermion attached to the last building block has a different flavour from the first two, i.e. we do \emph{not}
add diagrams with permutated external legs here. Notation $\bar y_k=1-y_k$, $y_1+y_2+y_3=1$.

\bea
    \lefteqn{ I_{aijkl}^{\mu\nu\alpha\beta\gamma\delta}(x,y_1,y_2)|_{(1)} } \nn\\
  &=& -\frac12\theta(\bar y_3-x)\left(\theta(x-y_2)\frac{\bar y_3-x}{y_1} + \theta(y_2-x)\frac{x}{y_2}\right)
 \frac{x-y_2}{\bar y_3(\bar y_3-x)} \delta^{\alpha\delta} t^b_{il}\nn\\
  && {} (t^at^b)_{kj} \left[\left(\frac{2g_\perp^{\mu\nu}}{x}+\frac{\gamma_\perp^\mu\gamma_\perp^\nu}{y_2-x}\right)\nps\right]^{\beta\gamma} 
    - (y_1,\beta,j\leftrightarrow y_3,\delta,l)\,,
\eea
\bea
    \lefteqn{ I_{aijkl}^{\mu\nu\alpha\beta\gamma\delta}(x,y_1,y_2)|_{(2)} } \nn\\
  &=& \frac12\theta(\bar y_3-x)\left(\theta(x-y_1)\frac{\bar y_3-x}{y_2} + \theta(y_1-x)\frac{x}{y_1}\right)
 \frac{x-y_1}{\bar y_3(\bar y_3-x)} \delta^{\alpha\delta} t^b_{il}\nn\\
  && {} (t^bt^a)_{kj} \left[\left(\frac{2g_\perp^{\mu\nu}}{x}+\frac{\gamma_\perp^\nu\gamma_\perp^\mu}{y_1-x}\right)\nps\right]^{\beta\gamma} 
    - (y_1,\beta,j\leftrightarrow y_3,\delta,l)\,,
\eea
\bea
    \lefteqn{ I_{aijkl}^{\mu\nu\alpha\beta\gamma\delta}(x,y_1,y_2)|_{(4)} } \nn\\
  &=& \frac{1}{8x} t^b_{il}(t^bt^a)_{kj} \Bigg\{ \frac{y_1+x}{y_1-x}[\gamma_\perp^\nu\gamma_\perp^\rho ]^{\alpha\delta} [\gamma_\perp^\rho \gamma_\perp^\mu \nps ]^{\beta\gamma}\nn\\
  && {} + [\gamma_\perp^\mu\gamma_\perp^\rho ]^{\alpha\delta} [ \gamma_\perp^\rho \gamma_\perp^\nu \nps ]^{\beta\gamma} 
  + g_\perp^{\mu\nu}[\gamma_\perp^\kappa\gamma_\perp^\rho ]^{\alpha\delta}  [ \gamma_\perp^\rho \gamma_\perp^\kappa \nps ]^{\beta\gamma} \Bigg\}\nn\\
  &&  \Bigg\{- \theta(x-y_1)\theta(\bar y_3-x)\frac{x^2\bar y_1+\bar x^2\bar y_3-\bar y_1\bar y_3}{\bar y_1 y_2\bar y_3} \nn\\
  && {} + \theta(y_1-x)\theta(x)\frac{x^2}{y_1\bar y_3} + \theta(\bar x)\theta(x-\bar y_3)\frac{\bar x^2}{\bar y_1 y_3}\Bigg\} 
   - (y_1,\beta,j\leftrightarrow y_3,\delta,l)\,,
\eea
\bea
    \lefteqn{ I_{aijkl}^{\mu\nu\alpha\beta\gamma\delta}(x,y_1,y_2)|_{(5)} } \nn\\
  &=& -\frac{1}{8x} t^b_{il}(t^at^b)_{kj} \Bigg\{ \frac{y_2+x}{y_2-x}[\gamma_\perp^\nu\gamma_\perp^\rho ]^{\alpha\delta} [ \gamma_\perp^\mu\gamma_\perp^\rho \nps ]^{\beta\gamma}\nn\\
  && {} + [\gamma_\perp^\mu\gamma_\perp^\rho ]^{\alpha\delta} [  \gamma_\perp^\nu\gamma_\perp^\rho \nps ]^{\beta\gamma} 
  + g_\perp^{\mu\nu}[\gamma_\perp^\kappa\gamma_\perp^\rho ]^{\alpha\delta}  [  \gamma_\perp^\kappa\gamma_\perp^\rho \nps ]^{\beta\gamma} \Bigg\}\nn\\
  &&  \Bigg\{- \theta(x-y_2)\theta(\bar y_3-x)\frac{x^2\bar y_2+\bar x^2\bar y_3-\bar y_2\bar y_3}{\bar y_2 y_1\bar y_3} \nn\\
  && {} + \theta(y_2-x)\theta(x)\frac{x^2}{y_2\bar y_3} + \theta(\bar x)\theta(x-\bar y_3)\frac{\bar x^2}{\bar y_2 y_3}\Bigg\} 
   - (y_1,\beta,j\leftrightarrow y_3,\delta,l)\,,
\eea
\bea
    \lefteqn{ I_{aijkl}^{\mu\nu\alpha\beta\gamma\delta}(x,y_1,y_2)|_{(6)} } \nn\\
  &=& -\frac{1}{4}t^b_{il} 
  \left(\theta(x-\bar y_3)\frac{\bar x}{y_3}+\theta(\bar y_3-x)\frac{x}{\bar y_3}\right)[\gamma_\perp^\nu\gamma_\perp^\rho]^{\alpha\delta} \nn\\
  && {}  \left[ \left( \frac{(t^at^b)_{kj}}{x-y_2}  \gamma_\perp^\mu \gamma_\perp^\rho 
  + \frac{(t^bt^a)_{kj}}{y_1-x}   \gamma_\perp^\rho \gamma_\perp^\mu \right) \nps \right]^{\beta\gamma}
   - (y_1,\beta,j\leftrightarrow y_3,\delta,l)\,,
\eea
\bea
     I_{aijkl}^{\mu\nu\alpha\beta\gamma\delta}(x,y_1,y_2)|_{(d,iii)} 
  &=&  \frac{i}{2}\theta(\bar y_3-x)\frac{\bar y_3-x}{\bar y_3^3} f^{cba} t^c_{kj} t^b_{il} g_\perp^{\mu\nu}  \delta^{\alpha\delta} \nps^{\beta\gamma} \nn\\
  && - (y_1,\beta,j\leftrightarrow y_3,\delta,l)\,,
\eea
\bea
    I_{aijkl}^{\mu\nu\alpha\beta\gamma\delta}(x,y_1,y_2)|_{(d,iv)} 
  &=&   -\frac12\frac{x\bar x}{\bar x-\bar y_3}(\gamma_\perp^\nu\gamma_\perp^\mu)^{\alpha\delta} t^b_{kj} (t^bt^a)_{il} \frac{1}{\bar y_3}\nps^{\beta\gamma}\nn\\
  && - (y_1,\beta,j\leftrightarrow y_3,\delta,l)\,,
\eea
\bea
    \lefteqn{ I_{aijkl}^{\mu\nu\alpha\beta\gamma\delta}(x,y_1,y_2)|_{(d,v)} } \nn\\
  &=& i\bar x \Bigg\{  \frac{(\gamma_\perp^\nu\gamma_\perp^\mu)^{\alpha\delta}}{2\bar x \bar y_3^2}\Bigg(\theta(x-\bar y_3)\frac{\bar x(x-\bar y_3+x\bar y_3)}{y_3} \nn\\
  && {}  +\theta(\bar y_3-x)\frac{x(2x-2\bar y_3+x\bar y_3)}{\bar y_3}\Bigg) + \delta^{\alpha\delta} g_\perp^{\mu\nu}\Bigg(\frac{(x+\bar y_3+x\bar y_3)\bar x}{x\bar y_3^2 y_3}\theta(x-\bar y_3) \nn\\
  && +\frac{x(3-4x)+\bar y_3(2x\bar x+1)}{2\bar x\bar y_3^3}\theta(\bar y_3-x)\Bigg)\Bigg\} f^{bac}t^b_{kj} t^c_{il}  \nps^{\beta\gamma} - (y_1,\beta,j\leftrightarrow y_3,\delta,l)\,, \nn\\
\eea
\bea
  \lefteqn{ I_{aijkl}^{\mu\nu\alpha\beta\gamma\delta}(x,y_1,y_2)|_{(d,vi)} } \nn\\
  &=& - \frac12\left(\frac{2}{x}\delta^{\alpha\delta}g_\perp^{\mu\nu}+\frac{1}{\bar x-\bar y_3}(\gamma_\perp^\nu\gamma_\perp^\mu)^{\alpha\delta}\right) \nn\\
  && \Bigg(\theta(x-y_3)\frac{\bar x^2}{\bar y_3^3}(2\bar x-(2+x) \bar y_3)  +\theta(y_3-x)\frac{\bar xx(\bar x-\bar y_3-x\bar y_3)}{y_3 \bar y_3^2}\Bigg) t^b_{kj} (t^bt^a)_{il} \nps^{\beta\gamma} \nn\\
 && - (y_1,\beta,j\leftrightarrow y_3,\delta,l) \,,
\eea
\bea
  \lefteqn{ I_{aijkl}^{\mu\nu\alpha\beta\gamma\delta}(x,y_1,y_2)|_{(d,vii)} } \nn\\
  &=&  \frac12 x\bar x^2\left(\frac{2}{x}\delta^{\alpha\delta} g_\perp^{\mu\nu}+\frac{1}{\bar x}(\gamma_\perp^\nu\gamma_\perp^\mu)^{\alpha\delta}\right)  t^b_{kj} (t^at^b)_{il}\, \frac{1}{\bar y_3^2} \nps^{\beta\gamma} - (y_1,\beta,j\leftrightarrow y_3,\delta,l) \,.
\eea

\section{Anomalous dimension of hermitian conjugated operators}
\label{app:cc}

In this appendix we collect rules for translating the anomalous dimensions given in the main text to operators for which \emph{all} fermion building 
blocks $\chi_{i_l}=W_i^\dag\xi_i(t_{i_l}\nnp{i})$ are 
replaced by $\bar\chi_{i_l}=\bar\xi_i W_i(t_{i_l}\nnp{i})$, and vice versa, 
for $1\leq l\leq n_i$ and all collinear directions $1\leq i\leq N$. 
We comment later on the case when only building blocks belonging to a 
single collinear direction are flipped. 
For a general $N$-jet operator $J_P(x)$ (in collinear momentum space) we 
denote the corresponding operator transformed in this way by $J_{\bar P}(x)$.
Note that  gluon building blocks ${\cal A}_{\perp i}$ and the sign of partial 
derivatives $i\partial_{\perp i}$ are \emph{not} changed. We also 
leave the ordering of (fermionic) operators the same. 
For example, for an operator $J_P=J_{{\cal A}_1\partial\chi_1}^{B2}(x)\times\chi_2\times \prod_{j>2}{\cal A}_j$ with gluonic leading-power operators 
in directions $j=3,4,\dots,N$, a fermionic leading power building block in 
direction $2$, and the $\lambda^2$-suppressed operator 
$J_{{\cal A}_1\partial\chi_1}^{B2}(x)={\cal A}_{\perp 1_1}^\mu(x) 
i\partial_{\perp 1}^\nu\chi_{1_2}(\bar x)$ in direction $1$, one has
$J_{\bar P}=J_{{\cal A}_1\partial\bar\chi_1}^{B2}(x)\times\bar\chi_2\times 
\prod_{j>2}{\cal A}_j$, where $J_{{\cal A}_1\partial\bar\chi_1}^{B2}(x)
={\cal A}_{\perp 1_1}^\mu(x) i\partial_{\perp 1}^\nu\bar\chi_{1_2}(\bar x)$.
For the moment we consider both $J_P$ and $J_Q$ to be current operators, 
and comment on operators containing time-ordered products further below.

The renormalization factor $\delta Z_{\bar P\bar Q}(x,y)$ is obtained 
from $\delta Z_{PQ}(x,y)$ by the operations summarized below. The same rules 
apply to the corresponding anomalous dimensions $\Gamma_{\bar P\bar Q}(x,y)$ 
and $\Gamma_{PQ}(x,y)$.
These relations can be obtained by interpreting Eq.\,\eqref{eq:rencond} as an 
operator-valued
equation, and applying a hermitian conjugation with respect to the Dirac and 
colour indices of the fermionic building blocks (i.e. the colour indices 
referring to
the $3$ and $\bar 3$ representations of $SU(3)_c$). For the spin structure, 
one needs to multiply the resulting equation with $\gamma^0$ for each open Dirac
index and use $\gamma^0(\gamma^\mu)^\dag\gamma^0=\gamma^\mu$. In addition, 
one needs to take into account the sign factors and ordering of fermionic 
field operators in the relation between  $J_{\bar P}$ and $J_P^\dag$, and 
analogously for $J_{\bar Q}$, which leads to the factors of $(-1)$ appearing 
below.

\paragraph {Spin and space-time structure:}
\begin{itemize}
\item Reverse ordering within strings of $\gamma$ matrices acting on the same 
fermionic collinear building block (for any $m\geq 2$ and $i=1,\dots,N$), 
\be
\gamma_{\perp i}^{\mu_1}\gamma_{\perp i}^{\mu_2}\cdots\gamma_{\perp i}^{\mu_m} \to \gamma_{\perp i}^{\mu_m}\cdots\gamma_{\perp i}^{\mu_2}\gamma_{\perp i}^{\mu_1}\,.
\ee
Note that when leaving Dirac indices implicit in the anomalous dimension, 
the products of $\gamma$ matrices are understood to act from the left on 
$\chi_{i_l}$ and from the right on $\bar\chi_{i_l}$, as usual. As mentioned 
above, the case $m>2$ could be reduced to linear combinations of terms with 
$m\leq 2$. However, in our results, we find it more convenient to keep also 
terms with $m>2$.
\item Factor of $(-1)^{d_P+d_Q}$, where $d_P$ denotes the number of partial derivatives $i\partial_{\perp j}$
contained in $J_P$ (for $j=1,\dots,N$), and $d_Q$ in $J_Q$. This factor arises because of our convention for the operator basis, and because
e.g. $i\partial_{\perp i}\bar\chi_{i_l}=-(i\partial_{\perp i}\chi_{i_l})^\dag\gamma_0$.
\item Additional factor of $(-1)^{a_P+a_Q}$, where $a_P$ denotes the number of fermion anticommutations required to bring the fermionic building blocks
in $J_P^\dag$ into the same order as in $J_{\bar P}$ (we use an operator basis in which building blocks appear in ascending order with respect to the
collinear directions $i=1,\dots,N$, and with respect to the building block labels $i_1,i_2,\dots$ in each direction).
\end{itemize}
For example, for the collinear contributions $\gamma^i_{PQ}$ to the anomalous dimension for $F_i=1$ at order $\lambda^2$ (see Eq.\,\eqref{sec:colllambda2}),
one has $d_P=1, a_P=0, d_Q=0, a_Q=0$ for $B$-to-$C$ mixing with $P={\cal A}_{i_1}^\mu i\partial_{\perp i}^\nu \chi_{i_2}$ and 
$Q={\cal A}_{i_1}^\mu{\cal A}_{i_2}^\nu\chi_{i_3}$, contributing a factor of $(-1)$ to $\gamma^i_{\bar P\bar Q}$, 
where $\bar P={\cal A}_{i_1}^\mu i\partial_{\perp i}^\nu\bar\chi_{i_2}$ and 
$\bar Q={\cal A}_{i_1}{\cal A}_{i_2}\bar\chi_{i_3}$. For the same $P$ but $Q=\chi_{i_1}\bar\chi_{i_2}\chi_{i_3}$ one has 
$d_P=1, a_P=0, d_Q=0, a_Q=3$, such that the factors of $(-1)$ due to the derivative and the fermion reordering compensate each other (note
that $\bar Q=\bar\chi_{i_1}\chi_{i_2}\bar\chi_{i_3}$). 
For all other contributions in Eq.\,\eqref{sec:colllambda2}, both $d_P+d_Q$ and $a_P+a_Q$ are even.

\paragraph {Colour structure:} The colour structure can be expressed either in 
terms of the usual $SU(3)_c$ generators $t^a=\frac12\lambda^a$ with Gell-Mann
$3\times 3$ matrices $\lambda^a$ (acting on the colour index of fermionic 
building blocks), as well as factors of $if^{abc}$, 
or alternatively using colour space operators. In the first case:
\begin{itemize}
\item Reverse ordering of generators acting on the same fermionic collinear 
building block (for any $m\geq 2$), 
\be
t^{a_1}t^{a_2}\cdots t^{a_m} \to t^{a_m}\cdots t^{a_2}t^{a_1}\,.
\ee
As for Dirac indices, the corresponding $3\times 3$ matrices are understood 
to act from the left on $\chi_i$ and from the right on $\bar\chi_i$.
\item Sign flip $if^{abc}\to -if^{abc}$.
\end{itemize}
Products of generators can equivalently be reduced to a linear combination of 
the $t^c$ and the $3\times 3$ unit matrix using (iteratively) the relation 
\be\label{eq:tatb}
t^at^b=if^{abc}t^c+\frac13\delta^{ab}+d^{abc}t^c\,.
\ee
In this case the replacement rules are $if^{abc}\to -if^{abc}$ and 
$d^{abc}\to d^{abc}$.

Colour operator notation for the collinear building block $l$ in direction $i$ 
can be obtained by using $t^a\chi_{i_l}= -\T_{i_l}^a\chi_{i_l}$, 
$\bar\chi_{i_l}t^a=\T_{i_l}^a\bar\chi_{i_l}$, 
$-if^{abc}{\cal A}_{\perp i_l}^{\mu c}=\T_{i_l}^b{\cal A}_{\perp i_l}^{\mu a}$.
When several generators act on the same fermionic building block, using the 
second relation iteratively reverses the order. For example, 
$\bar\chi_{i_l}t^at^b=\T_{i_l}^b\T_{i_l}^a\bar\chi_{i_l}$. This compensates 
the change in the ordering due to hermitian conjugation.
Therefore, in colour operator notation, the ordering of colour operators in 
$\delta Z_{\bar P\bar Q}(x,y)$ and $\delta Z_{PQ}(x,y)$ is identical.
However, the definition of the colour operator contains a different sign for 
$\chi_{i_l}$ and $\bar\chi_{i_l}$, respectively. When 
$\delta Z_{\bar P\bar Q}(x,y)$ is expressed in colour operator notation, this 
leads to a sign flip for every colour space operator acting on fermionic 
fields. In addition, for the (hermitian) gluonic building block, the sign 
flip $if^{abc}\to -if^{abc}$ obtained from hermitian conjugation is inherited 
by the corresponding result expressed in terms of colour space operators 
$\T_{i_l}^a$. Finally, the colour operator ${\bf D}_{i_l}^a$ acts only on 
gluonic building blocks, and is defined by 
$d^{abc}{\cal A}_{\perp i_l}^{\mu c}=
{\bf D}_{i_l}^b{\cal A}_{\perp i_l}^{\mu a}$ \cite{Beneke:2017ztn}.
Inspection of Eq.~\eqref{eq:tatb} shows that, when translating to colour 
operator space notation, also this operator is odd. Therefore, the rules from 
above translate to colour operator notation
in the following way:
\begin{itemize}
\item \emph{No} change in ordering of colour operators.
\item Sign flip $\T_{i_l}^a\to -\T_{i_l}^a$ for all colour space operators acting on both fermionic and gluonic building blocks.
\item Sign change of the symbol 
$(\T_{i_1}\times\T_{i_2})^a=if^{abc}\T_{i_1}^b\T_{i_2}^c 
\to -(\T_{i_1}\times\T_{i_2})^a $ 
introduced in Ref.~\cite{Beneke:2017ztn}.
\item Sign change for the colour operator ${\bf D}_{i_l}^a\to -{\bf D}_{i_l}^a$.
\item Sign flip $if^{abc}\to -if^{abc}$ for remaining factors of $if^{abc}$ 
that have not been expressed in terms of colour space operators.
\end{itemize}
The rules given above can be used in order to translate the collinear contributions $\gamma^i$ for operators with $F_i=1,2,3$ to those with $F_i=-1,-2,-3$.
Note that, since $\gamma_{PQ}^i$ does not depend on collinear directions $j\not= i$, the rules from above can also be used to
obtain $\gamma^i_{\bar P\bar Q}$ from $\gamma^i_{PQ}$ for the case in which $\bar P (\bar Q)$ is related to $P(Q)$ by flipping the fermion number
in direction $i$ \emph{only}, while leaving the contributions to the $N$-jet operator from all other directions untouched.

For the soft contributions $\gamma^{ij}_{PQ}$ we need to consider operators $J_P$ involving time-ordered 
products with insertions of the power-suppressed Lagrangian  $i{\cal L}^{(1)}$ and $i{\cal L}^{(2)}$. 
We define the operator $J_{\bar P}$ analogously as above, i.e. by replacing  \emph{all} fermionic building blocks
$\chi_{i_l}\to\bar\chi_{i_l}$, but no changes otherwise. Due to the time-ordered product $J_{\bar P}$ \emph{cannot} be related
to $J_{P}^\dag$ up to sign factors, since hermitian conjugation would turn time- into anti-time-ordering.
Nevertheless, we find that $\gamma^{ij}_{\bar P\bar Q}$ can be obtained from $\gamma^{ij}_{PQ}$ using the \emph{identical} set of rules as for
the current-current mixing given above.\footnote{This can be understood when using an alternative description of operator mixing, based on the
``interaction picture'' for which also the power-suppressed contributions to the Lagrangian are kept in the time-evolution operator.
This is not our preferred option because it allows current operators to mix into currents with a higher level of power suppression.}

For the soft contributions $\gamma^{ij}_{PQ}$ for $F_i=F_j=1$ summarized in Eq.\,\eqref{eq:gammasoft}, the rules from above can therefore be used to
obtain the anomalous dimension for $F_i=F_j=-1$. 
For example, all entries of $\gamma^{ij}_{PQ}$ with $P=\left(J_{\chi,\xi}^{T1}\right)_i\left(J_{\chi,\xi}^{T1}\right)_j$ can be obtained by applying the replacement
rules to Eq.\,\eqref{eq:L1L1toA1A1}, giving the operator mixing
\bea
\left(J_{\bar\chi_{\alpha},\xi}^{T1}\right)_i 
\left(J_{\bar\chi_{\beta},\xi}^{T1}\right)_j 
&\to& - \frac{2\alpha_s}{\pi\epsilon} \,G_{\lambda\kappa}^{ij}  
\nn\\
&& \times \left[ \T_i^a J^{A1}_{\partial^\lambda\bar\chi_\alpha} 
-\frac12 \int dy  \left(\gamma_{\perp i}^\mu\gamma_{\perp i}^\lambda
+\frac{\gamma_{\perp i}^\lambda\gamma_{\perp i}^\mu}{\bar y}\right)_{\gamma\alpha
}\T_i^b\T_i^a J^{B1}_{{\cal A}_{\mu }^b\bar\chi_{\gamma}}(y) \right]_i 
\nn\\
&& \times \left[\T_j^a J^{A1}_{\partial^\kappa\bar\chi_\beta} 
-\frac12 \int dy'  \left(\gamma_{\perp j}^\nu\gamma_{\perp j}^\kappa
+\frac{\gamma_{\perp j}^\kappa\gamma_{\perp j}^\nu}{\bar{y}'}
\right)_{\delta\beta}\T_j^c\T_j^a J^{B1}_{{\cal A}_{\nu }^c\bar\chi_{\delta}}(y') \right]_j \,.
\nn\\
\eea
Recall that in our convention $J^{A1}_{\partial^\lambda\bar\chi_\alpha}=
+i\partial_{\perp i}^\lambda\bar\chi_\alpha$. Compared to 
Eq.~\eqref{eq:L1L1toA1A1}, the order of gamma matrices is reversed, and the 
Dirac indices are flipped, while all factors of $(-1)$ cancel out.

For the case $F_i=1$, $F_j=-1$ we find that the anomalous dimension 
$\gamma^{ij}$ can be extracted from Eqs.~\eqref{eq:L1L1toA1A1} to 
\eqref{eq:L1YML1YM} in a straightforward way by applying the rules from above 
to the contributions to the loop amplitude related to direction $j$ only, given by the last lines in each of these equations, respectively.
This can be traced back to the ``factorization'' of soft loops into contributions from directions $i$ and $j$, respectively, as discussed
in the main text. However, we note that the cusp part of $\Gamma_{PQ}(x,y)$ 
in the first line of Eq.~\eqref{eq:GammaPQ} is not part of the 
$\gamma^{ij}$ above, and the first line of Eq.~\eqref{eq:GammaPQ} 
applies to all 
$N$-jet operators, involving arbitrary combinations of building blocks 
containing both $\chi$ and $\bar\chi$ as well as gluonic fields.

\bibliography{paper}

\end{appendix}

\end{document}